\makeatletter
\def\input@path{{/Users/fladias/Desktop/Report//}}
\makeatother
\documentclass[11pt,english]{spie}
\usepackage[T1]{fontenc}
\usepackage[latin9]{inputenc}
\usepackage{geometry}
\usepackage{float}
\usepackage{amsmath}
\usepackage{graphicx}

\makeatletter

\providecommand{\tabularnewline}{\\}

\usepackage{babel}
\makeatother

\begin{document}

\title{LO vs NLO comparisons for $Z+jets$: MC as a tool for background
determination for NP searches at LHC}

\maketitle
\noindent \begin{center}
Flavia de Almeida Dias%
\footnote{flavia.dias@cern.ch%
}
\par\end{center}

\begin{center}
Instituto de Fisica Teorica, Universidade Estadual Paulista
\par\end{center}

\noindent \begin{center}
Emily Nurse%
\footnote{nurse@hep.ucl.ac.uk%
}
\par\end{center}

\begin{center}
Dept. of Physics and Astronomy, University College London
\par\end{center}

\noindent \begin{center}
Gavin Hesketh%
\footnote{gavin.hesketh@cern.ch%
}
\par\end{center}

\begin{center}
Dept. of Physics and Astronomy, University College London
\par\end{center}

\begin{abstract}
The leptonic decays of the heavy gauge bosons \emph{W} and/or \emph{Z}
provide a clear experimental signature at hadron colliders. The production
of accompanying jets is an excellent signal to probe QCD, while also
being the main background to many searches for new physics. Describing
the complex final state of \emph{W} or \emph{Z} + jets is a theoretical
challenge with most existing calculations combining matrix elements
for high energy jet production with a parton shower for lower energy
jet production. We focus on two models: ${\tt SHERPA}$, which uses
Leading Order matrix elements for boson and jet production; and ${\tt POWHEG}$
with ${\tt HERWIG++}$, which uses a Next-To-Leading Order Matrix
element for \emph{Z} production. In order to isolate the impact of
the matrix elements for jet production, it is first essential to constrain
the differences in the rest of the calculation in each case: specifically,
the Multiple Parton Interaction models, and the tuning of the
parton shower interfaced to the matrix elements. We test all three
aspects of these models against data from the Tevatron, and perform
a study of some basic kinematic variables at the LHC energy.

\newpage{}
\end{abstract}

\section{Introduction}

The leptonic decays of the heavy gauge bosons $W$ and/or $Z$ accompanied
by jets, $W\to\ell\nu+\text{jets}$ and $Z\to\ell\ell+\text{jets}$,
offer search channels for new interactions of particles in high energy
collisions. Many extensions of the Standard Model predict new particles
with electroweak (EWK) couplings that decay into the SM gauge bosons
$W$, $Z$, and $\gamma$, accompanied by jets. Any production of new heavy particles with quantum
numbers conserved by the strong interaction and EWK couplings is likely
to contribute to signatures with one or more EWK gauge bosons; additional
jets will always be present at some level from initial-state radiation,
and may also be products of cascade decays of new heavy particles.
$W+N\;\text{jets}$ and $t\bar{t}+\text{jets}$ are backgrounds to
most supersymmetry searches in final states with leptons, jets and
missing transverse energy. The $Z(\to\nu\bar{\nu})+\;\text{jets}$ signal
is an irreducible background to inclusive hadronic searches of Dark
Matter, based on jets and missing energy (MET). W or Z + jets is also a major background for Higgs searches.

In the context of the SM, the study of the production of electroweak
bosons with $N\;\text{jets}$ allows for tests of perturbative QCD.
The production cross section scales approximately with the strong
coupling constant for each additional jet. While current theoretical
predictions at leading order (LO) and next to leading order (NLO)
are in good agreement with data at the Tevatron, comparison at the
higher energy of the LHC, and at higher jet multiplicities are needed.
Inclusive and differential cross sections access the parameters of
the perturbative expansion, such as the renormalization and the factorization
scales, as well as the parton density functions (PDFs), through
the $p_{T}$ and $\eta$ distributions of the vector bosons and the
associated jets. Predicting all these quantities and comparing them
to the Tevatron data has already produced several improvements in
the calculation and generation techniques, such as the introduction
of generators based on the LO calculations of matrix-element (ME)
for the associated jet production, and the definition of several matching
procedures to the parton-shower generators.

In this paper we make a comparative study of ME+PS Monte Carlo generators
(${\tt SHERPA}$ and ${\tt HERWIG++}$) with an NLO calculation
 (${\tt POWHEG}$), validated with Tevatron D0 and CDF
data, and generated at the LHC energy. This will serve as a preparatory
MC study of the major SM backgrounds for the regions that are relevant
to New Physics searches.

\section{Matrix Elements Corrections / Parton Shower}

The Parton Shower (PS) technique is a collinear approximation of the
description of parton splittings in QCD radiation that accompanies
hard scattering processes.

However, although it provides a good description
of low $p_{\perp}$ observables, it usually fails to fill the phase
space (can't emit anything above the factorisation scale by construction) and in the hard region (close to the factorisation scale) this approximation is simply no good anymore. One way to improve the description of kinematical
observables is to add a Matrix Element (ME) correction for the extra
emissions. This can be implemented in different ways, as will be described
in each generator section. 

\subsection{${\tt SHERPA}$}

The ${\tt SHERPA}$ generator uses the improved CKKW Matrix Element
- Parton Shower merging {[}1] that relies on a separation of the event
phase space in two separated regions, defined by a choice of scale.
Above the chosen scale, all radiation must be produced by the Matrix
Elements, and below the chosen scale all radiation produced by the
Parton Shower. In matching these two regions to provide full phase
space coverage, overlaps such as a parton shower emission in the scale
range covered by the Matrix Element, must be removed {[}2]. 

The basic steps in the algorithm implementation are {[}2]:

\begin{itemize}
\item Events are generated based on the matrix elements for \emph{W/Z} +
k jets, where k = 0, ... N. The Matrix Element jets are required to
be above the merging scale, $Q_{cut}$.
\item The event is then reconstructed back from the final state particles.
Particles are combined in the most likely combinations according to
the parton shower probabilities. 
\item Events are then passed to the parton shower, which is allowed to generate
radiation from any part of the process.
\item The shower is then analyzed, and any events in which parton shower produces an emission above the scale $Q_{cut}$ are vetoed, leading to a rejection of the full event. This process is called Sudakov rejection.
\end{itemize}

The ${\tt SHERPA}$ generator automatizes the generation of inclusive
samples, combining ME for different parton multiplicities with PS
and hadronization. Some parameters related to the ME and PS calculations
have to be set accordingly.

\subsection{${\tt HERWIG++}$}

The aim of the ME corrections in the PS on ${\tt HERWIG++}$ is to
correct for two deficiencies in the shower algorithm: to populate
the uncovered region of high $p_{\perp}$ in the phase space (non-soft
non-collinear), and to correct the populated region, where the extrapolation
away from the soft and collinear limits is not perfect. These corrections
are called the hard and soft matrix element corrections respectively
{[}3].

\subsubsection*{Soft Matrix Element Corrections}

The soft correction is derived by comparing the probability density
that the $i^{th}$ resolvable parton is emitted into a region of the
phase space in the PS approximation (quasi-collinear limit), and the
exact ME calculation. A simple veto algorithm is then be applied to
the parton shower to reproduce the matrix element distribution, which
relies on there always being more parton shower emissions than matrix
element emissions. This is ensured simply by enhancing the emission
probability of the parton shower with a constant factor {[}3].

The correction is applied to the hardest emission so far in the shower,
to ensure that the leading order expansion of the shower distribution
agrees with the leading order matrix element, and that the hardest
(i.e. furthest from the soft and collinear limits) emission reproduces
it {[}4].

\subsubsection*{Hard Matrix Element Corrections}

The hard ME corrections aim to populate the high $p_{\perp}$ region
that the PS leaves uncovered. This domain of the phase space should
have radiation distributed according to the exact tree level ME for
this extra emission, and as the Parton Shower does not populate this
region, a different approach is needed to achieve this. Prior to any
showering, the algorithm checks if the required ME is available for
the hard process. Then, a point is generated in the appropriate region
of phase space, with a probability based on a sampling of the integrand.
The differential cross section associated with this point is then
calculated and multiplied by a phase space volume factor, giving the
event weight. The emission is accepted if the weight is less than
a uniformly distributed random number in the {[}0,1] interval, and
the momenta of the new parton configuration is processed by the shower
as normal {[}3].

\section{Next to Leading Order Methods}

Both ${\tt SHERPA}$ and ${\tt HERWIG++}$ described above include
dominant QCD effects with leading order matrix elements combined with
a leading logarithmic parton shower. However, higher order calculations
are required to match the precision of current data measurements.
Going even one step beyond the leading order is already a complex
task: the initial hard process should be implemented in NLO; and shower
development would have to be improved in next to leading logarithmic
accuracy in collinear and soft structure. An intermediate step has
already been developed: keeping the leading logarithmic order for
the shower approximation, while improving the treatment of the hard
emission to NLO accuracy (NLO+PS approach). We test one such generator:
${\tt POWHEG}$.

\subsection{${\tt POWHEG}$}

In the ${\tt POWHEG}$ (Positive Weight Hardest Emission Generator)
formalism, the generation of the hardest emission is performed first,
using full NLO accuracy, and using the ${\tt HERWIG++}$ parton shower
to generate subsequent radiation. The ${\tt POWHEG}$ cross section
for the generation of the hardest event has the following properties:

\begin{itemize}
\item At large $k_{T}$ (momentum of incoming particle) it coincides with
the pQCD cross section up to (O$(\alpha_{S}^{2})$) terms. 
\item It reproduces correctly the value of infrared safe observables at
the NLO. Thus, also its integral around the small $k_{T}$ region
has NLO accuracy. 
\item At small $k_{T}$ it behaves as well as standard shower Monte Carlo
generators. 
\end{itemize}
The ${\tt POWHEG}$ formula can be used as an input to any parton
shower program to perform all subsequent (softer) showers and hadronization.
However, as $k_{T}$ is used to define the matrix element scale, the
parton shower must also be ordered in $p_{\perp}$. The shower is
then initiated with an upper limit on the scale equal to the $k_{T}$
of the ${\tt POWHEG}$ event, and fills in all radiation below that
scale (a truncated shower). For the case of a virtuality or angular
ordered shower, emissions at higher $k_{T}$ may be produced, and
must subsequently be vetoed to avoid double counting with the ME emission:
a vetoed truncated shower, which is not possible with current parton
shower programs like ${\tt PYTHIA}$ and ${\tt HERWIG}$. We point
out, however, that the need of vetoed truncated showers is not specific
to the ${\tt POWHEG}$ method. It also emerges naturally when interfacing
standard matrix element calculations with parton shower. At present,
there is no evidence that the effect of vetoed truncated showers may
have any practical importance {[}5].

The ${\tt POWHEG}$ method solves the problem of negative event weights
that arise in other NLO methods, such as in ${\tt MC@NLO}$. It also
defines how the highest $p_{\perp}$ emission may be modified to include
the logarithmically enhanced effects of soft wide-angle radiation.
In the ${\tt POWHEG}$ framework, positive weight events distributed
with NLO accuracy can be showered to resume further logarithmically
enhanced corrections by {[}3]:

\begin{itemize}
\item Generating an event according to the ${\tt POWHEG}$ formula; 
\item Hadronizing non-radiating events directly;
\item Mapping the radiative variables parametrizing the emission into the
evolution scale, momentum fraction and azimuthal angle, from which
the parton shower would reconstruct identical momenta; 
\item Evolve, using the original LO configuration, the leg emitting the
extra radiation from the default initial scale, determined by the
colour structure of the N-body process, down to the hardest emission
scale such that the $p_{\perp}$ is less than that of the hardest
emission, the radiation is angular-ordered and branchings do not change
the flavour of the emitting parton; 
\item When the evolution scale reaches the hardest emission scale, insert
a branching with parameters into the shower; 
\item From all external legs, generate $p_{\perp}$ vetoed showers. 
\end{itemize}
This procedure allows the generation of the truncated shower with
only a few changes to the normal ${\tt HERWIG++}$ shower algorithm
{[}3].

\section{Comparisons to Tevatron Data}

The inclusion of Matrix Elements is only one component of the simulation.
In order to describe real data, other aspects
of the generators must also be accurate: the Parton Shower (PS), and
the model of Multiple Parton Interactions (MPI). The choice of 
Parton Distribution Function (PDF)
may also play a role, or at least be highly coupled to the tuning
of these phenomenological models. Finally, there are some settings
unique to the generators themselves, such as the choice of matching
scale between the parton shower and matrix elements. So, in order
to isolate the impact of using LO or NLO matrix elements, we must
also constrain all the other aspects of these models.

\subsection{Selection of Events and Kinematic Cuts}

We use the most recent data from the Tevatron experiments, CDF and
D0, to test the generator performance. The Tevatron is a proton anti-proton
collider with a center of mass energy $\sqrt{s}$ =1.96 TeV, located
at Fermilab, USA.

For each of the analyses, Monte Carlo events are selected according to the corresponding
data selection, described in their papers. But, in general, the lepton
pair invariant mass is required to be between some mass range (around the
\emph{Z} mass peak) to enhance the contribution of pure \emph{Z} exchange
over $\gamma*$ exchange and \emph{Z}/$\gamma*$ interference terms,
and pseudorapidity cuts on the (CDF or D0) detector acceptance.

The generators were configured to produce inclusive \emph{Z}/$\gamma*$
particles decaying into lepton pairs (electrons or muons), with the
invariant mass constraint. Also some generator parameters and inputs are changed
in order to obtain the best description of the data: the intrinsic momentum of the partons
in the incoming (anti)protons (${\tt K\text{\_}PERP}$), MPI model parameters and
the input PDFs.

The comparison plots were made using the ${\tt RIVET}$ framework
{[}6], version 1.2.1. The ${\tt RIVET}$ project (Robust Independent
Validation of Experiment and Theory) allows validation of Monte Carlo
event generators. It provides a set of experimental standard validated analyses
useful for generator validation and tuning, as well as a convenient infrastructure
for adding user's own analyses.

\subsection{\emph{Z} Transverse Momentum}

The main benefit to using \emph{Z} events to probe the underlying
process is that the \emph{Z} can be fully and unambiguously reconstructed.
The \emph{Z} $p_{\perp}$ is generated by the momentum balance against
initial state radiation (ISR) and the primordial/intrinsic $p_{\perp}$
of the \emph{Z}'s parent partons in the incoming hadrons. Within an
event generator, this recoil may be generated by a hard matrix element
at high $p_{\perp}$, or by the parton shower or intrinsic momentum of the partons
at low $p_{\perp}$. The inclusive \emph{Z} $p_{\perp}$ is
therefore an excellent first test for the generators, before looking
at more exclusive \emph{Z} + jet final states. 

We look at three measurements of the \emph{Z} $p_{\perp}$: the Run
I measurement by CDF, and two Run II measurements by D0, in the electron
and muon channels.

\paragraph{CDF Run I result}

The \emph{Z} $p_{\perp}$ analysis from CDF
Run I {[}7] is a measurement of the cross section as a function of
the transverse momentum of $e^{+}e^{-}$ pairs in the \emph{Z} boson
mass region of $66\;\text{GeV}<m_{ee}<116\;\text{GeV}$ from $p\bar{p}$
collisions at $\sqrt{s}$ = 1.8 TeV, with the measured lepton acceptance
extrapolated to $4\pi$ with no $p_{\perp}$ cut. The analysis is
also subject to ambiguities in the experimental \emph{Z} $p_{\perp}$
definition {[}7]. 

\begin{figure}[!h]
\begin{centering}
\includegraphics[width=0.5\columnwidth]{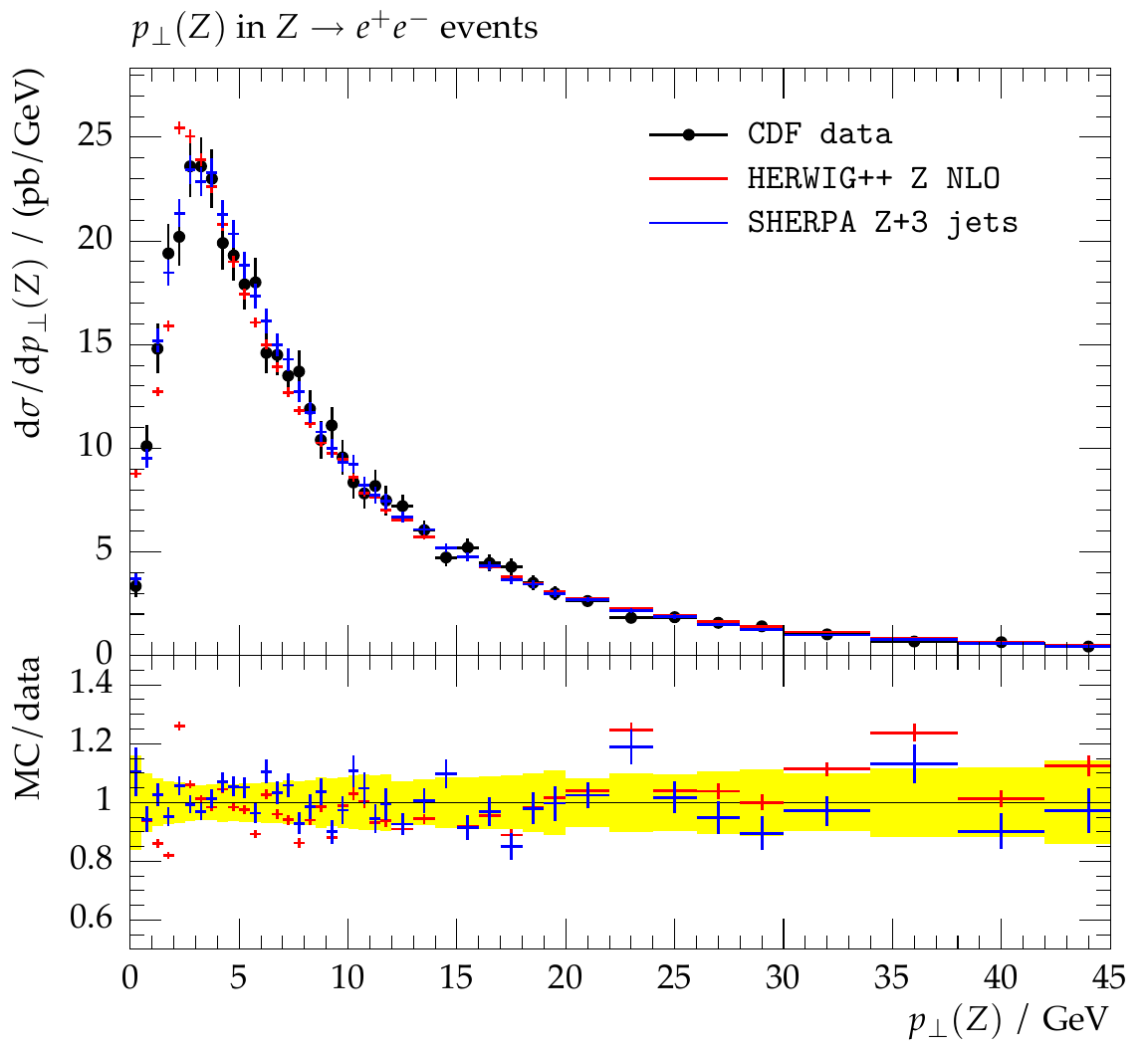}
\par\end{centering}

\caption{\label{fig:pTRunI}\emph{Z} $p_{\perp}$ analysis for Tevatron Run
I. The yellow band in the ratio plot correspond to the uncertainty
of the data.}

\end{figure}

Fig. \ref{fig:pTRunI} shows that the MC description for both
${\tt HERWIG++}$ \emph{Z} NLO generation in the ${\tt POWHEG}$ formalism
and ${\tt SHERPA}$ \emph{Z}+3 jets in the PS+ME formalism agree with
CDF data, inside the experimental uncertainties, except for the low
\emph{Z} $p_{\perp}$ region for ${\tt HERWIG++}$. 

\paragraph{D0 Run II results}

The D0 \emph{Z} ($\rightarrow e^{+}e^{-}$) $p_{\perp}$ analysis
{[}8] is based on $p\bar{p}$ collisions at $\sqrt{s}$ = 1.96 TeV,
with a looser lepton pair mass cut of $40<m_{ee}<200$ GeV. The measured
differential spectrum is normalized to the total \emph{Z} cross section,
to reduce overall systematic uncertainties. The electrons were measured
in |$\eta$| < 1.1 or 1.5 < |$\eta$| < 2.5, with $p_{\perp}$ > 25
GeV. The result was extrapolated to $4\pi$ with no $p_{\perp}$ cut.

\begin{figure}[!h]
\includegraphics[width=0.5\columnwidth]{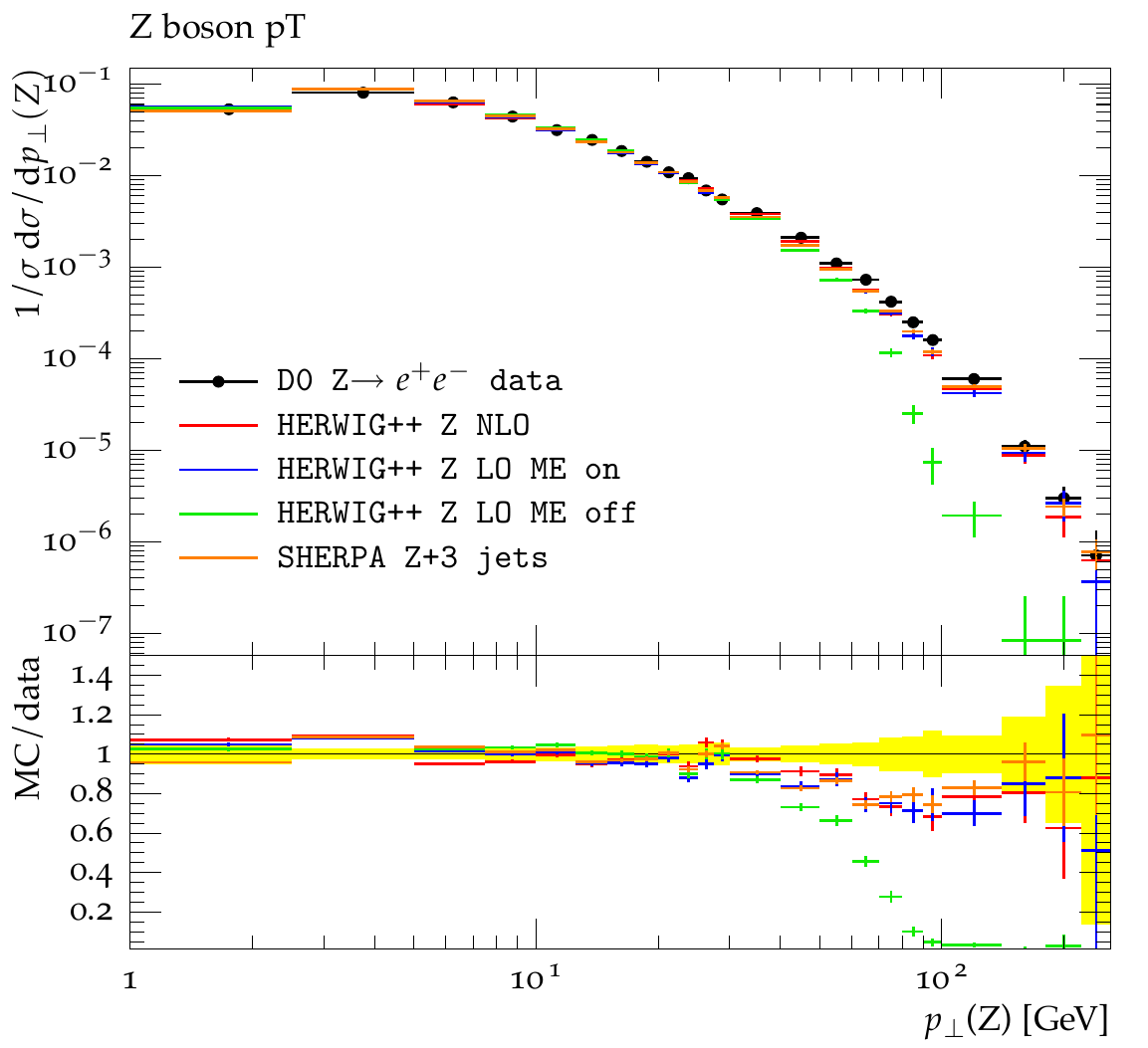}
\includegraphics[width=0.5\columnwidth]{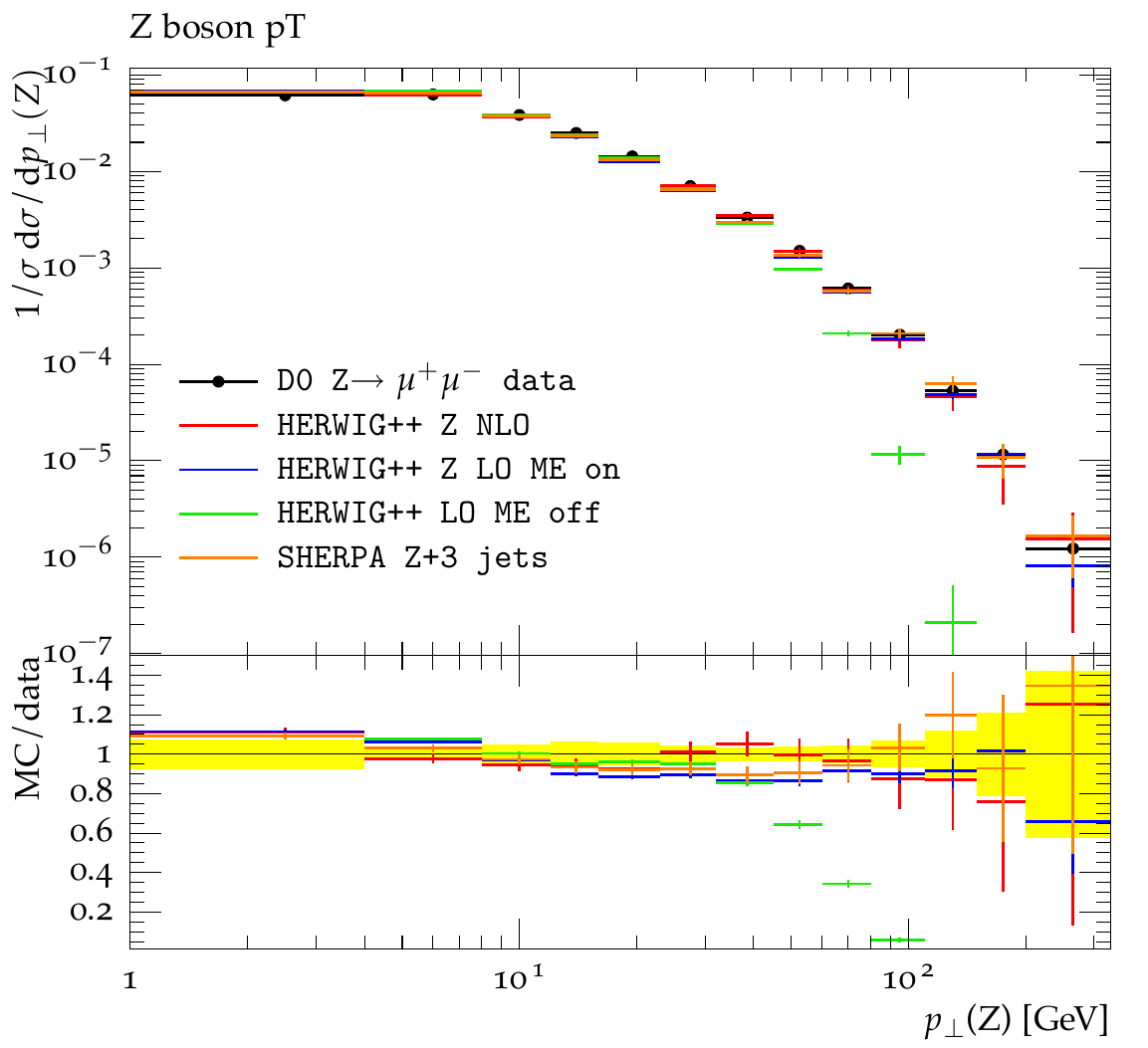}

\caption{\label{fig:NLO123}Comparison of boson transverse momentum for \emph{Z}
production at: NLO ${\tt HERWIG++}$ (${\tt POWHEG}$ formalism),
LO ${\tt HERWIG++}$ (ME corrections on and off) and LO ${\tt SHERPA}$
\emph{Z} + 3 jets, in the \emph{Z} $\rightarrow e^{+}e^{-}$ channel
(left) and \emph{Z} $\rightarrow\mu^{+}\mu^{-}$ channel (right).
Both plots are scaled by 1 over the total cross section,
to reduce systematic uncertainties. }

\end{figure}

We can see in Fig. \ref{fig:NLO123} (left) that the D0 analysis
in the electron \emph{Z} decay channel has a systematic behaviour
in the region of medium \emph{Z} $p_{\perp}$, for both ${\tt HERWIG++}$
\emph{Z} NLO (${\tt POWHEG}$ formalism), ${\tt HERWIG++}$ \emph{Z}
LO with ME formalism and ${\tt SHERPA}$ \emph{Z} + 3 jets. The NLO
plot shows slightly better behaviour in the mid region (30-60 GeV)
than the LO with ME corrections, but no significant differences
in the low region. The high $p_{\perp}$ region lacks the statistics
to make a detailed comparison.

The middle region suggests a problem either with the 
the Monte Carlo or something in the analysis. 
We therefore looked also at a new D0 \emph{Z} $p_{\perp}$ analysis {[}9] in the muon
channel. The muons were measured in |$\eta$| < 1.7 and $p_{\perp}$
> 15 GeV. This new analysis has as an important development: the definition
of the final observable is made at the level of particles entering the detector,
while previous measurements have applied theoretical factors correcting
for any  final state radiation and for the measured lepton
acceptance to full $4\pi$ coverage. This new approach minimises the dependence
on theoretical models, and therefore any biases in comparisons. The
differential cross section is normalized to the total \emph{Z} cross
section, to reduce overall systematic uncertainties, as in the electron
channel analysis {[}8]. 

We can see in Fig. \ref{fig:NLO123} (right) that in the muon
channel the discrepancy in the medium \emph{Z} transverse momentum
region is gone, and the ratio of MC to data does not show significant
systematic discrepancies for both generators.
So the Monte Carlo can better reproduce the analysis with the more
limited lepton acceptance and without the model dependent corrections.
For ${\tt HERWIG++}$ LO without ME corrections, the expected behaviour
is to produce fewer events in the high transverse momentum region,
as this is not fully populated by the PS formalism alone. 

\subsubsection{PDF and ${\tt K\text{\_}PERP}$ choice on ${\tt SHERPA}$}

For the sake of using the best parameters for the description of Tevatron
data, some other parameters were studied in the ${\tt SHERPA}$ generator:
the impact of the use of different PDFs and different values
for intrinsic transverse momentum of the partons within the hadrons
 (${\tt K\text{\_}PERP\text{\_}MEAN}$) and its gaussian 
 width (${\tt K\text{\_}PERP\text{\_}SIGMA}$). 

The same \emph{Z} ($\rightarrow\mu^{+}\mu^{-}$) $p_{\perp}$ analysis
{[}9] of the previous section is used, for the reasons discussed above.

\paragraph{Intrinsic Transverse Momentum of Partons Within the Hadrons}

\begin{figure}[!h]
\includegraphics[width=0.5\columnwidth]{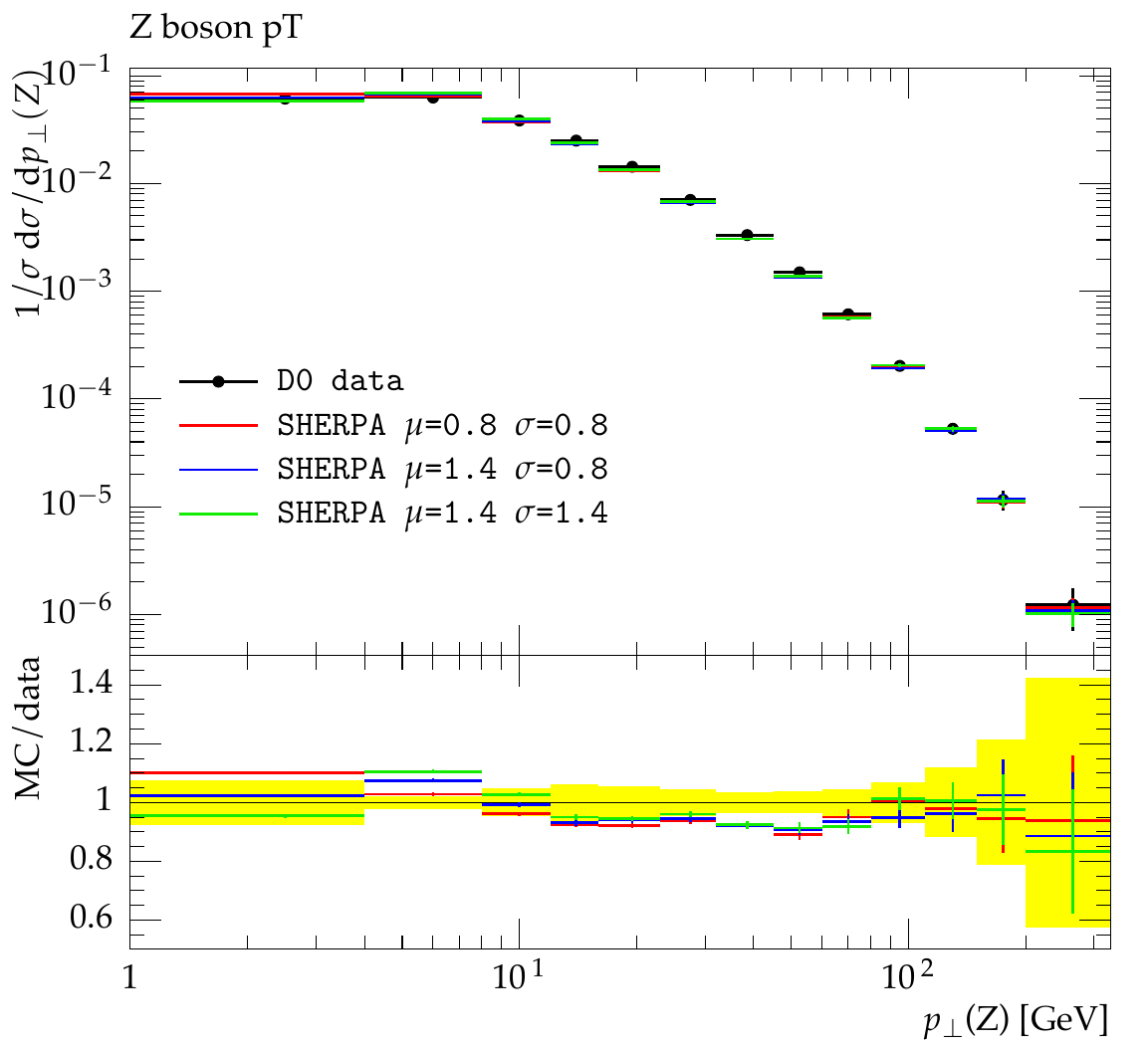}
\includegraphics[width=0.5\columnwidth]{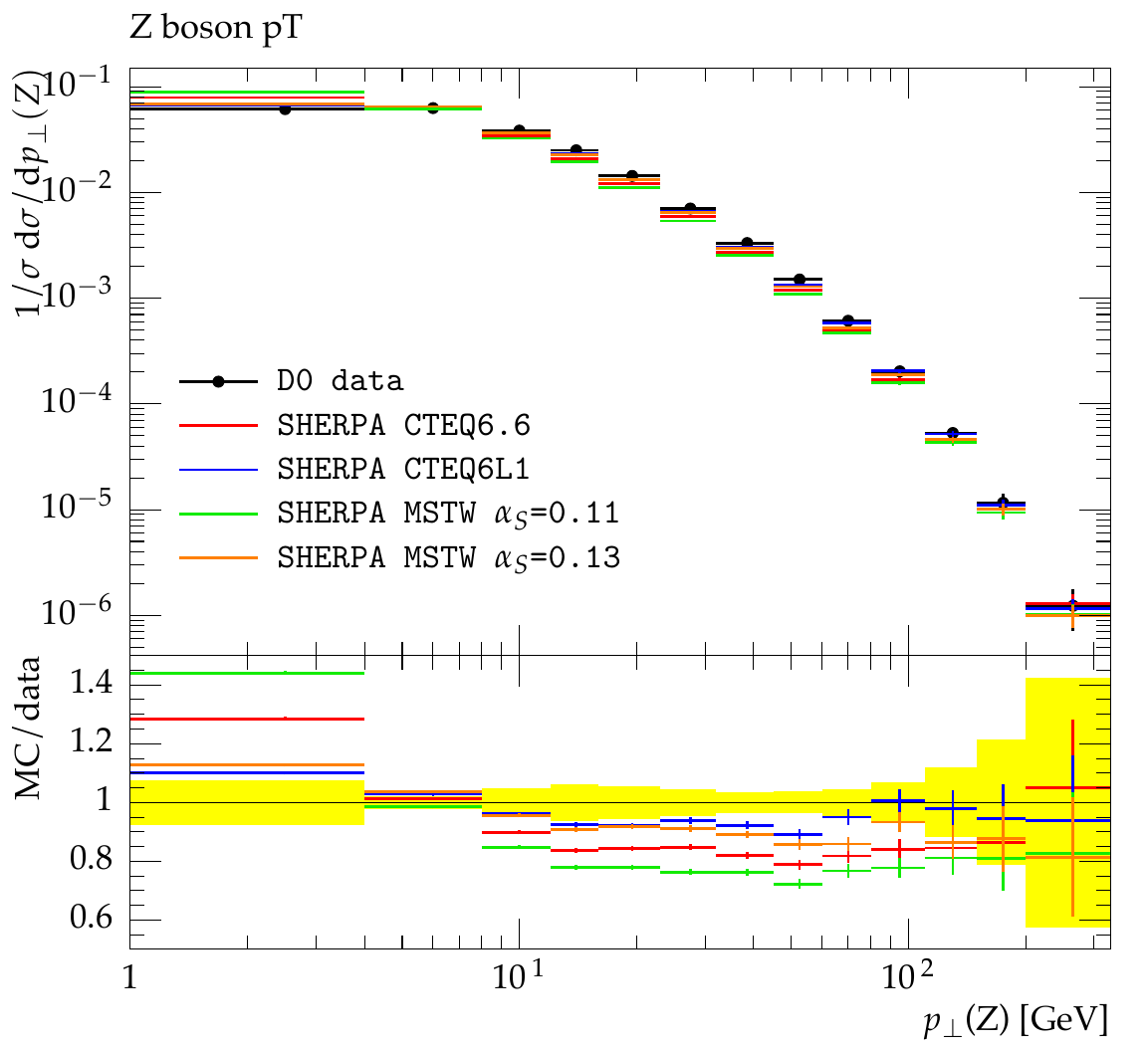}

\caption{\label{fig:KperpZpT}The \emph{Z} $p_{\perp}$ in ${\tt SHERPA}$
for several different parameters of ${\tt K\text{\_}PERP}$ and its
gaussian width (left), using ${\tt CTEQ6L1}$ PDF, and for
different PDFs (right), using default values of ${\tt K\text{\_}PERP}$
(0.8) and its width (0.8), equivalent to the blue curve on the left
graphic. In this plots, no MPI model was simulated. The plots
are scaled by 1 over the total cross section, to reduce systematic
uncertainties.}

\end{figure}

In Fig. \ref{fig:KperpZpT} (left) we show the transverse momentum
of the \emph{Z} boson in \emph{Z}+3 jets production with different
values for the transverse momentum of the partons within the hadrons.
We can see that the ${\tt K\text{\_}PERP\text{\_}MEAN}$ = 1.4 (the default value is 0.8)
and ${\tt K\text{\_}PERP\_SIGMA}$ = 0.8 (default value) shows a better
agreement in the lowest \emph{Z} $p_{\perp}$ bin.

\paragraph{PDF Set With Different $\alpha_{S}$ Values}

In order to show the impact of changing the PDF on the shape of
the \emph{Z} $p_{\perp}$, the analysis was run with the ${\tt MSTW08}$
{[}10] set, that is a set of PDFs fitted with different values
of the strong coupling constant in the \emph{Z} mass pole, $\alpha_{S}(M_{Z})=0.11,\;0.13$.
Fig. \ref{fig:KperpZpT} shows the change introduced by varying $\alpha_{S}$. 
These were compared to
two other PDFs: the default in ${\tt SHERPA}$, ${\tt CTEQ6.6}$,
and to ${\tt CTEQ6L1}$. The last provides a better description of
the \emph{Z} $p_{\perp}$ data, especially in the low transverse momentum
region. For ${\tt HERWIG++}$ with ${\tt POWHEG}$, a NLO PDF is necessary, and all the
simulations used the default one, ${\tt MRST'02\; NLO}$ {[}11].

\subsection{Multiple Parton Interactions}

With both generators providing a reasonable description of the \emph{Z}
$p_{\perp}$, we can study the rest of the event in more detail. However,
before looking at the jets recoiling against the \emph{Z}, we must
first constrain the other source of hadronic activity, MPI.

\subsubsection{Constraining UE/MPI}

To study the model of the MPI in each of the generators used,
the CDF underlying event analysis {[}12] was compared to both the ${\tt HERWIG++}$
and ${\tt SHERPA}$ generators. 

The analysis was made for Drell-Yan events with $Z/\gamma*\rightarrow e^{+}e^{-}$
and $Z/\gamma*\rightarrow\mu^{+}\mu^{-}$. A mass cut $ 70 < m_{ll} < 110$ GeV was applied at the generator level. The analysis
is based on the observation that the hard interaction in an event
typically falls along an axis, and activity from MPI is completely
uncorrelated with this axis. Each event is therefore decomposed into
regions in the azimuthal angle, $\phi$. The \char`\"{}toward\char`\"{}
region defined by the direction of the \emph{Z}, which is used to
set $\phi$ = 0. The opposite direction, the \char`\"{}away\char`\"{}
region, is then dominated by the recoil to the \emph{Z}. The \char`\"{}transverse\char`\"{}
regions, defined by 60 < $\phi$ < 120, generally have little activity
from the hard interaction, and so are most sensitive to MPI
and the underlying event. In the analysis, the transverse region is
defined for |$\eta$| < 1 {[}12].

\begin{figure}[!h]
\includegraphics[width=0.5\columnwidth]{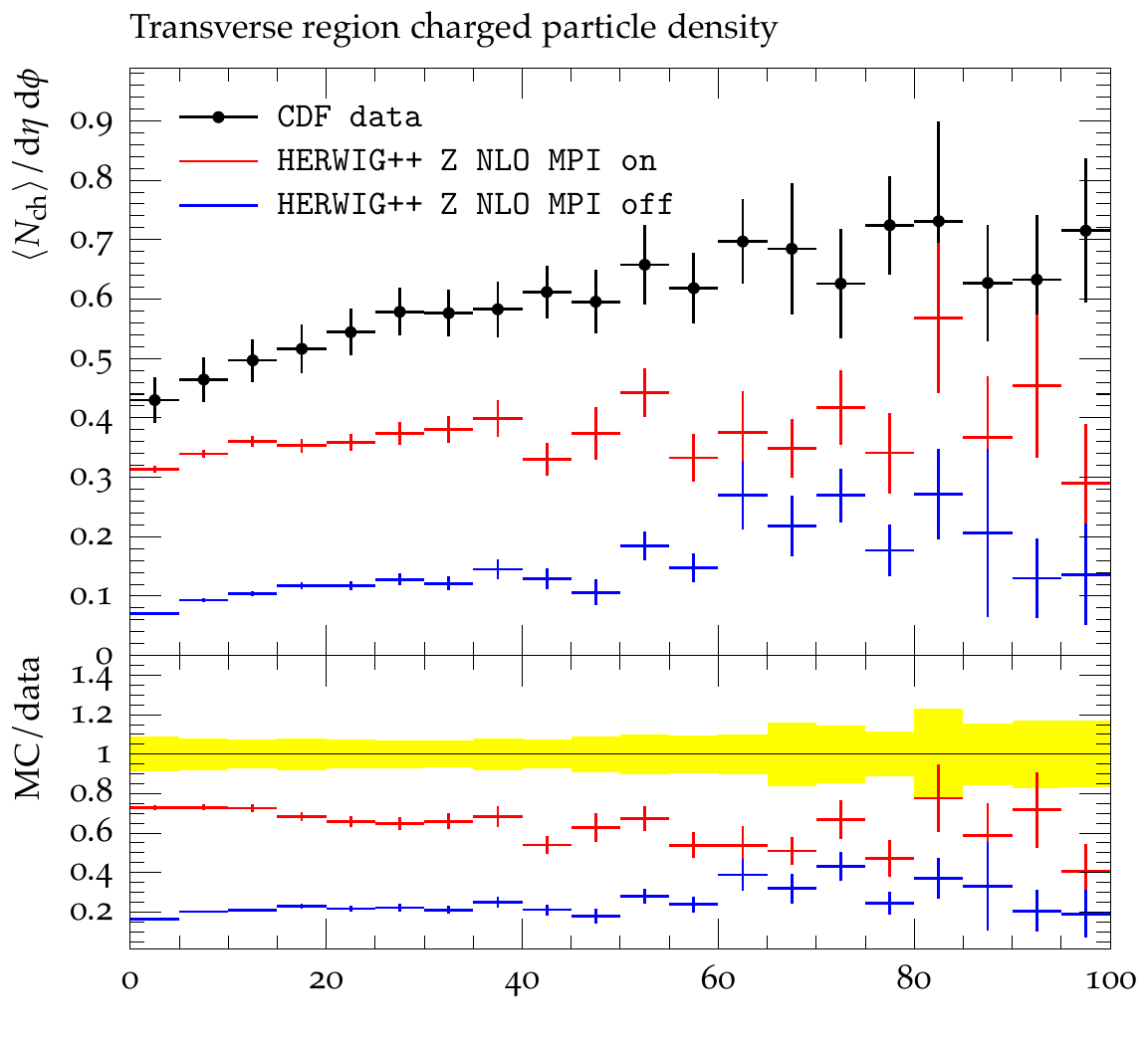}
\includegraphics[width=0.5\columnwidth]{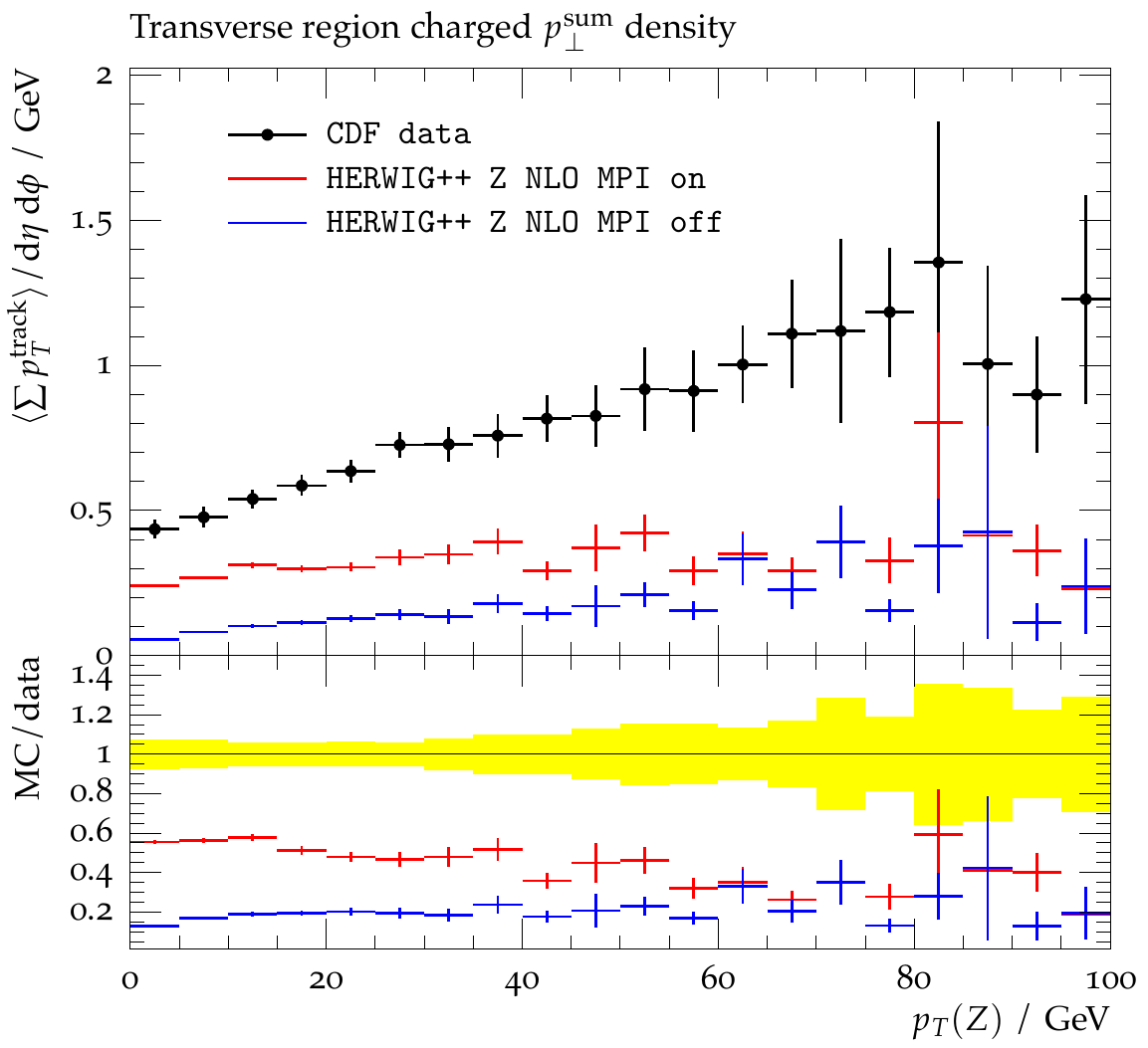}

\caption{\label{fig:UE/MPI}The transverse region charged particle density
(left) and the transverse region charged $p_{\perp}^{sum}$ density
(right) in Underlying Event analysis for ${\tt HERWIG++}$ \emph{Z}
NLO with MPI on and off. }

\end{figure}

Fig. \ref{fig:UE/MPI} shows the comparison of ${\tt HERWIG++}$ to the
CDF data with the MPI model turned on and off. There
appears to be a disagreement in the observed UE, with ${\tt HERWIG++}$
always producing less activity than the data in the transverse region.
The default settings for the MPI model in ${\tt HERWIG++}$
were tuned to provide the best fit to the jet data from Run I and
Run II. Adjusting some of these parameters does not yield a significant
improvement in the description of the \emph{Z} data, however {[}3]. 

\begin{figure}[!h]
\includegraphics[width=0.5\columnwidth]{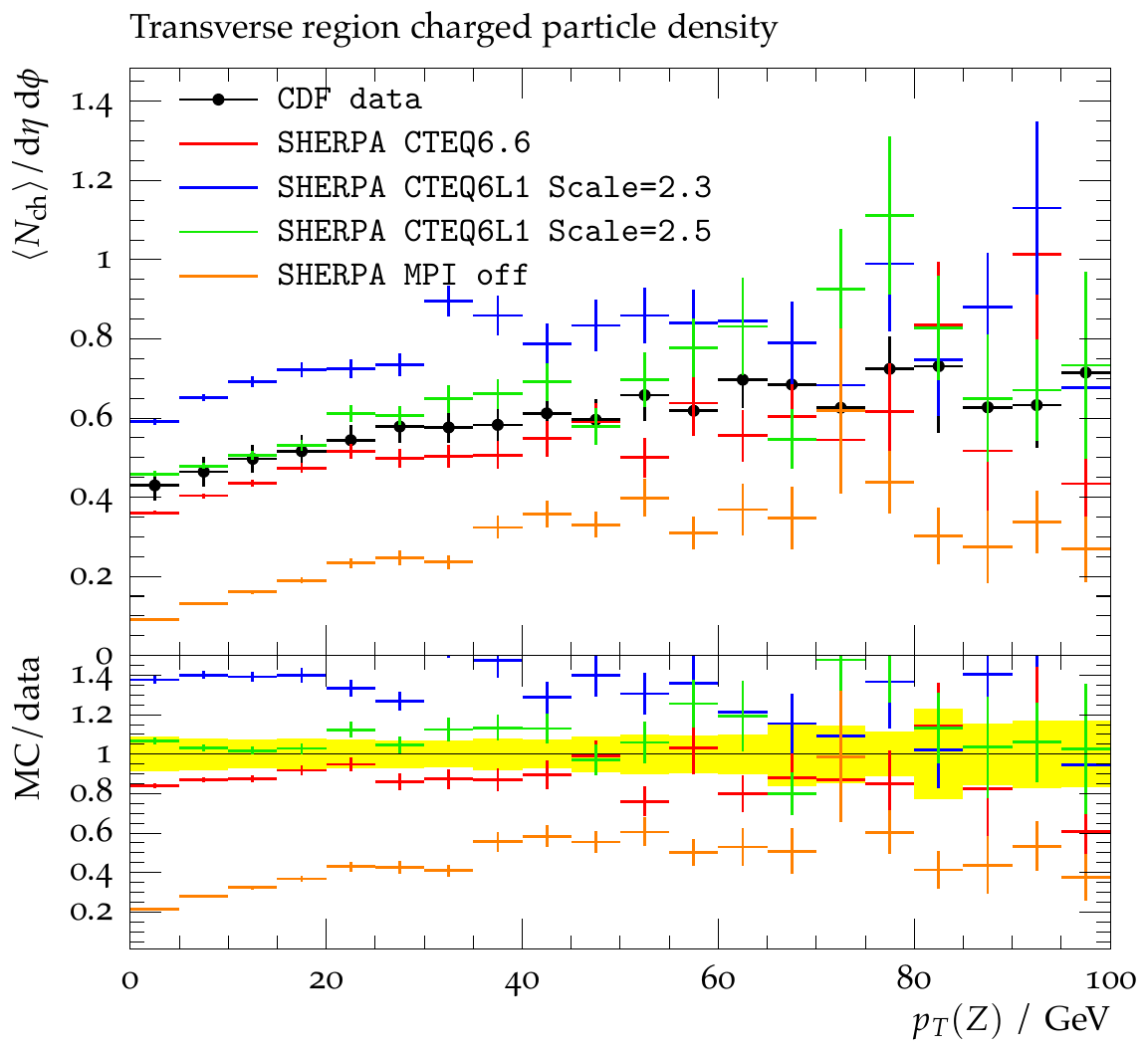}
\includegraphics[width=0.5\columnwidth]{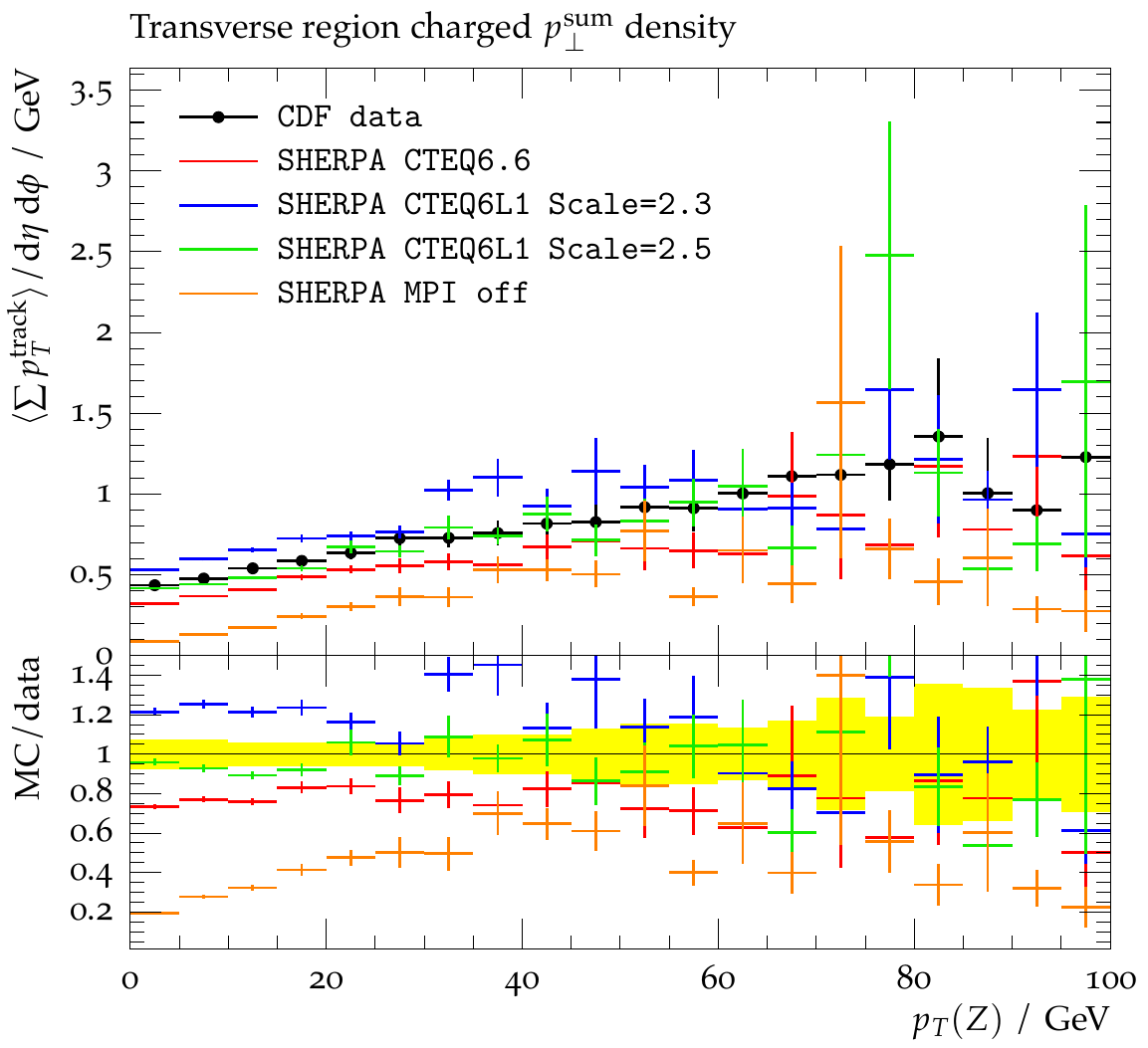}

\caption{\label{fig:UE/MPISherpa}The transverse region charged particle density
(left) and the transverse region charged $p_{\perp}^{sum}$ density
(right) in Underlying Event analysis for ${\tt SHERPA}$ \emph{Z}
+ 3 jets with MPI on, and different PDFs: ${\tt CTEQ6.6}$
with standard MPI scale tuning or ${\tt CTEQ6L1}$ with scale
parameter equals to 2.3 and 2.5 GeV. It also shows the performance
when the MPI is turned off.}

\end{figure}

For the ${\tt SHERPA}$ generator, comparisons are made with the default
settings for MPI {[}13]. Changing the PDF to ${\tt CTEQ6L1}$,
which provided the best description of the \emph{Z} $p_{\perp}$,
significantly degrades the MPI model performance, most probably
due to the different value of $\alpha_{S}$ ($M_{Z}$) and running
of $\alpha_{s}$ between the two PDF sets. The most important
parameter in tuning the MPI model is the scale of the transverse
momentum cutoff, and three values are tested with the ${\tt CTEQ6L1}$
PDF: 2.1, 2.3 and 2.5 GeV. The plots in Fig. \ref{fig:UE/MPISherpa}
show that the best settings for ${\tt SHERPA}$ are the ${\tt CTEQ6L1}$ PDF with the scale set to 2.5 GeV. 

\subsection{\emph{Z} + jets - ${\tt HERWIG++}$ and ${\tt SHERPA}$ LO vs ${\tt POWHEG}$
NLO}

We now return to our main aim: to assess the impact of NLO matrix
elements in the simulation of \emph{Z} (+ jets) events. We use the
following generator configurations: LO without ME correction, LO with
ME correction and NLO (${\tt POWHEG}$ formalism), for the ${\tt HERWIG++}$
generator; and \emph{Z} + 3 jets at LO for the ${\tt SHERPA}$ generator.

Table \ref{tab:xSections} shows a comparison of the total \emph{Z} $\rightarrow e^{+}e^{-}$
cross section measured at the Tevatron CDF experiment {[}14]. As expected, the NLO simulation
has the best prediction of the cross section, while at LO, ${\tt HERWIG++}$
has a slightly better prediction than ${\tt SHERPA}$. Further comparisons
between the ME+PS merging in ${\tt HERWIG++}$ and ${\tt SHERPA}$
(besides ${\tt ALPGEN}$ and ${\tt PYTHIA}$) can be found in reference
{[}15].

\begin{table}[!h]
\begin{centering}
\begin{tabular}{|c|c|c|}
\hline 
 & Total $\sigma_{Z}$ {[}pb] & Uncertainty {[}pb]\tabularnewline
\hline
\hline 
CDF data & 256.0 & 2.1\tabularnewline
\hline 
${\tt HERWIG++}$ LO ME on & 185.1 & 0.7\tabularnewline
\hline 
${\tt HERWIG++}$ LO ME off & 185.2 & 0.7\tabularnewline
\hline 
${\tt HERWIG++}$ NLO & 230.4 & 0.9\tabularnewline
\hline 
${\tt SHERPA}$ \emph{Z} + 1 jet & 171.5 & 0.3\tabularnewline
\hline 
${\tt SHERPA}$ \emph{Z} + 3 jets & 172.6 & 0.4\tabularnewline
\hline
\end{tabular}
\par\end{centering}

\caption{\label{tab:xSections}The total cross sections for \emph{Z}$\rightarrow e^{+}e^{-}$ production
in data, ${\tt SHERPA}$ and ${\tt HERWIG++}$ Monte Carlo generators.
The parameters for ${\tt HERWIG++}$ are the default, with MPI
simulation. For ${\tt SHERPA}$, PDF ${\tt CTEQ6L1}$, MPI
with scale 2.5 GeV, and optmized ${\tt K\text{\_}PERP}$ parameters.}
 
\end{table}

\paragraph{\emph{Z} boson rapidity}

\begin{figure}[!h]

\begin{centering}
\includegraphics[width=0.5\columnwidth]{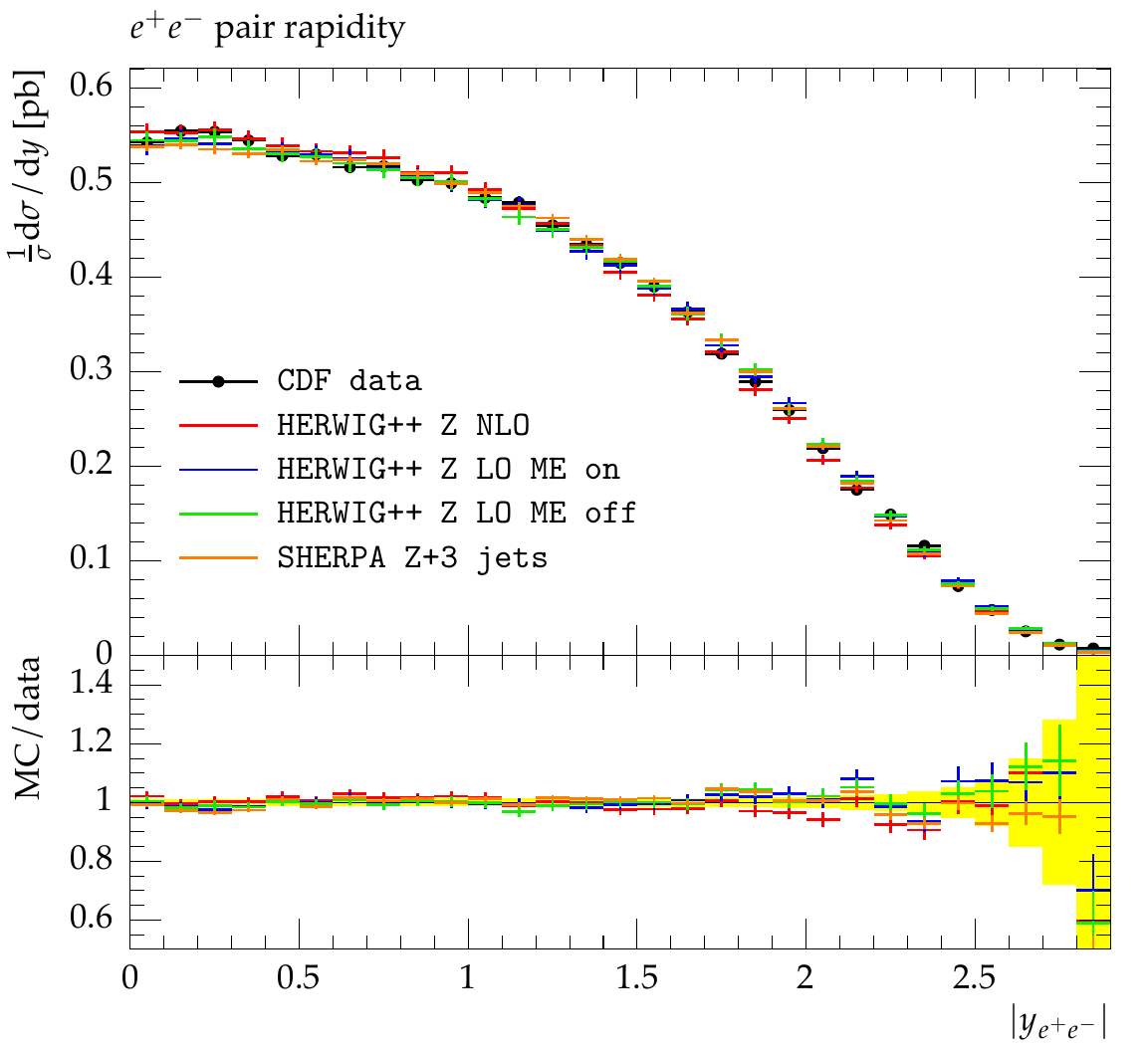}
\par\end{centering}

\caption{\label{fig:Zrapidity}Comparison plots for \emph{Z} production
at: LO (ME correction off), LO (ME correction on) and NLO for ${\tt HERWIG++}$,
and \emph{Z}+3 jets LO for ${\tt SHERPA}$: the cross section
as a function of the \emph{Z} boson rapidty. The plot is normalized to its integral. The
parameters for ${\tt HERWIG++}$ are the defaults, with MPI
simulation. For ${\tt SHERPA}$, PDF ${\tt CTEQ6L1}$, MPI
with scale 2.5 GeV, and optmized ${\tt K\text{\_}PERP}$ parameters.}

\end{figure}

Fig. \ref{fig:Zrapidity} shows the  D0 measurement of the  \emph{Z} $\rightarrow e^{+}e^{-}$  cross section as a function of the \emph{Z} boson rapidity {[}16].
All generators describe the data well.
The differential cross section is also normalized to the total cross
section for \emph{Z} $\rightarrow e^{+}e^{-}$ production.

\paragraph{Jet multiplicity}

\begin{figure}[!h]
\includegraphics[width=0.5\columnwidth]{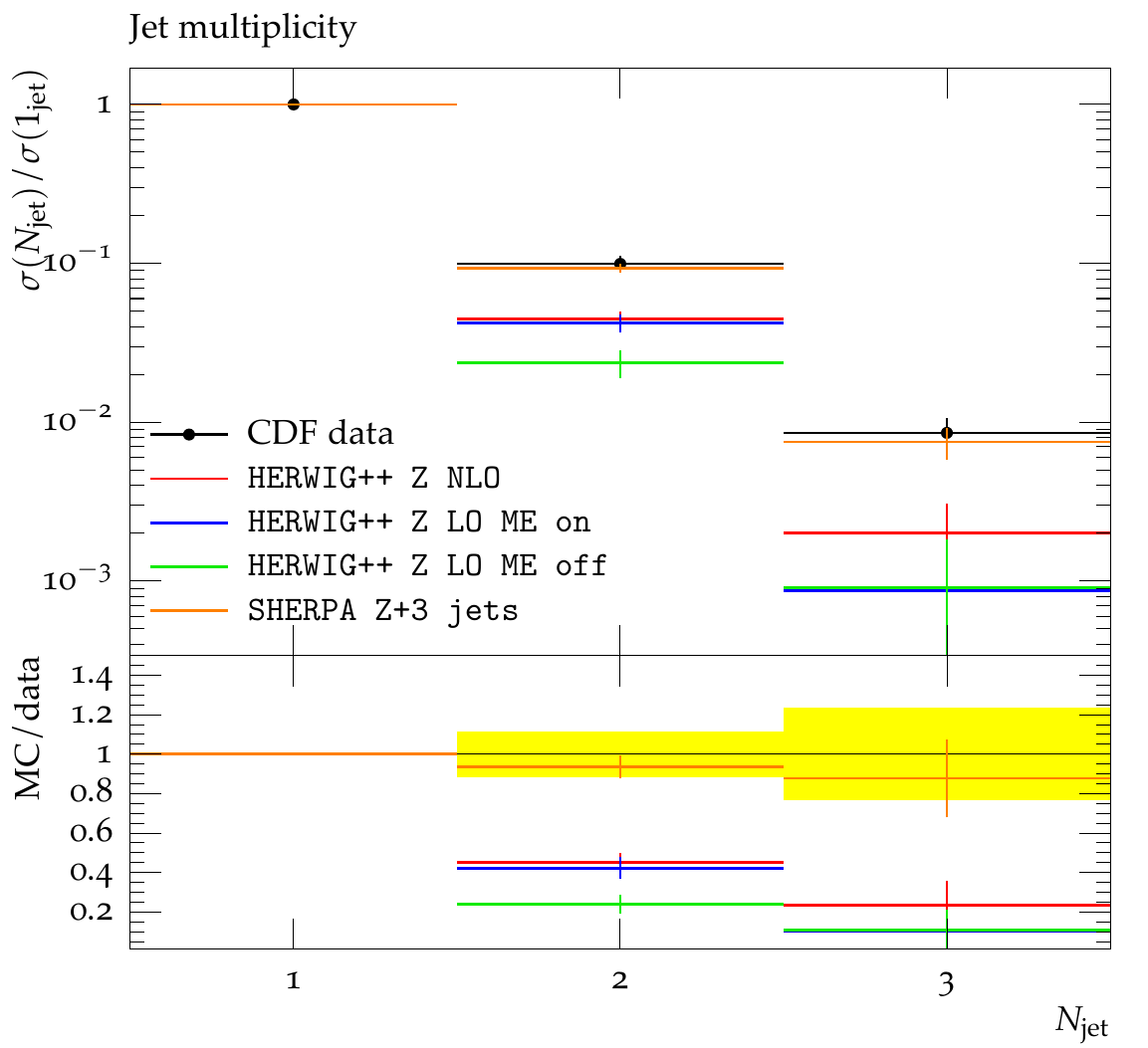}
\includegraphics[width=0.5\columnwidth]{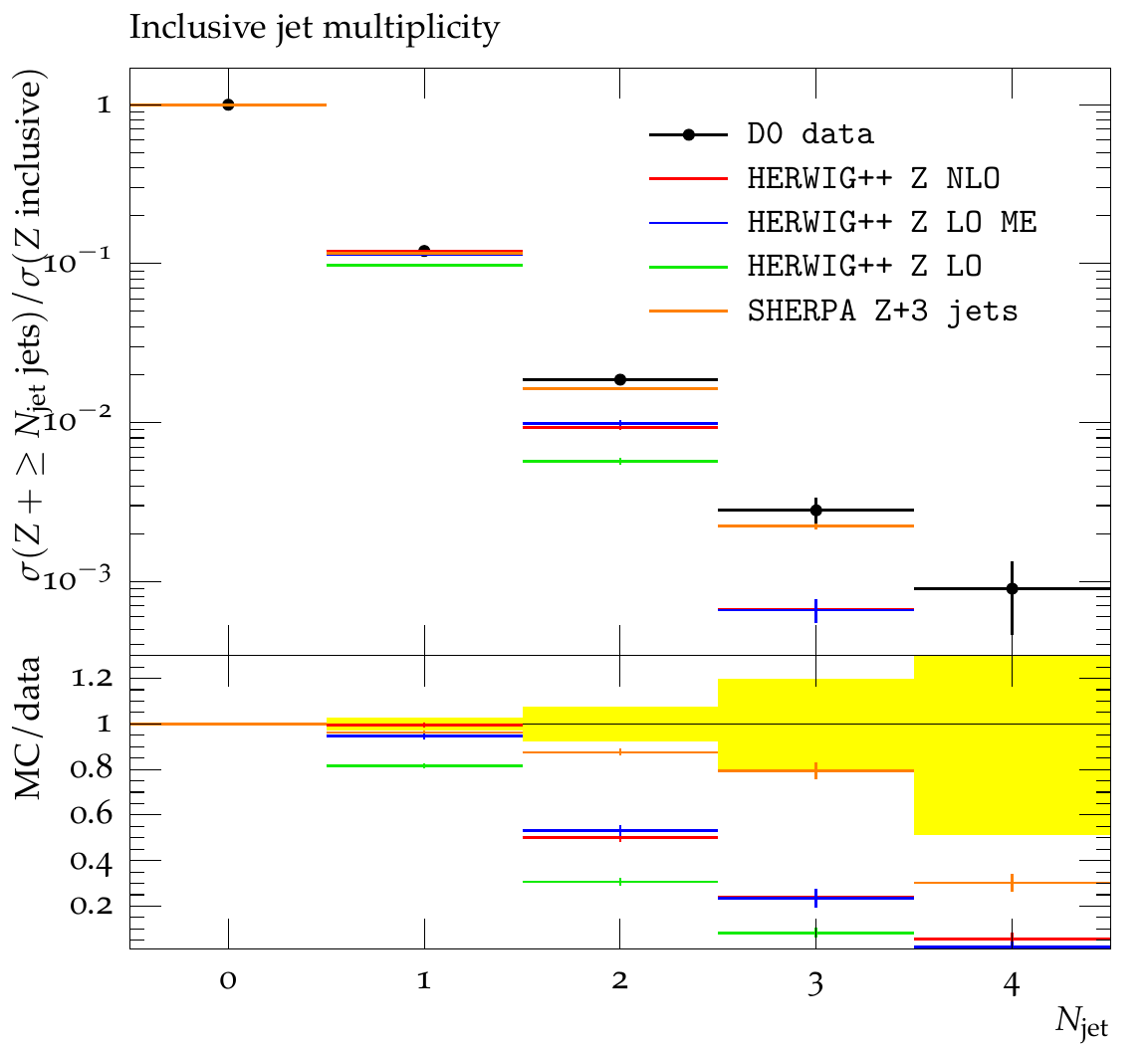}

\caption{\label{fig:Comparison-plots-for}Comparison plots for \emph{Z} production
at: LO (ME correction off), LO (ME correction on) and NLO, on ${\tt HERWIG++}$,
and \emph{Z}+3 jets LO on ${\tt SHERPA}$. In the left, the cross
section prediction as a function of jet multiplicities, normalized to the first bin (the data cross
section for one jet in the event). The right plot shows
the ratios of the expected $n$-jets cross sections to the \emph{Z}
total production cross section (first bin). The parameters for ${\tt HERWIG++}$
are the defaults, with MPI simulation. For ${\tt SHERPA}$,
PDF ${\tt CTEQ6L1}$, MPI with scale 2.5 GeV, and
optimized ${\tt K\text{\_}PERP}$ parameters.}

\end{figure}

In Fig. \ref{fig:Comparison-plots-for} (left) the CDF \emph{Z}
$\rightarrow e^{+}e^{-}$ {[}10] analysis shows the cross section
as a function of jet multiplicity. The jets are required to have $p_{\perp}$ > 30 GeV
and  to be within the rapidity range, $|y_{jet}|<2.1$. The plot is then normalized to the cross section of production of Z+1 jet (first bin) in data. This is useful to test the relative fraction of two and three jet events. We can see that the ${\tt SHERPA}$ \emph{Z}+3 jets generator can describe,
inside data uncertainties, the expected number of events with two
and three jets, while ${\tt HERWIG++}$ fails for jet multiplicities
higher than one. 

The D0 \emph{Z} $\rightarrow e^{+}e^{-}$ analysis {[}17] measures the $n$-jet
cross section ratios and is shown in Fig. \ref{fig:Comparison-plots-for} (right).
The distribution is inclusive and normalized to the total \emph{Z} $\rightarrow e^{+}e^{-}$  data cross section (i.e., the bin of zero or more jets). Here we can see that the ${\tt HERWIG++}$
\emph{Z} NLO , ${\tt HERWIG++}$ LO with ME corrections and ${\tt SHERPA}$ \emph{Z} + 3 jets predictions describe the first jet bin well. 
For higher jet multiplicities, the ${\tt SHERPA}$ \emph{Z} + 3 jets generator
describes the data better, however it fails for four
jets, as one may expect. ${\tt HERWIG++}$ LO without ME corrections does not describe the data in any bins.

\paragraph{Jet transverse momentum}

\begin{figure}[!h]
\includegraphics[width=0.5\columnwidth]{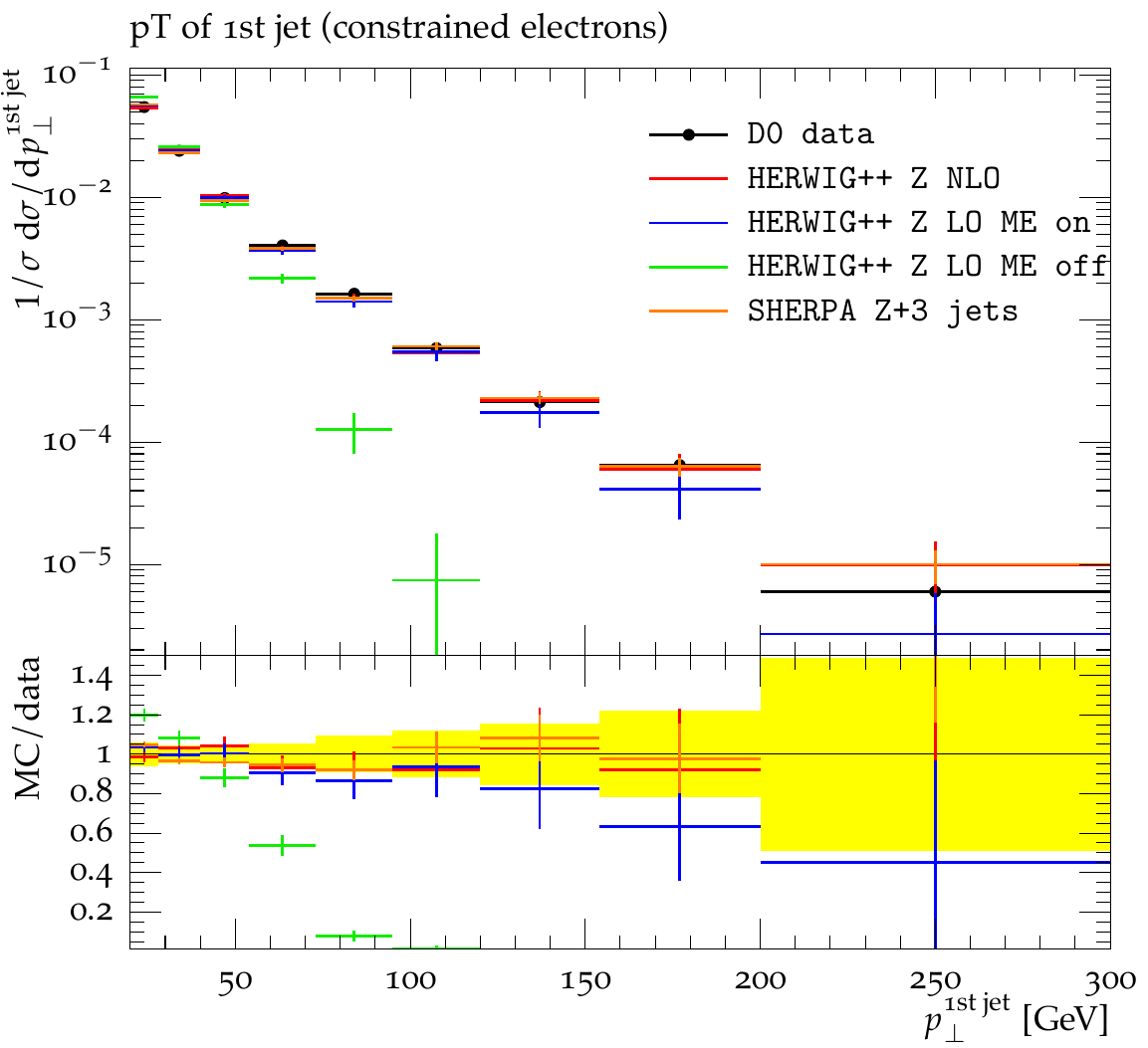}
\includegraphics[width=0.5\columnwidth]{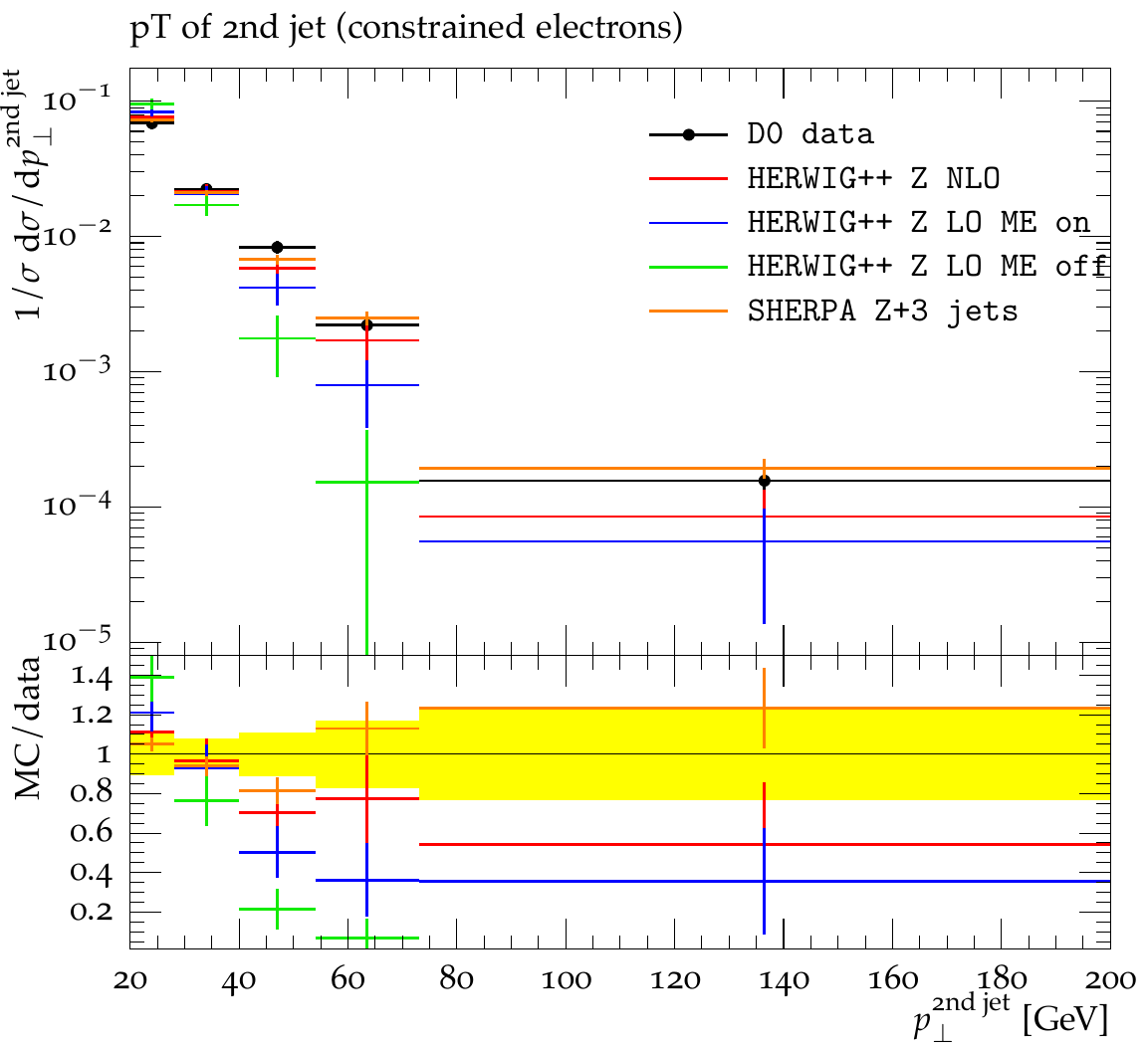}

\begin{centering}
\includegraphics[width=0.5\columnwidth]{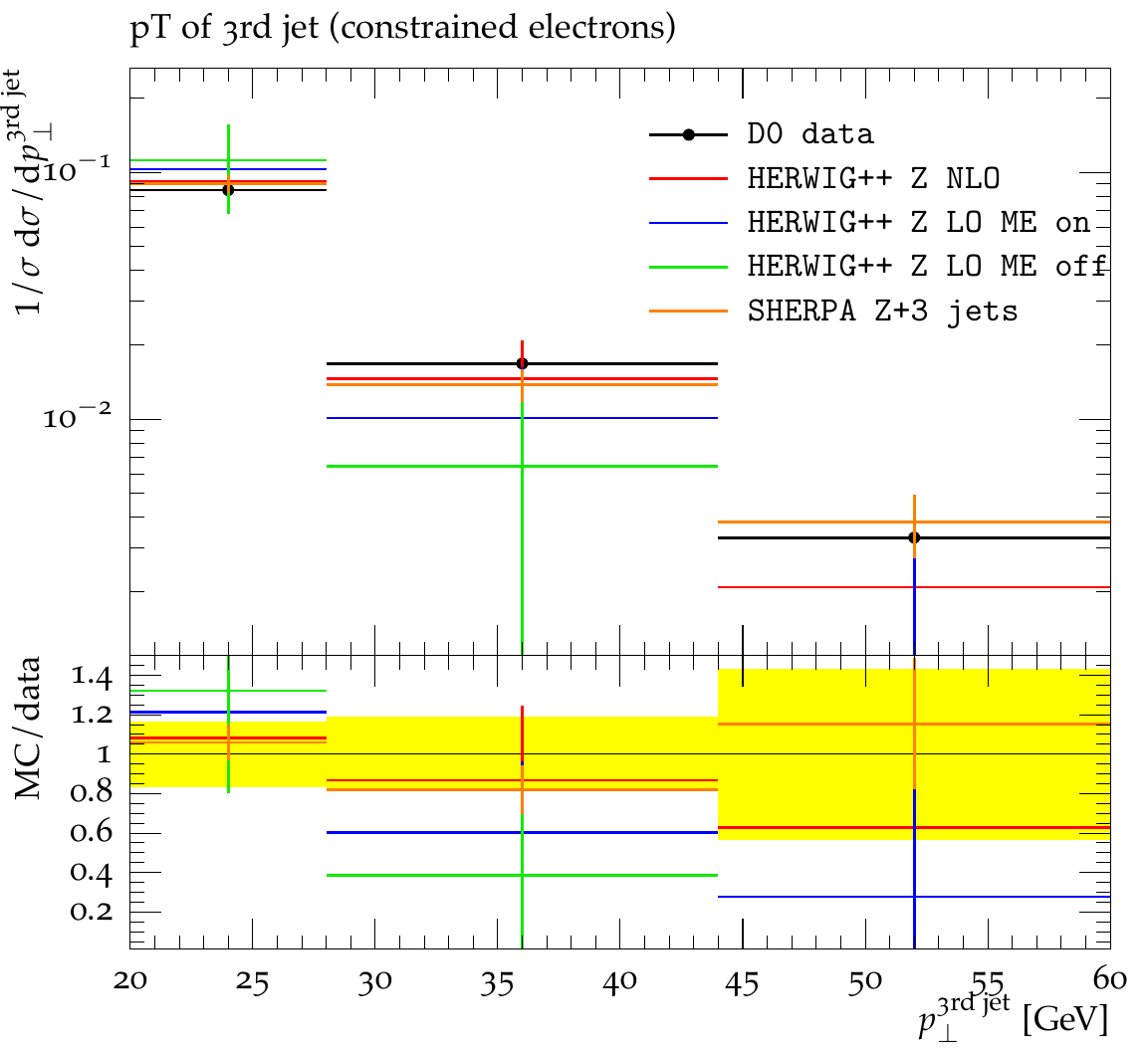}
\par\end{centering}

\caption{\label{fig:LOvsNLOHerwig}Comparison plots for \emph{Z} production
at: LO (ME correction off), LO (ME correction on) and NLO for ${\tt HERWIG++}$,
and \emph{Z}+3 jets LO for ${\tt SHERPA}$. The transverse momentum of the leading, second and third leading jet respectively. All the plots are normalized to their integrals. The
parameters for ${\tt HERWIG++}$ are the defaults, with MPI
simulation. For ${\tt SHERPA}$, PDF ${\tt CTEQ6L1}$, MPI
with scale 2.5 GeV, and optmized ${\tt K\text{\_}PERP}$ parameters.}

\end{figure}

The D0  analysis in {[}18]
measures the differential cross section as a function of the transverse
momentum, 
of the three leading jets in the production of $Z/\gamma*\rightarrow e^{+}e^{-}+\text{jets}+X$,
normalized to the total cross section of $Z/\gamma*\rightarrow e^{+}e^{-}$ production.
Fig. \ref{fig:LOvsNLOHerwig} shows comparisons with the ${\tt HERWIG++}$ generator at
NLO (POWHEG formalism) and at LO with and without ME corrections,
and with ${\tt SHERPA}$ \emph{Z} + 3 jets. The LO without ME corrections
shows the expected failure of the parton shower alone to populate
the high $p_{\perp}$ region, because it corresponds to the phase
space where the PS can't fill properly. When the ME correction is
turned on, the effect in correcting the high transverse momentum region
can be seen. The behaviour for the NLO plot is more close to the data
in the higher $p_{\perp}$ region (above 130 GeV), while it has similar
results to LO with ME corrections at lower $p_{\perp}$. However,
for the second and third leading jets, the description is also not
good - NLO \emph{Z} production includes the LO matrix element for
one jet production, but the second and third jets are produced only
by the parton shower, which again underestimates the data. For the
${\tt SHERPA}$ \emph{Z} + 3 jets, the matrix element for the further
jet emissions allows for a description similar to the NLO curve for
the leading jet, and does a better job for the second and third leading
jets. 

\subparagraph{D0 $\text{Z}\rightarrow\mu^{+}\mu^{-}$ Analysis}

To further study the jet recoil and \emph{Z} boson $p_{\perp}$, the
D0 analysis with \emph{Z} ($\rightarrow\mu^{+}\mu^{-}$) + jets
{[}19] was used. It measures the cross sections as a function of the
boson momentum, and as a function of momentum and rapidity of the
leading jet. It is required at least one jet in the event, with $p_{\perp}$ > 20 GeV and $|\eta|$ < 2.8. This analysis doesn't normalize
the data by the \emph{Z} total cross section production - the shape
of the plots are normalized by their integral.

\begin{figure}[!h]
\includegraphics[width=0.5\columnwidth]{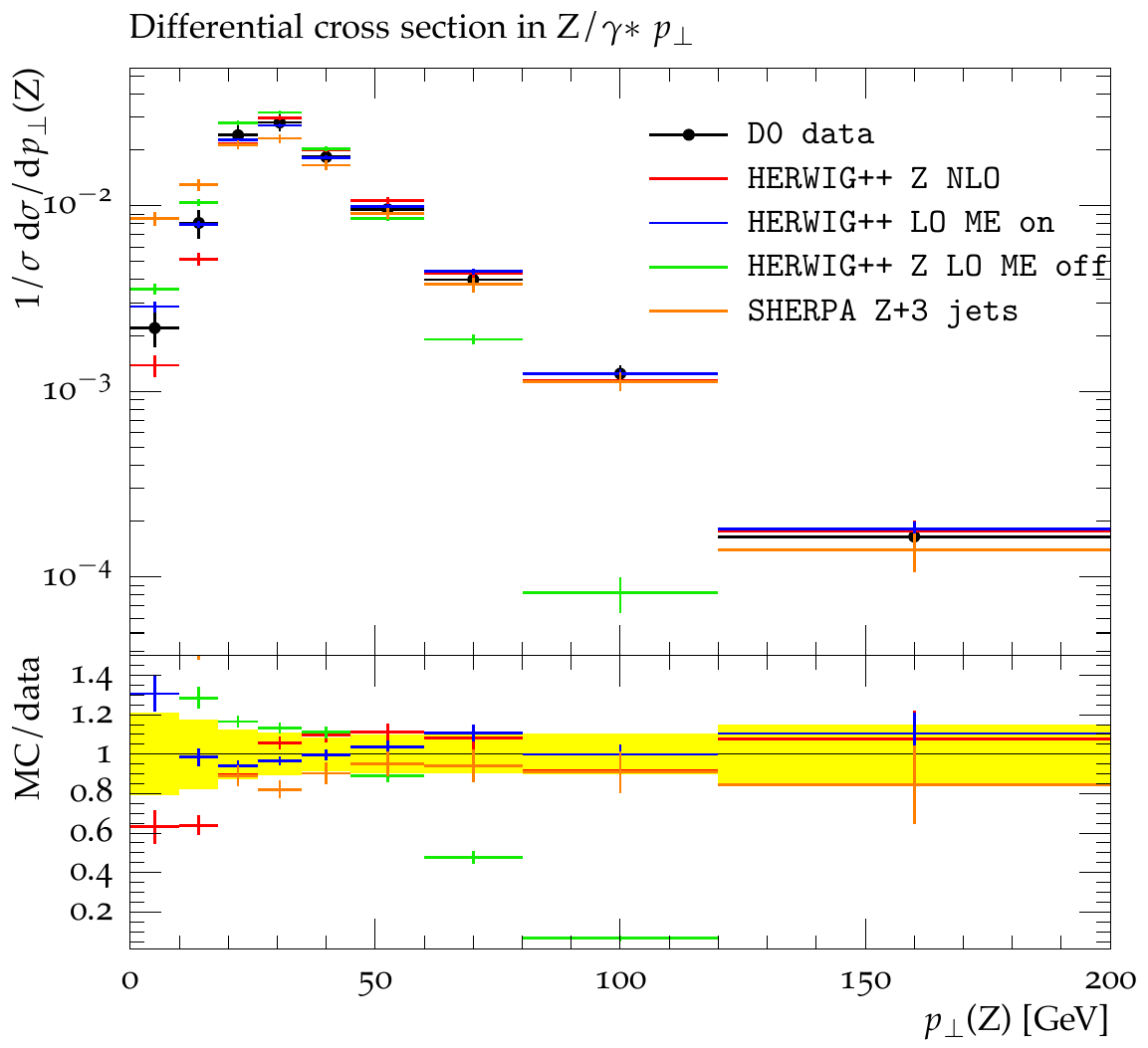}
\includegraphics[width=0.5\columnwidth]{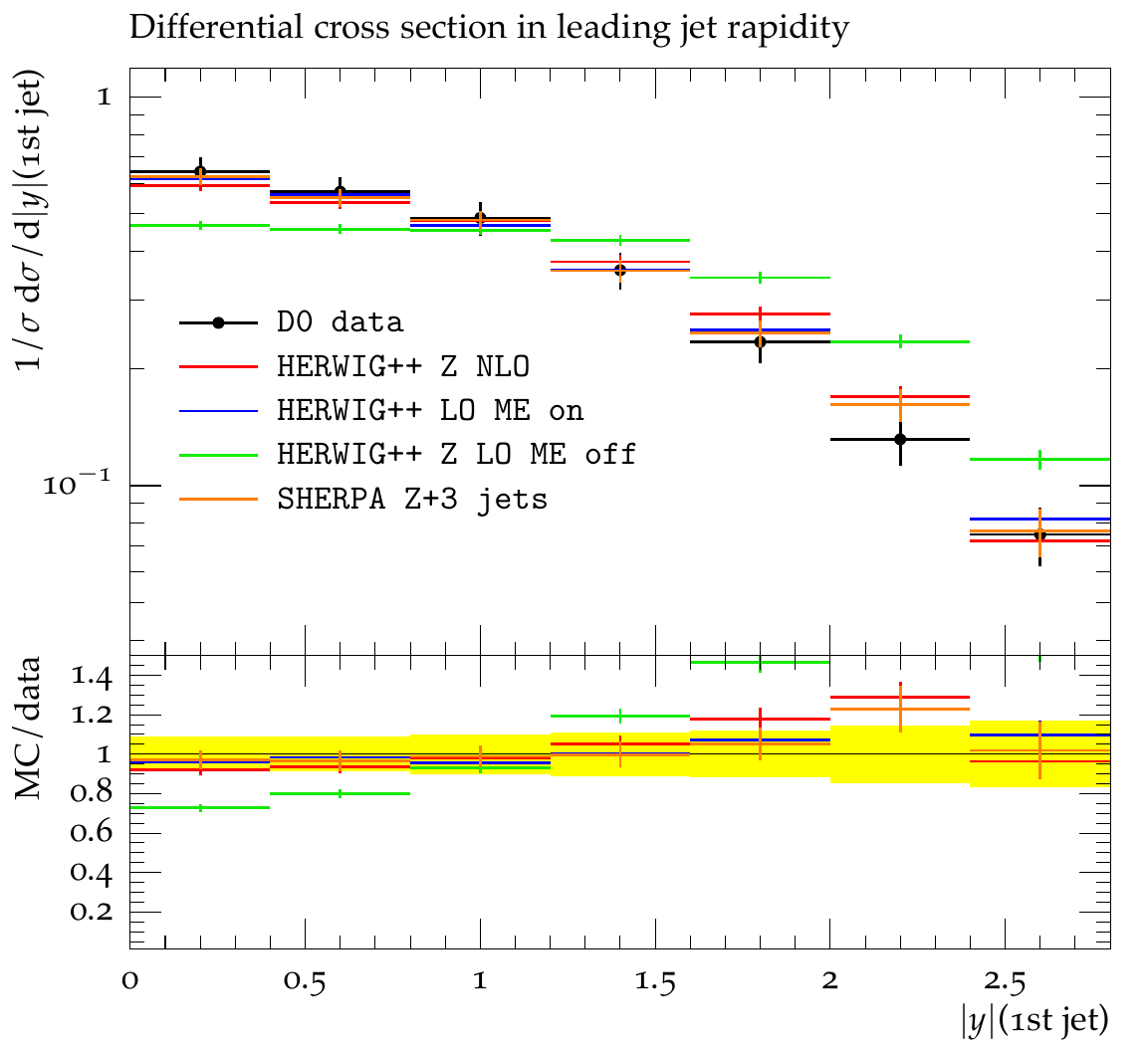}

\begin{centering}
\includegraphics[width=0.5\columnwidth]{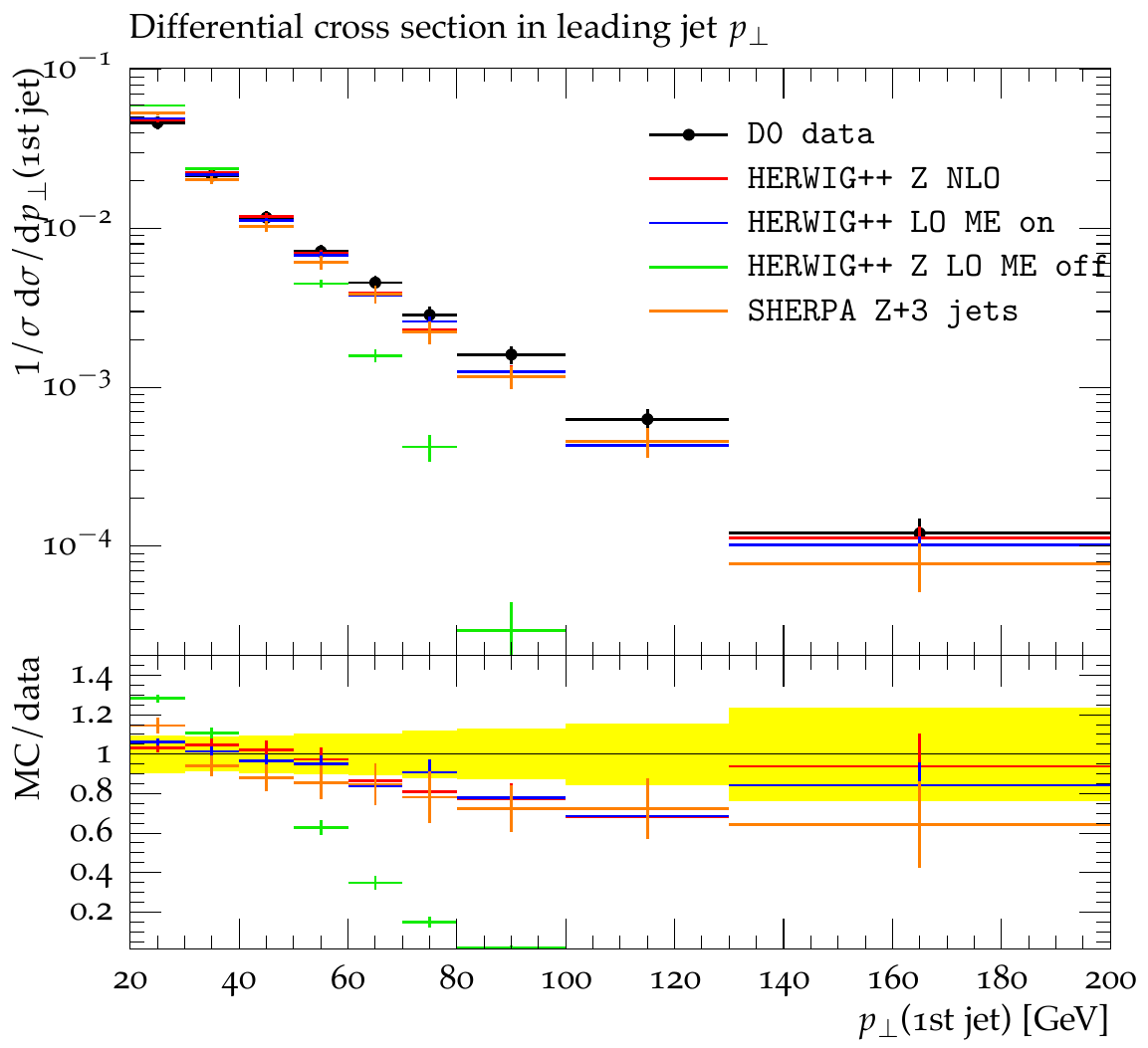}
\par\end{centering}

\caption{\label{fig:MuonAnal}Comparison plots for \emph{Z} production at
LO and NLO for ${\tt HERWIG++}$ and LO for ${\tt SHERPA}$ \emph{Z}+3
jets, in the muon channel: upper left, the \emph{Z} $p_{\perp}$, upper
right, the leading jet rapidity, and lower, the leading jet $p_{\perp}$.
All plots are normalized to their integrals. For both generators MPI is included. The analysis asks for at least one jet with $p_{\perp}$ > 20 GeV and $|\eta|$ < 2.8.
 ${\tt HERWIG++}$ has the
default parameters. The ${\tt SHERPA}$ PDF is ${\tt CTEQ6L1}$,
the MPI scale parameter is 2.5 GeV, and optmized values of ${\tt K\text{\_}PERP}$ are used.}

\end{figure}

As can be seen in Fig. \ref{fig:MuonAnal}, the cross section in leading
jet rapidity is well behaved inside the uncertainties, however ${\tt HERWIG++}$
\emph{Z} LO without ME corrections tends to produce a wider distribution. 
The \emph{Z} $p_{\perp}$ shows imprecisions in the low momentum region: for ${\tt HERWIG++}$
\emph{Z} NLO the production in low \emph{Z} $p_{\perp}$ is around
40\% lower than data, and in ${\tt SHERPA}$ it's greater than data.
This region is particularly sensitive to jets produced by MPI,
which are not recoiling against the \emph{Z} boson.

The leading jet $p_{\perp}$ does not show any significant discrepancy in both
cases. However, a behaviour that one can see is the deficit
of events in the Monte Carlo compared to the data in the range of
50 < $p_{\perp}$ < 120 GeV. The \emph{Z} $p_{\perp}$ shows no such
deficit (see Fig. \ref{fig:NLO123}), suggesting the MC is not fully
describing the hadronic recoil of the \emph{Z}, and this is studied
in detail in the appendix. 

${\tt HERWIG++}$ LO with ME corrections show similar behaviour
to the NLO prediction, and without the ME corrections has a completely different
shape in all plots, due to the previously discussed issues in populating
the high $p_{\perp}$ regions. 

\section{LHC Analyses Cuts}

After choosing the Monte Carlo parameters for the ${\tt HERWIG++}$
\emph{Z} NLO (in ${\tt POWHEG}$ formalism) and ${\tt SHERPA}$ \emph{Z}
+ 3 jets (LO), comparison plots for $Z\rightarrow e^{+}e^{-}$ in
the LHC energy (first Run - 7 TeV) were performed, using the following
kinematic cuts (taken from {[}20]): 

\begin{itemize}
\item Transverse momentum of the lepton $p_{\perp}(l)>15\;\text{GeV}$; 
\item Absolute value of the lepton pseudorapidity $|\eta(l)|<2.4$;
\item Transverse momentum of the jet $p_{\perp}(j)>20\;\text{GeV}$;
\item Absolute value of the jet pseudorapidity $|\eta(j)|<4.5$;
\item Lepton isolation criteria: $\Delta R_{ll}>0.2$; $\Delta R_{lj}>0.4$;
\end{itemize}
The jets are reconstructed with the $antik_{T}$ clustering algorithm,
with cone radius R=0.7. A mass cut on the leptons invariant mass $60<M_{ll}<110\;\text{GeV}$
was applied. The events generated correspond to a integrated luminosity
of 1 $fb^{-1}$. 

\begin{figure}[!h]
\includegraphics[width=0.32\columnwidth]{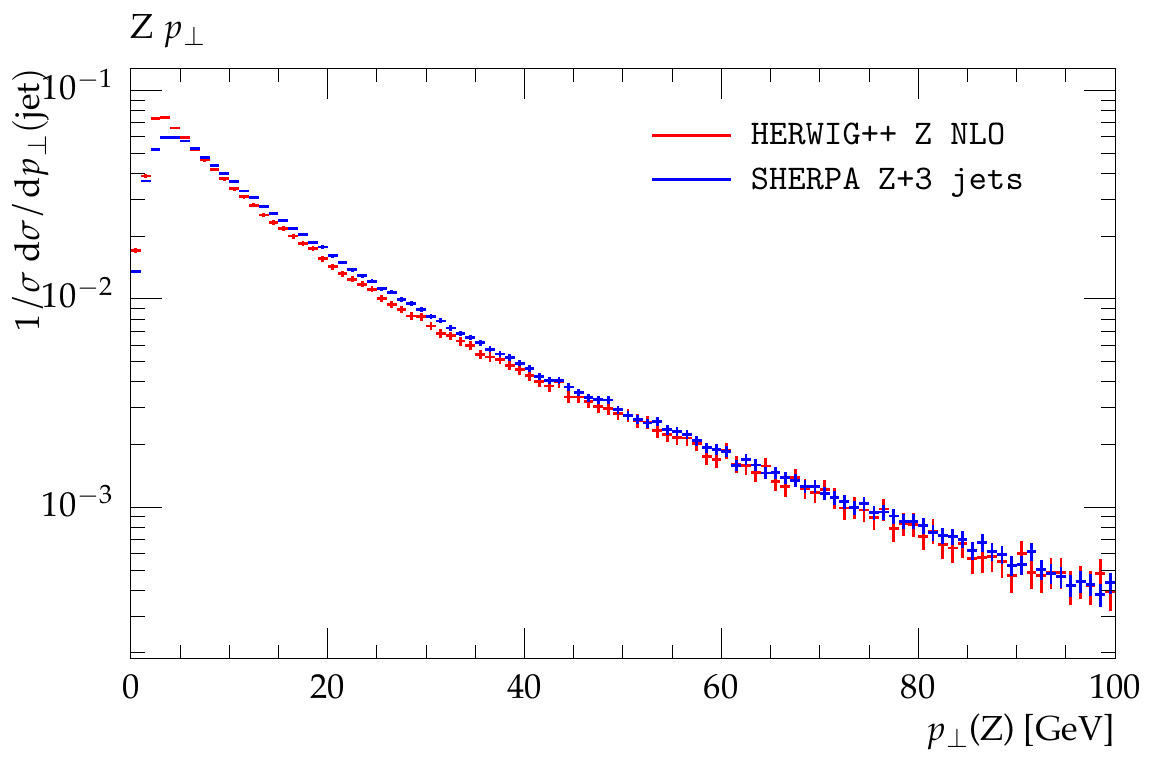}
\includegraphics[width=0.32\columnwidth]{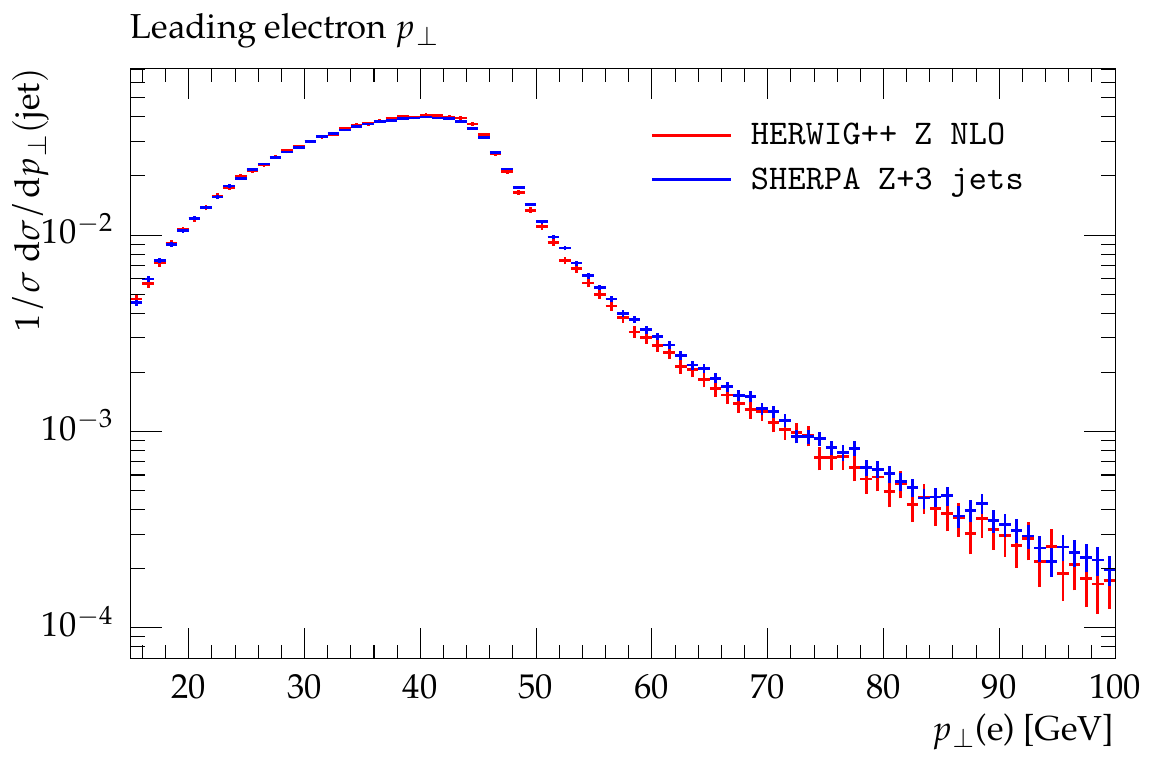}
\includegraphics[width=0.32\columnwidth]{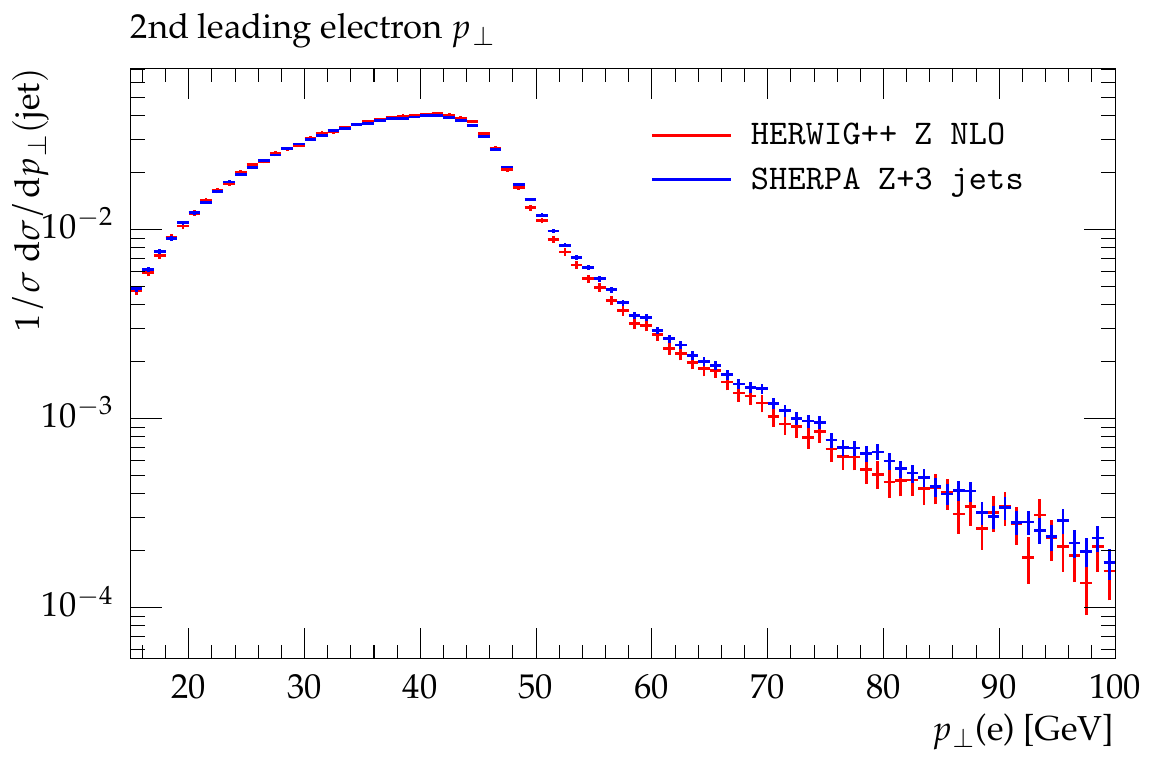}

\caption{\label{fig:LHC1}Comparison plots for LHC (7 TeV) energy, for \emph{Z}
production in electron channel: the \emph{Z} $p_{\perp}$ (left),
leading electron $p_{\perp}$ (center) and 2nd leading electron $p_{\perp}$
(right), for ${\tt HERWIG++}$ \emph{Z} NLO and ${\tt SHERPA}$ \emph{Z}+3
jets, both with MPI simulation, and ${\tt SHERPA}$ with
optimized ${\tt K\text{\_}PERP}$ and ${\tt CTEQ6L1}$ PDF,
scale parameter 2.5 GeV. }

\end{figure}

\begin{figure}[!h]
\includegraphics[width=0.32\columnwidth]{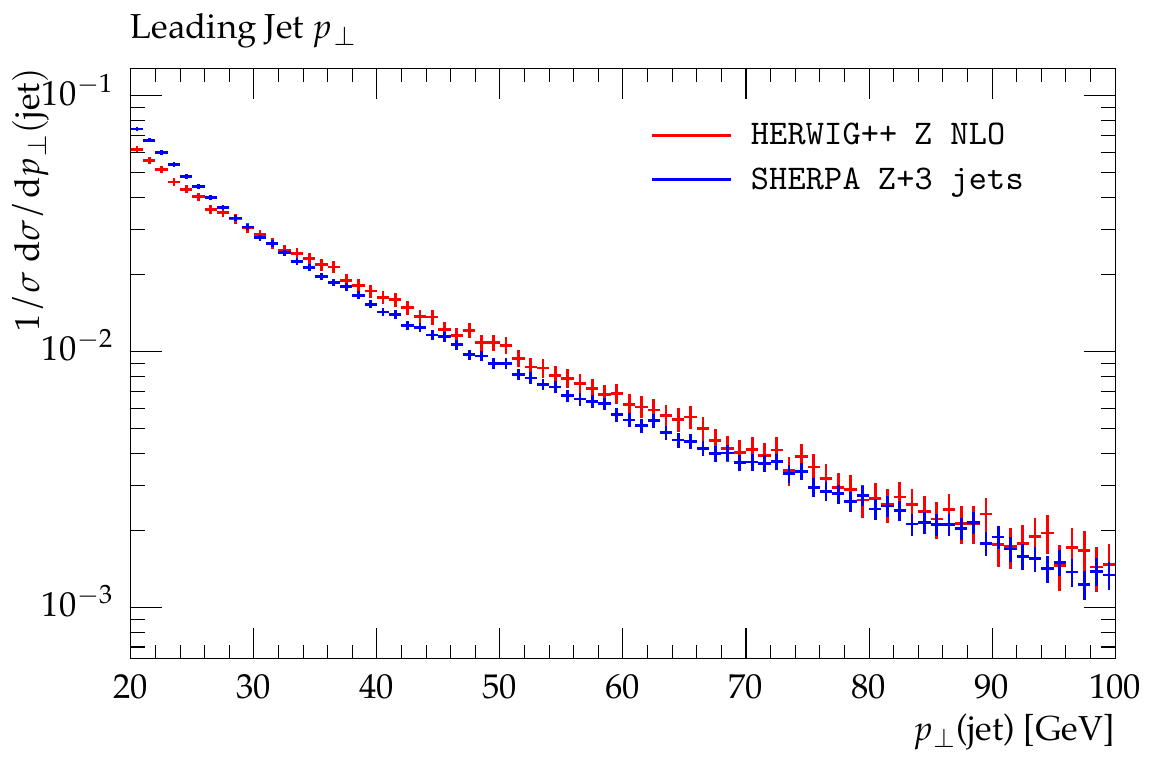}
\includegraphics[width=0.32\columnwidth]{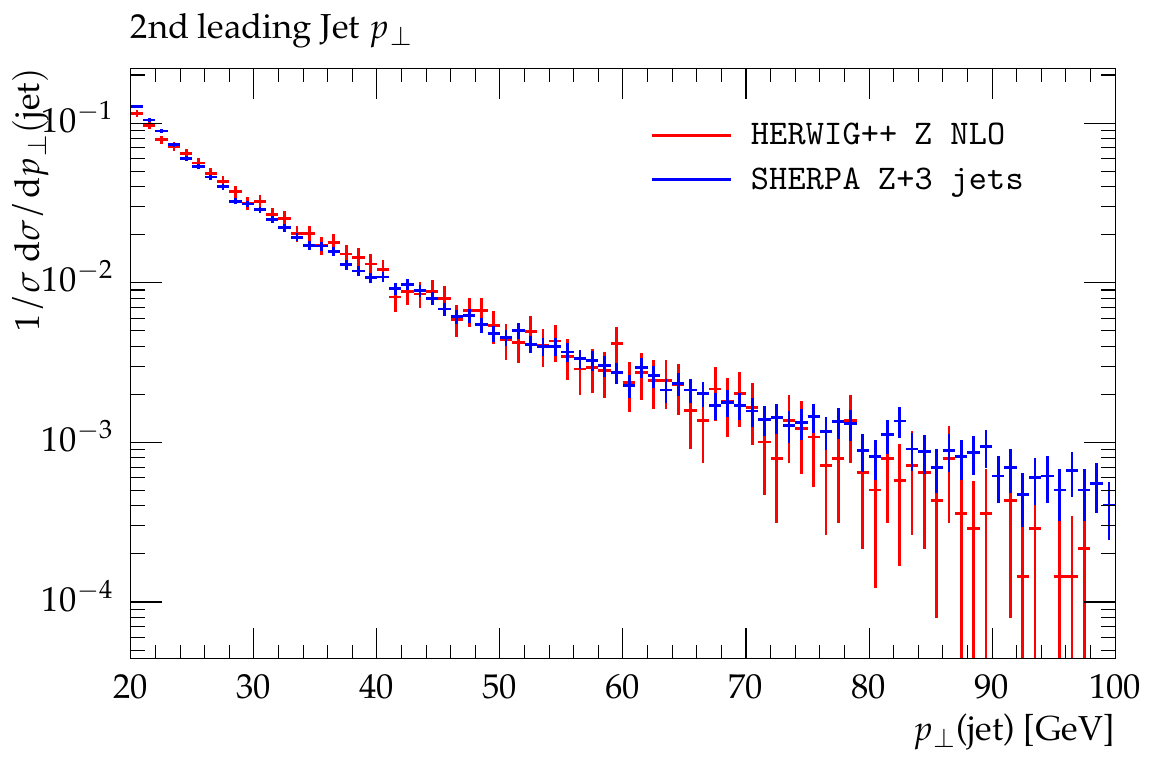}
\includegraphics[width=0.32\columnwidth]{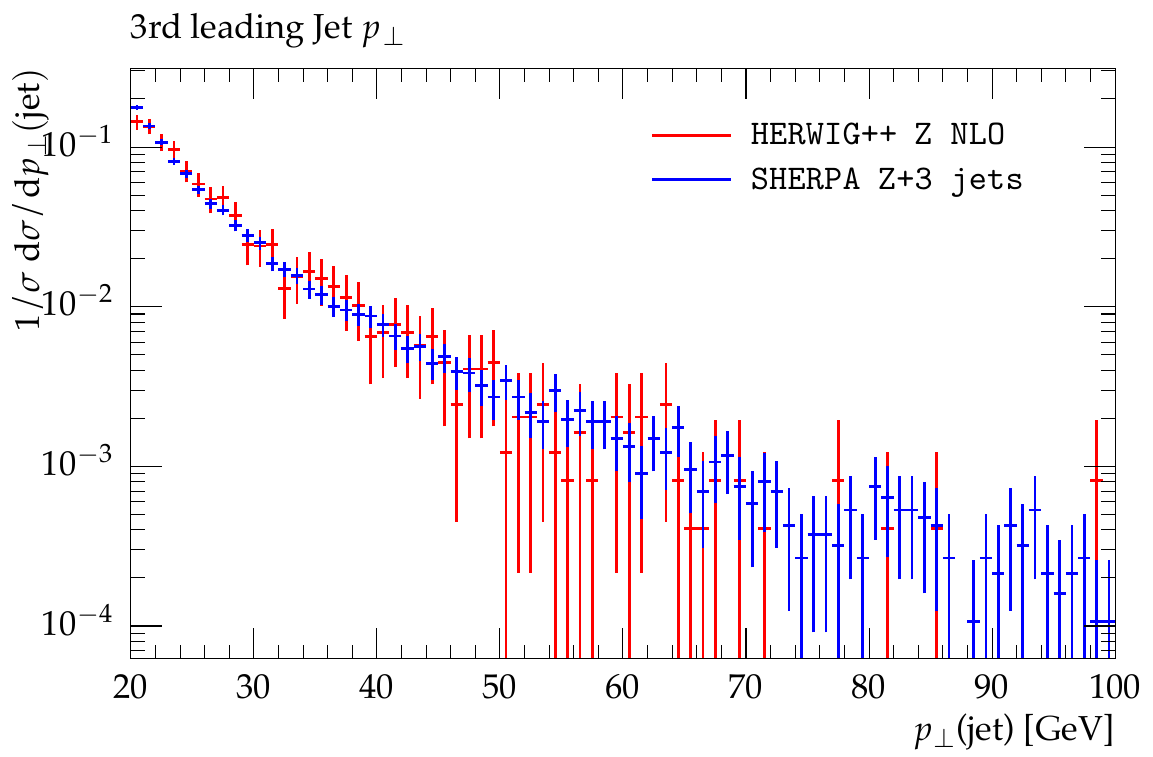}

\caption{\label{fig:LHC2}Comparison plots for LHC (7 TeV) energy, for \emph{Z}
production in electron channel: the leading (left), second (center)
and third (right) leading jet $p_{\perp}$, for ${\tt HERWIG++}$
\emph{Z} NLO and ${\tt SHERPA}$ \emph{Z}+3 jets, both with MPI
simulation, and ${\tt SHERPA}$ with optimized ${\tt K\text{\_}PERP}$
and ${\tt CTEQ6L1}$ PDF, scale parameter 2.5 GeV. }

\end{figure}

\begin{figure}[!h]
\includegraphics[width=0.32\columnwidth]{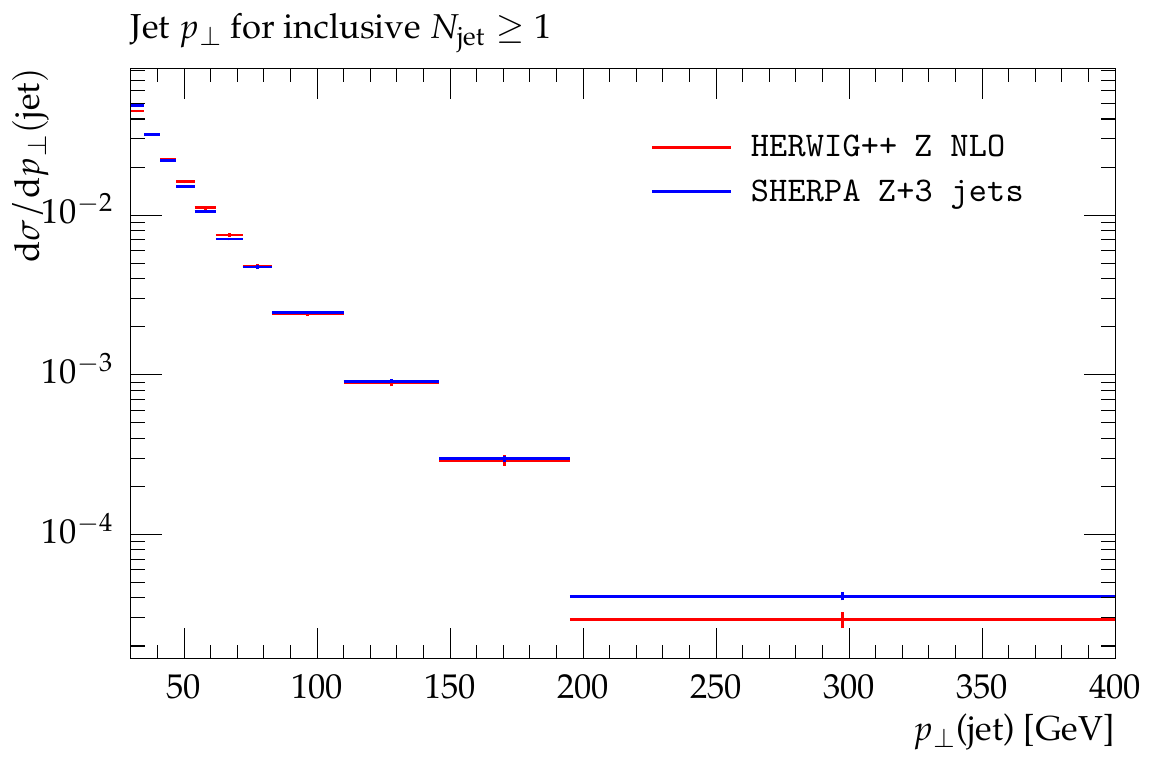}
\includegraphics[width=0.32\columnwidth]{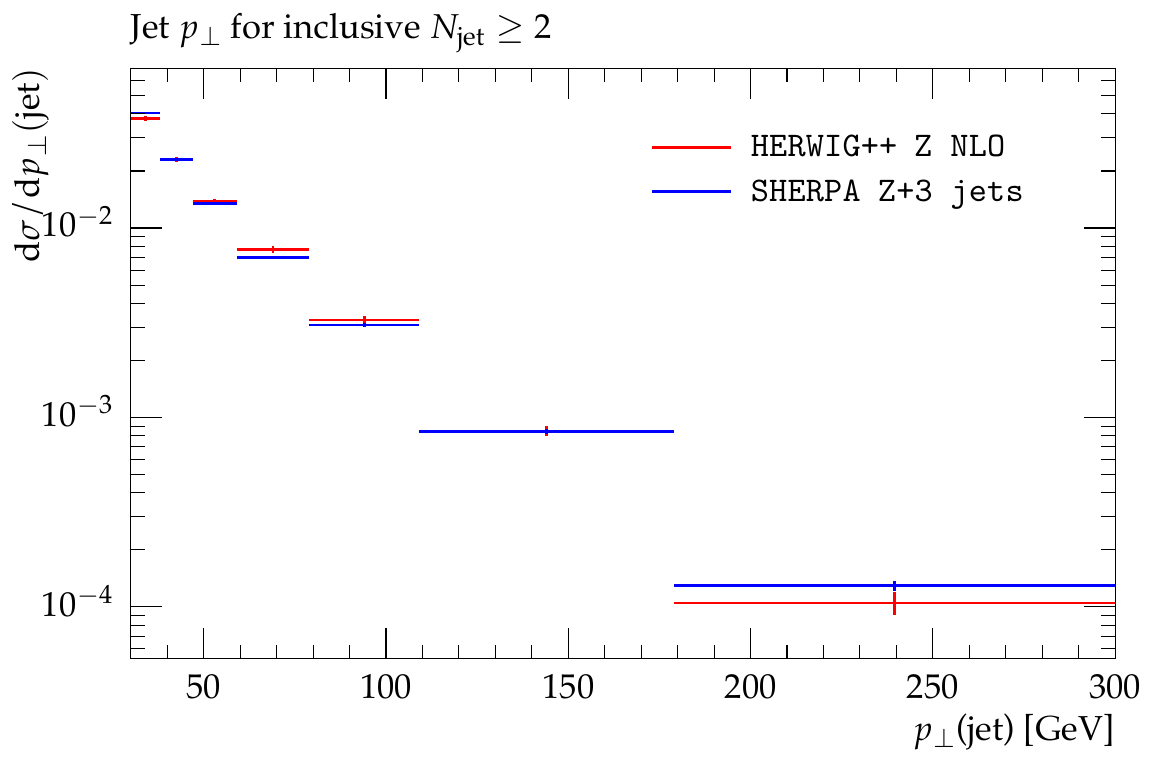}
\includegraphics[width=0.32\columnwidth]{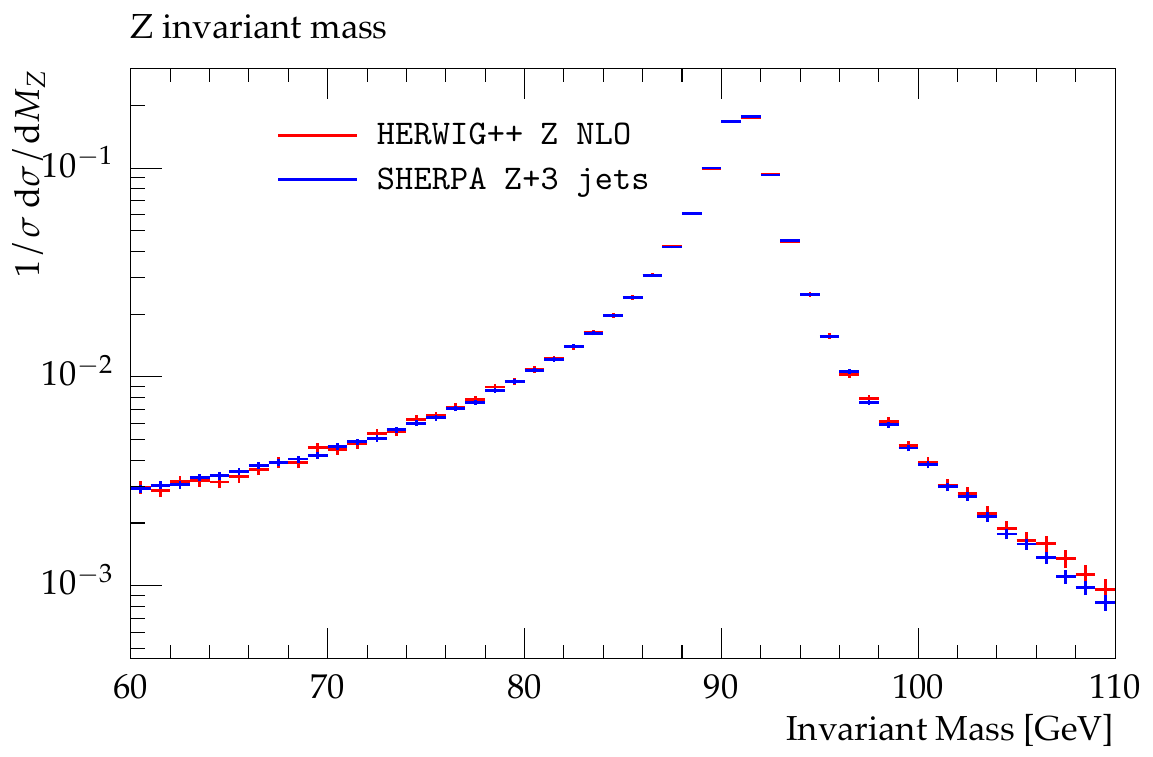}

\caption{\label{fig:LHC3}Comparison plots for LHC (7 TeV) energy, for \emph{Z}
production in electron channel: the jet $p_{\perp}$ for inclusive
$N_{jet}\geq1$ (left), jet $p_{\perp}$ for inclusive $N_{jet}\geq2$
(center) and \emph{Z} invariant mass (right), for ${\tt HERWIG++}$
\emph{Z} NLO and ${\tt SHERPA}$ \emph{Z}+3 jets, both with MPI
simulation, and ${\tt SHERPA}$ with optimized ${\tt K\text{\_}PERP}$
and ${\tt CTEQ6L1}$ PDF, scale parameter 2.5 GeV. }

\end{figure}

\begin{figure}[!h]
\includegraphics[width=0.32\columnwidth]{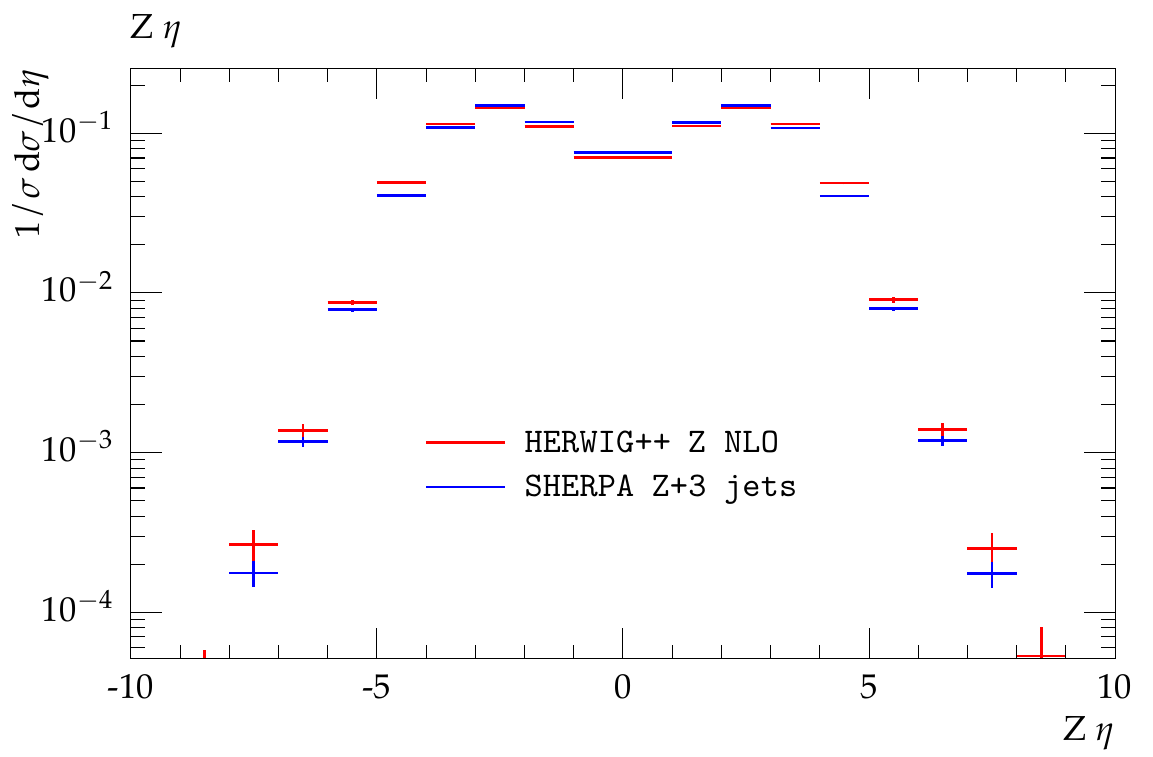}
\includegraphics[width=0.32\columnwidth]{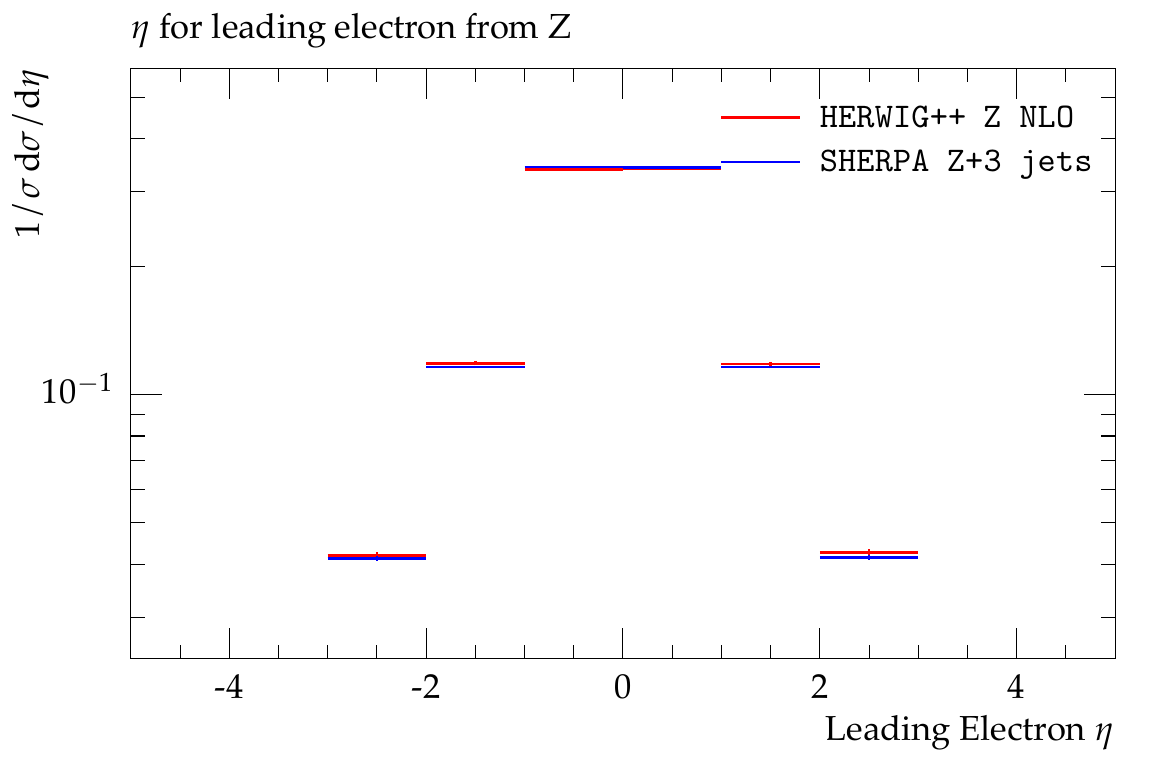}
\includegraphics[width=0.32\columnwidth]{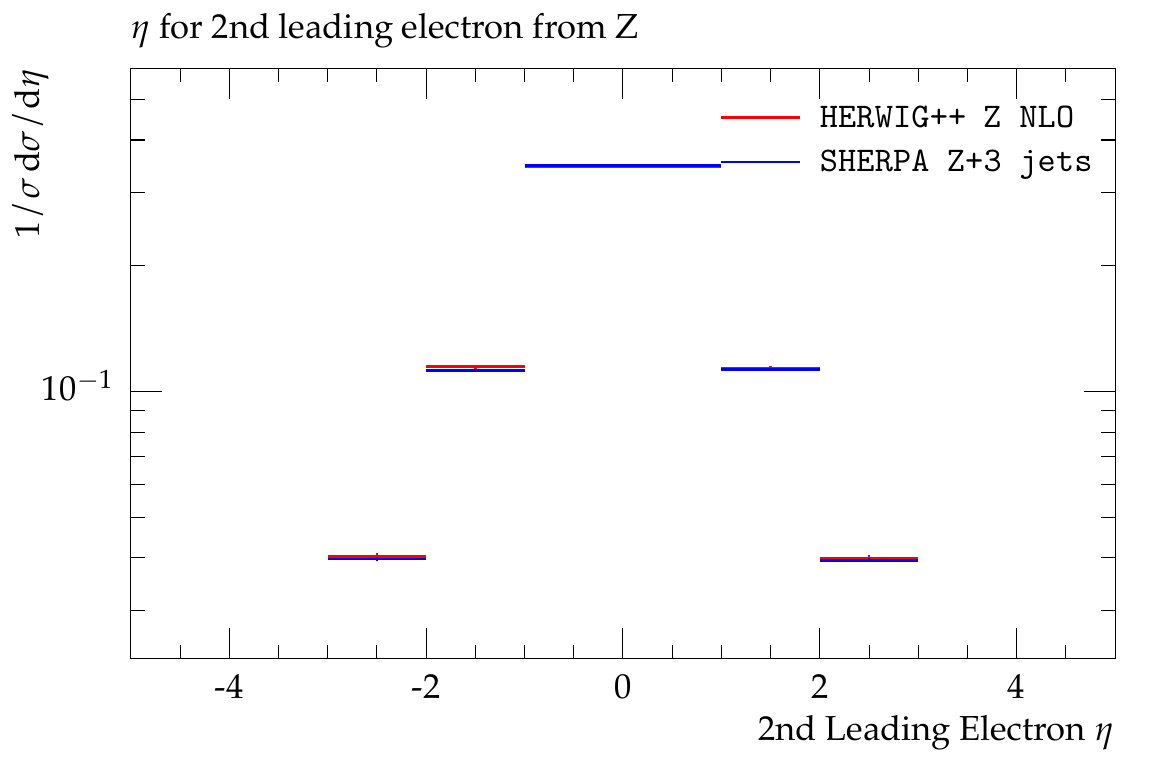}

\caption{\label{fig:LHC4}Comparison plots for LHC (7 TeV) energy, for \emph{Z}
production in electron channel: the \emph{Z} $\eta$ (left), leading
electron $\eta$ (center) and 2nd leading electron $\eta$ (right),
for ${\tt HERWIG++}$ \emph{Z} NLO and ${\tt SHERPA}$ \emph{Z}+3
jets, both with MPI simulation, and ${\tt SHERPA}$ with
optimized ${\tt K\text{\_}PERP}$ and ${\tt CTEQ6L1}$ PDF,
scale parameter 2.5 GeV. }

\end{figure}

\begin{figure}[!h]
\includegraphics[width=0.32\columnwidth]{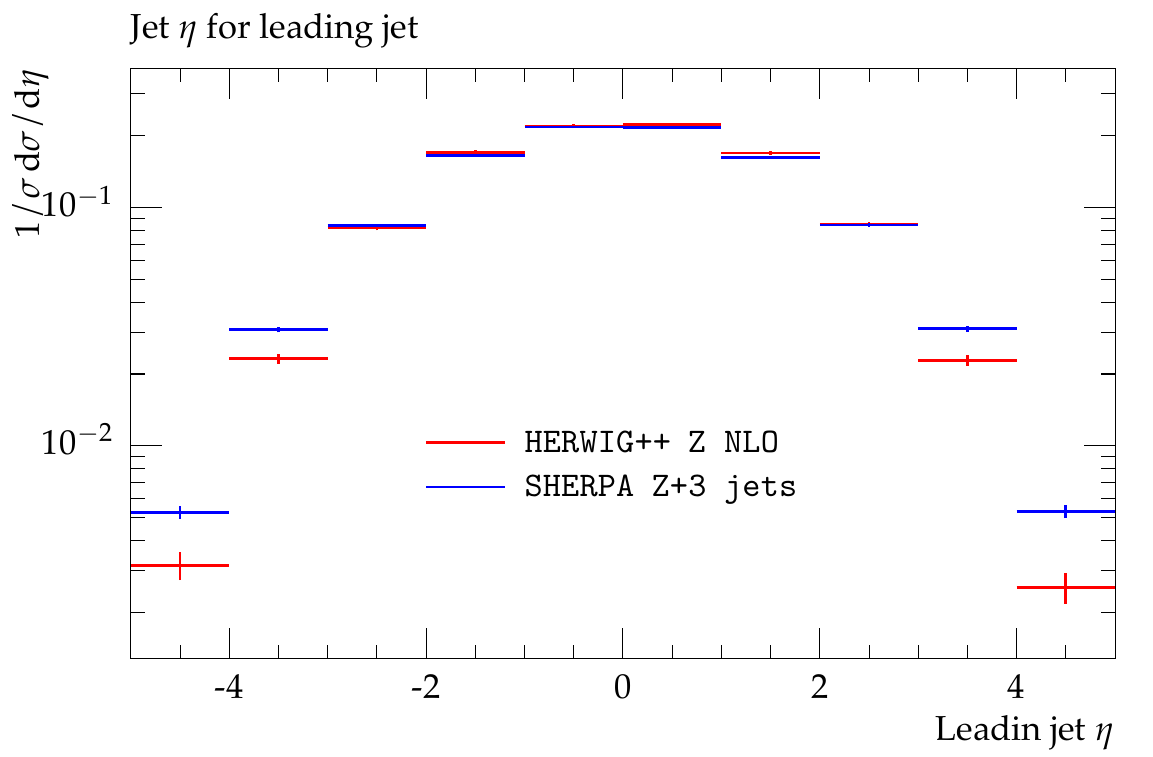}
\includegraphics[width=0.32\columnwidth]{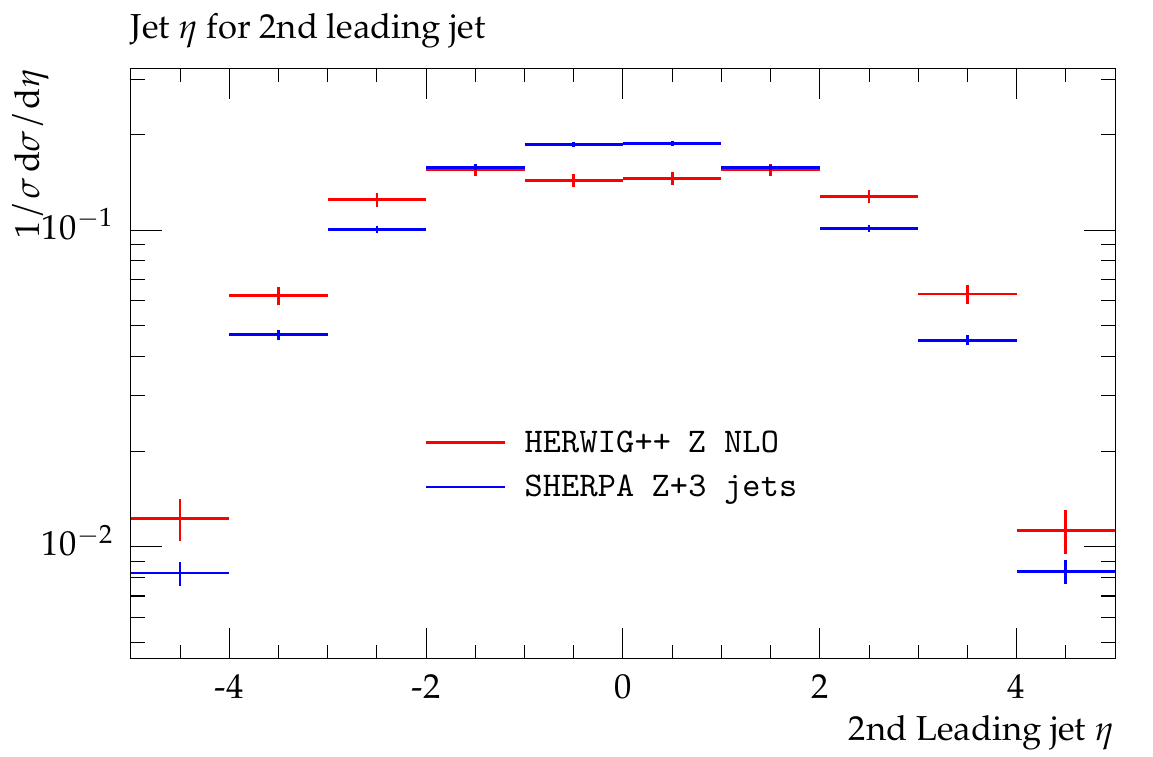}
\includegraphics[width=0.32\columnwidth]{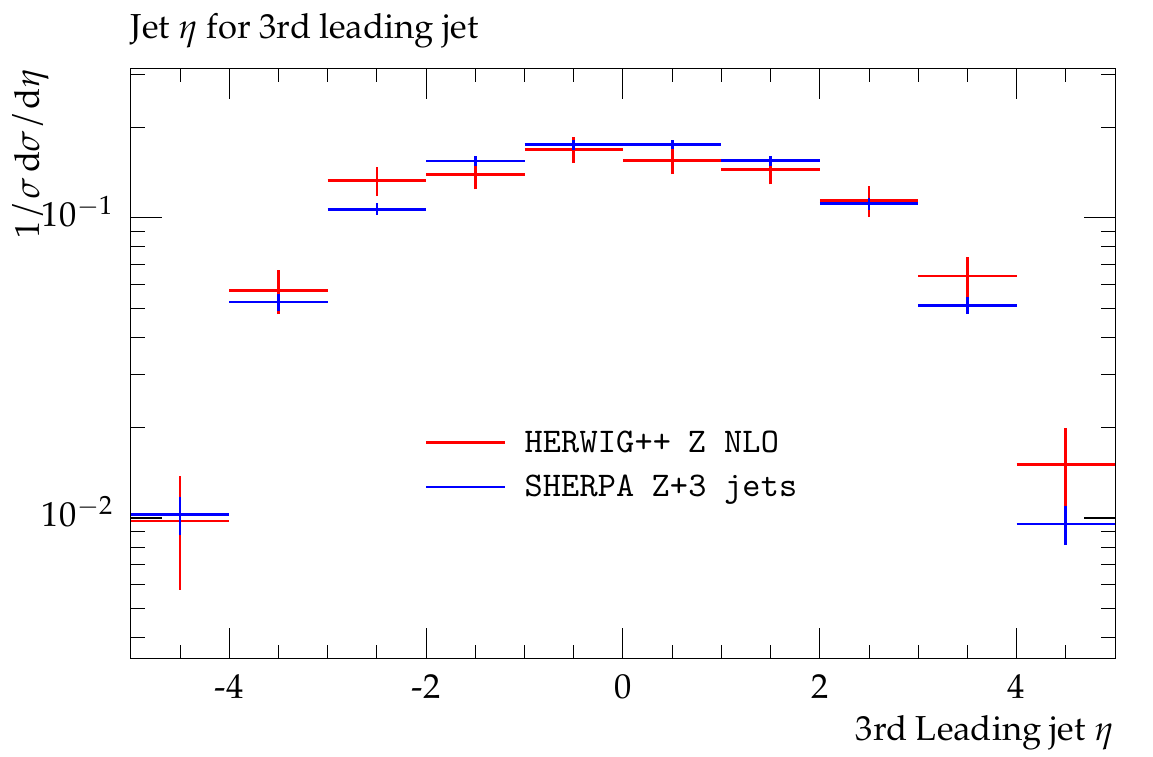}

\caption{\label{fig:LHC5}Comparison plots for LHC (7 TeV) energy, for \emph{Z}
production in electron channel: the leading (left), second (center)
and third (right) leading jet $\eta$, for ${\tt HERWIG++}$ \emph{Z}
NLO and ${\tt SHERPA}$ \emph{Z}+3 jets, both with MPI simulation,
and ${\tt SHERPA}$ with optimized ${\tt K\text{\_}PERP}$ and ${\tt CTEQ6L1}$
PDF, scale parameter 2.5 GeV. }

\end{figure}

\begin{figure}[!h]
\includegraphics[width=0.32\columnwidth]{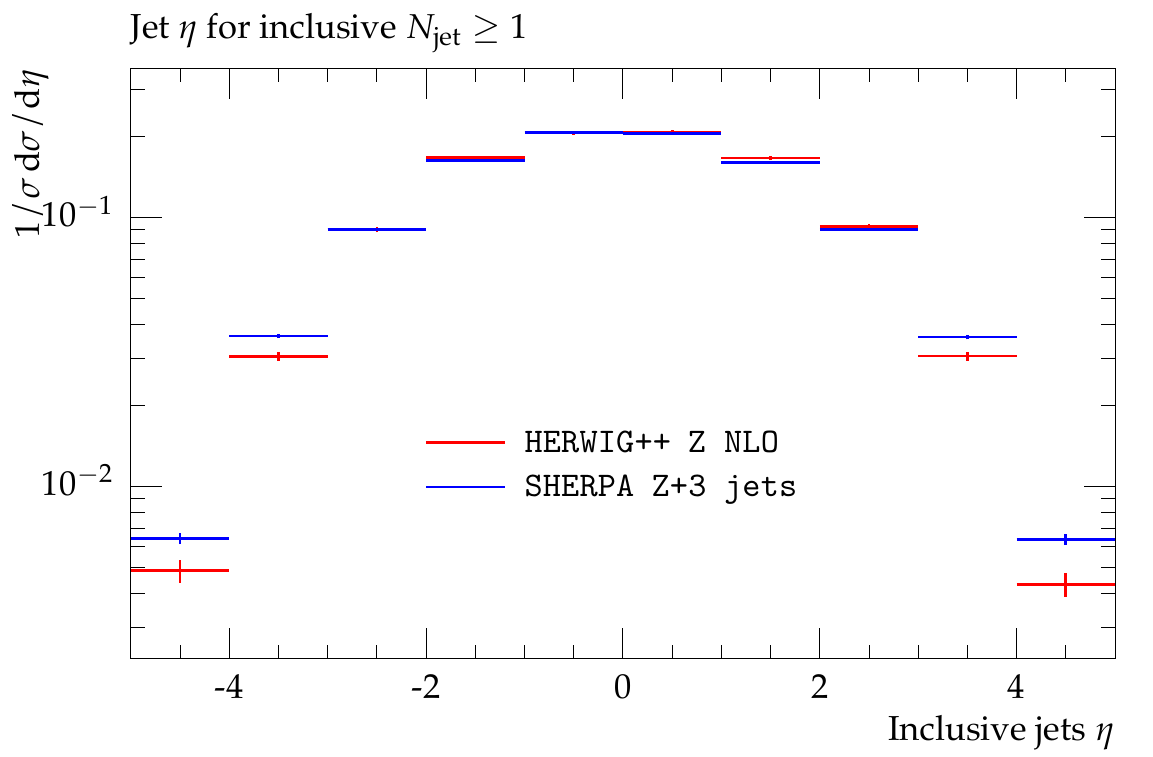}
\includegraphics[width=0.32\columnwidth]{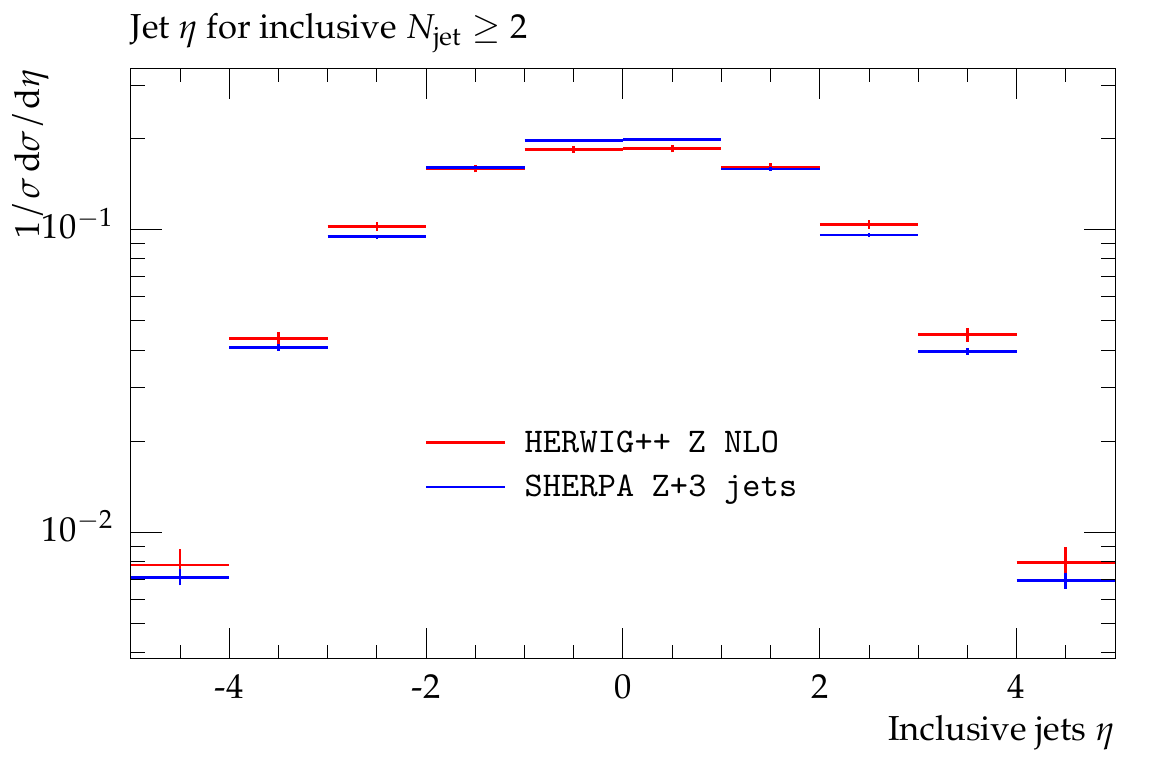}
\includegraphics[width=0.32\columnwidth]{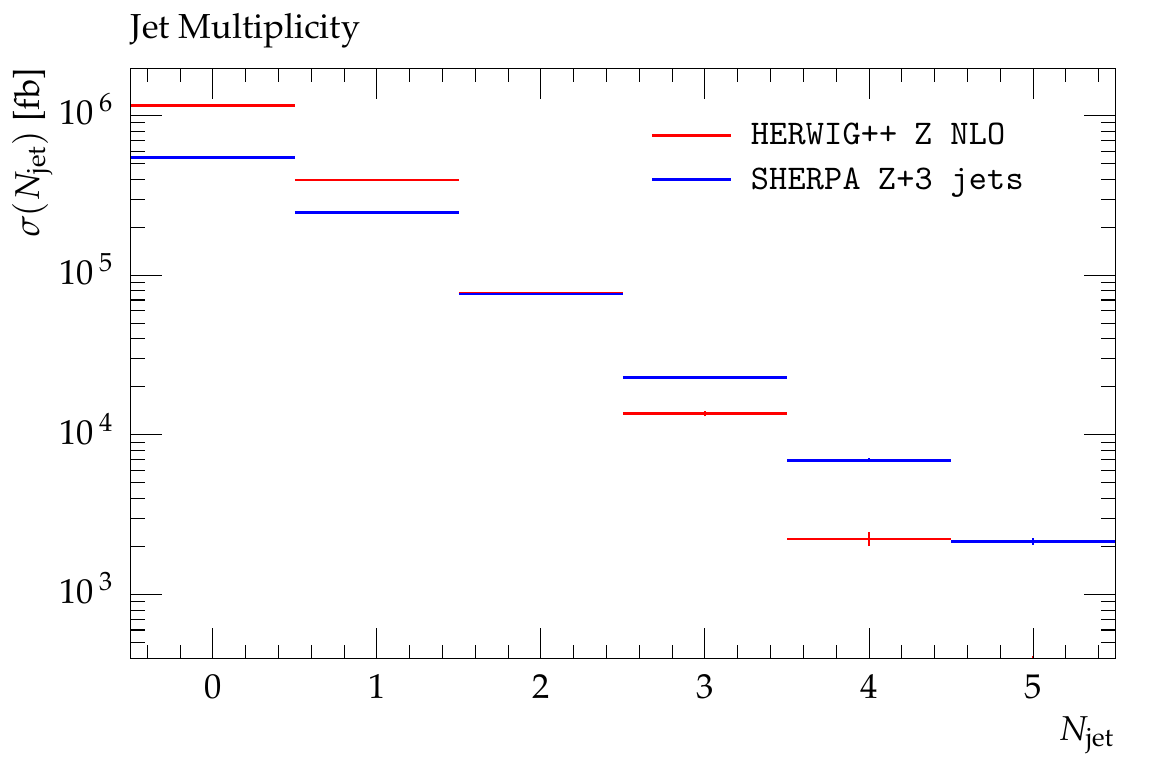}

\caption{\label{fig:LHC6}Comparison plots for LHC (7 TeV) energy, for \emph{Z}
production in electron channel: the jet $\eta$ for inclusive $N_{jet}\geq1$
(left), jet $\eta$ for inclusive $N_{jet}\geq2$ (center) and jet
multiplicity (right), for ${\tt HERWIG++}$ \emph{Z} NLO and ${\tt SHERPA}$
\emph{Z}+3 jets, both with MPI simulation, and ${\tt SHERPA}$
with optimized ${\tt K\text{\_}PERP}$ and ${\tt CTEQ6L1}$ PDF,
scale parameter 2.5 GeV. }

\end{figure}

The kinematic variables plotted are shown in Figs. \ref{fig:LHC1}
to \ref{fig:LHC6}: the transverse momentum of the first, second and
third leading jets, and the inclusive $p_{\perp}$ for 1 and 2 or
more jets in the event; the transverse momentum of the two leptons
used to reconstruct the \emph{Z} boson, and of the \emph{Z}; The
pseudorapidity of \emph{Z} and jets; the invariant mass of the \emph{Z};
the jet multiplicity in the event. All the plots are normalized to
their integral, except the jet multiplicity, that is the cross section
of production of the event with $n$-jets. 

The \emph{Z} $p_{\perp}$ (Fig. \ref{fig:LHC1} left) shows a discrepancy
in the low transverse momentum region: ${\tt SHERPA}$ simulates fewer
events in this region, compared to ${\tt HERWIG++}$. However, for
the leading electrons $p_{\perp}$ of the event (Fig. \ref{fig:LHC1}
center and right), the discrepancy isn't as much as for the boson $p_{\perp}$.

For the leading jet $p_{\perp}$ (Fig. \ref{fig:LHC2} left), the
behaviour is also discrepant -
the ${\tt SHERPA}$ generator simulates more events in the low transverse
momentum region, and fewer events in medium and high transverse momentum,
although in the later ones it agrees with ${\tt HERWIG++}$ inside
the statistical uncertainties. For the second and third leading jets
$p_{\perp}$ (Fig. \ref{fig:LHC2} center and right), there is good
agreement inside MC uncertainties. 

The jet $p_{\perp}$ for $N_{jet}\geq1$ and $N_{jet}\geq2$, that is the transverse
momentum of all jets in the event that pass the cuts $p_{\perp}$ > 30 GeV and 
$|\eta|$ < 2.1 for events with at least one or two jets 
(Fig. \ref{fig:LHC3} left and center, respectively) agree in both
generators inside the errors, as well as the invariant mass of the
\emph{Z} boson (Fig. \ref{fig:LHC3} right). 

The pseudorapidity for the \emph{Z} boson and leading electrons (Fig.
\ref{fig:LHC4}) agree for the generators, inside the errors. However,
for the leading jets, the behaviour is different: while for the leading
jet (Fig. \ref{fig:LHC5} left) the region in which $|\eta|$ < 3
is agreed, the region in the range $3<|\eta|<5$ shows fewer events
for the ${\tt HERWIG++}$ generator than for ${\tt SHERPA}$. For
the second leading jet (Fig. \ref{fig:LHC5} center), ${\tt SHERPA}$
has more events in the central region and fewer events in the range
$3<|\eta|<5$. For the third leading jet (Fig. \ref{fig:LHC5}
right), due to high statistical errors, both generators agree in full
range. In the jet inclusive $\eta$ for $N_{jet}\geq1$ (Fig. \ref{fig:LHC6}
left) the range of $3<|\eta|<5$ shows more events for ${\tt SHERPA}$
generator, while for $N_{jet}\geq2$ (Fig. \ref{fig:LHC6} center)
the generators agree inside errors. 

For the cross section as a function of jet multiplicity (Fig. \ref{fig:LHC6}
right), the simulation of \emph{Z} production at NLO (${\tt POWHEG}$
formalism) has a greater cross section than ${\tt SHERPA}$ \emph{Z}+3
jets at LO for the production of one and two jets. However, because
${\tt SHERPA}$ takes into account matrix elements up to 3 jets, the
description for higher jet multiplicities have a greater cross section
for the ${\tt SHERPA}$ generator. The data will usually have greater cross
sections than predicted with the generators, because for having
equal cross sections one should simulate up to infinite orders of
QCD. 

\newpage{}

\section{Conclusions}

We studied the effects of adding a next to leading order term in the
Monte Carlo simulation for \emph{Z} production in hadron colliders.
In addition, the influence of adding a matrix element correction in
the leading order calculations for improving the showering in the
parton shower formalism, and the influence of some theoretical parameters
that enter as input in the Monte Carlo programs. 

We could see that the use of the NLO term, studied here in the ${\tt POWHEG}$
formalism as implemented in the ${\tt HERWIG++}$ generator, improves
the prediction of the cross sections of the processes, as well as
the behaviour of the physics observables, specially in the region
of higher transverse momentum. The implementation of the matrix element
correction in the parton shower formalism also improves the description
of the data in the region of high transverse momentum, compared to
the calculations in leading order without the correction. 

It was also seen that both ${\tt SHERPA}$ and ${\tt HERWIG++}$ generators
show a systematic behaviour lower than the Tevatron data in the region
of mid-range transverse momentum of the \emph{Z} boson. The D0 analysis
performed in the muon channel has a better Monte Carlo description
than the one performed in the electron channel, and doesn't use any
kind of extrapolation based in theory. According to the new muon analysis,
if the same theory based corrections are applied, the muon data agree
with the electron channel data. So, these theoretical dependent corrections
could be responsible for the disagreement and the systematic behaviour
between the Monte Carlo and the Tevatron data in mid range transverse
momentum of the \emph{Z}. 

The underlying event analysis showed that it is possible to choose
a good tune for the parameters in the multiple parton interactions
model (Amisic) for the ${\tt SHERPA}$ generator, using the PDF
that best describes the \emph{Z} $p_{\perp}$ data (${\tt CTEQ6L1}$).
However, for the ${\tt HERWIG++}$ generator, there was no parameter
selection that could describe well the underlying event data, and
the setup used was the standard one, based in the best description
of other physics observables by the authors. 

In the analysis of the balance of the \emph{Z} $p_{\perp}$ against
the leading jet $p_{\perp}$ and the sum of $p_{\perp}$ of all jets
in the event was possible to see the effects that the MPI
model have in the region of low transverse momentum. This is done in the appendix.  

The analysis for LHC first run energy (7 TeV) shows that some kinematic
quantities have a different prediction from ${\tt HERWIG++}$ \emph{Z}
NLO when compared to ${\tt SHERPA}$ \emph{Z}+3 jets simulation, such as the
low region of the \emph{Z} boson transverse momentum and leading jet
$p_{\perp}$ and leading jets pseudorapidity in the high absolute
value of $\eta$ range. For the cross section of jet production, the
${\tt HERWIG++}$ \emph{Z} NLO generator predicts a higher cross section
for the first and second jets, while the ${\tt SHERPA}$ \emph{Z}+3
jets show higher cross sections for higher jet multiplicities. A full
comparison once LHC has enough luminosity will show some features
that the Monte Carlo generators will have to be able to deal with,
implementing in their processes new information that the new data
will provide.

\section*{Acknowledgements}

We would like to thank Frank Siegert, Prof. F. Krauss, Prof. Peter
Richardson, for many useful discussions. 

This work was supported by the Marie Curie Research Training Network
\textquotedblleft{}MCnet\textquotedblright{} (contract number MRTN-CT-2006-035606).

\section*{Appendix}

\paragraph{Jet Recoil}

Because primarily the \emph{Z} $p_{\perp}$ should be balanced in
the event with the $p_{\perp}$ of the leading jet (or all jets in
the event), the study of the shape of the jet $p_{\perp}$ should
give some information about the systematic difference in the description
of the $Z\ \rightarrow e^{+}e^{-}$ $p_{\perp}$, and the leading
jet $p_{\perp}$ in \emph{Z} events. With the data, it is impossible
to isolate the different effects of MPI and hard scatters,
or extend below the $p_{\perp}$ cuts used in analysis. So, we perform
some studies to check the $p_{\perp}$ balance in the MC samples,
to see if the \emph{Z} $p_{\perp}$ was summing up to zero when added
to the $p_{\perp}$ of the leading jet, the sum of the $p_{\perp}$
of all jets and, for a more fundamental check of balance, with the
sum of the $p_{\perp}$ of all particles found in the event. 

The first sanity check was done to assure that the sum of the transverse
momentum of all particles in the event was balanced, so that both
MC generators and ${\tt RIVET}$ were dealing with the events correctly.
After this check, the \emph{Z} $p_{\perp}$ balance against the sum
of the $p_{\perp}$ of all jets in the event and leading jet was analyzed,
for both ${\tt SHERPA}$ and ${\tt HERWIG++}$ generators, with multiple
parton interactions (MPI) turned on and off. 

For the study of the balance of the leading jet, the Tevatron analysis cuts are used ($p_{\perp}$(jet) > 20 GeV and |$\eta$| < 2.1), so it can be comparable to the jet $p_{\perp}$ and Z $p_{\perp}$ spectra. In the other hand, very loose cuts ($p_{\perp}$(jet) > 5 GeV and |$\eta$| < 10) were applied to analyse the balance of the Z $p_{\perp}$ against the sum of all jets in the events, so it's possible to truly test the balance between jets and Z $p_{\perp}$ at generator level. 

\begin{figure}[!h]
\includegraphics[width=0.5\columnwidth]{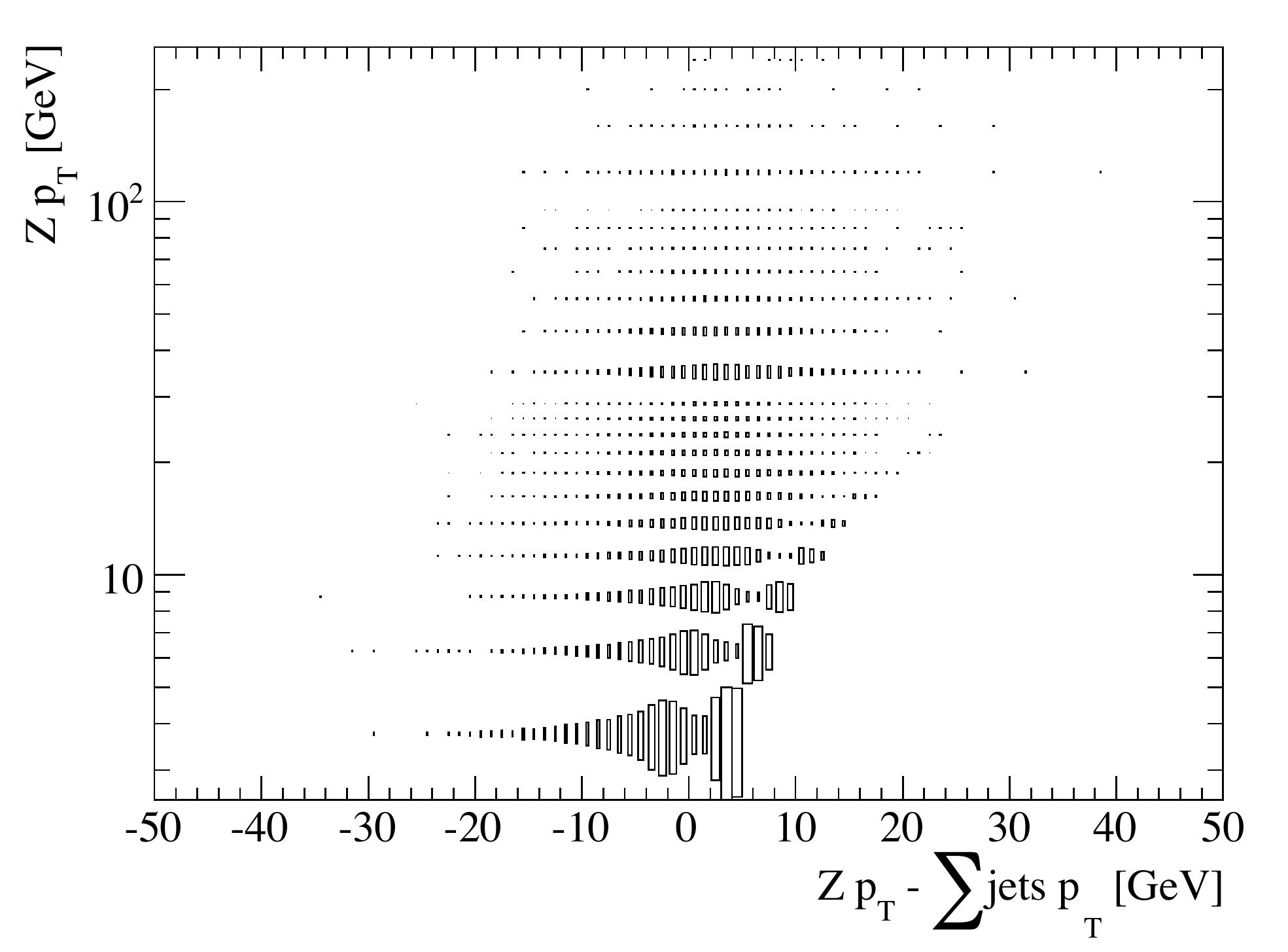}
\includegraphics[width=0.5\columnwidth]{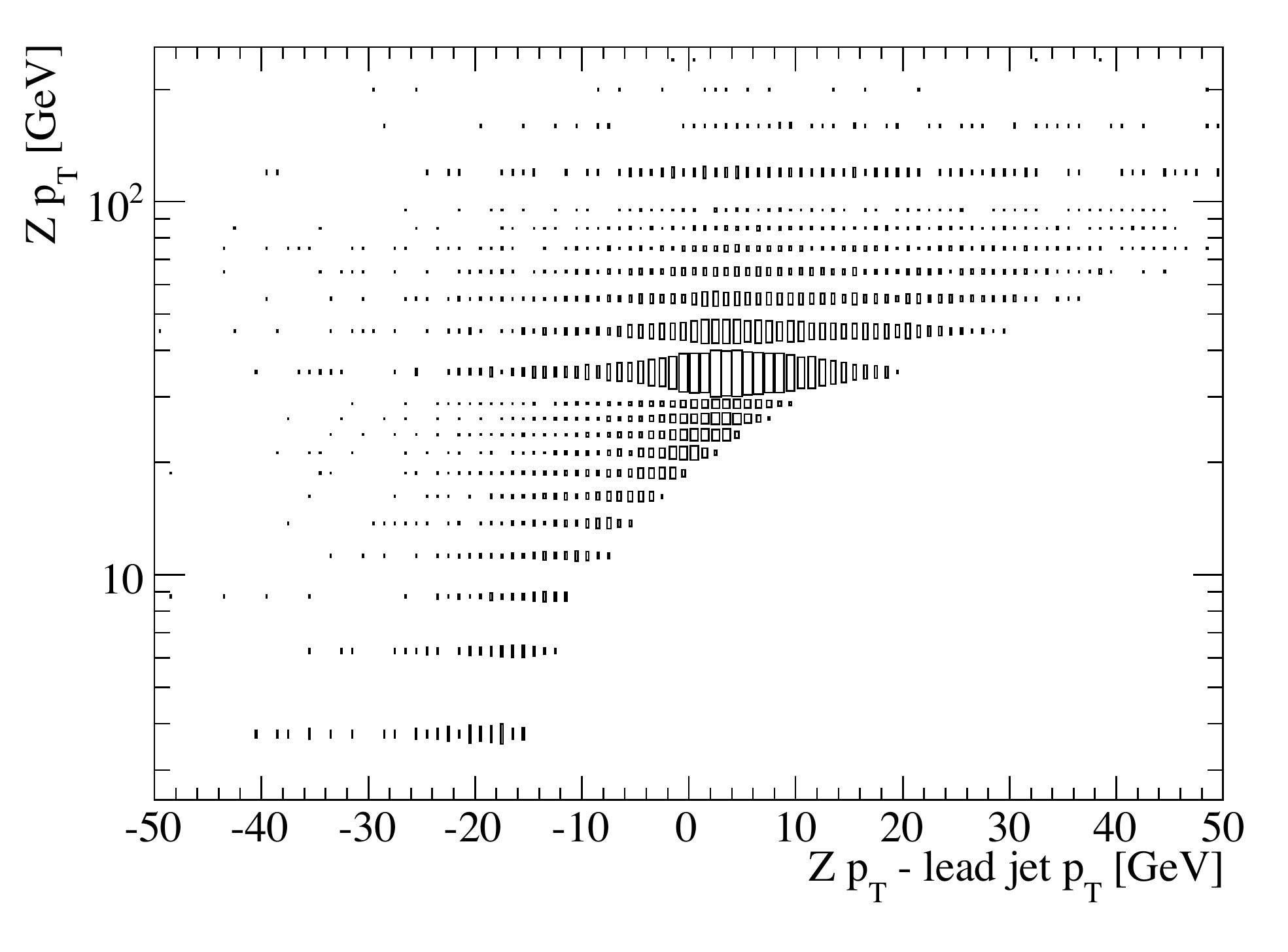}

\includegraphics[width=0.5\columnwidth]{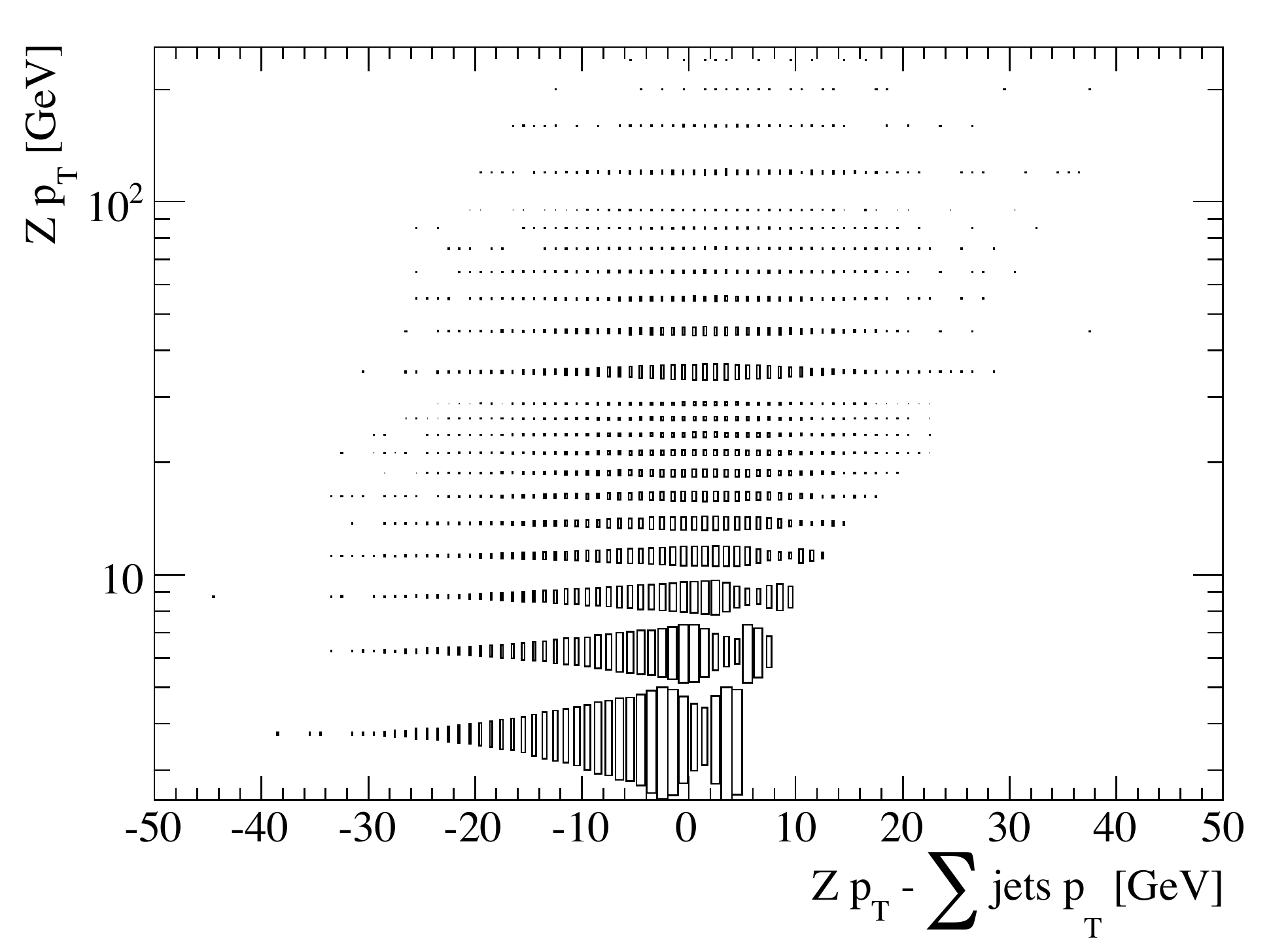}
\includegraphics[width=0.5\columnwidth]{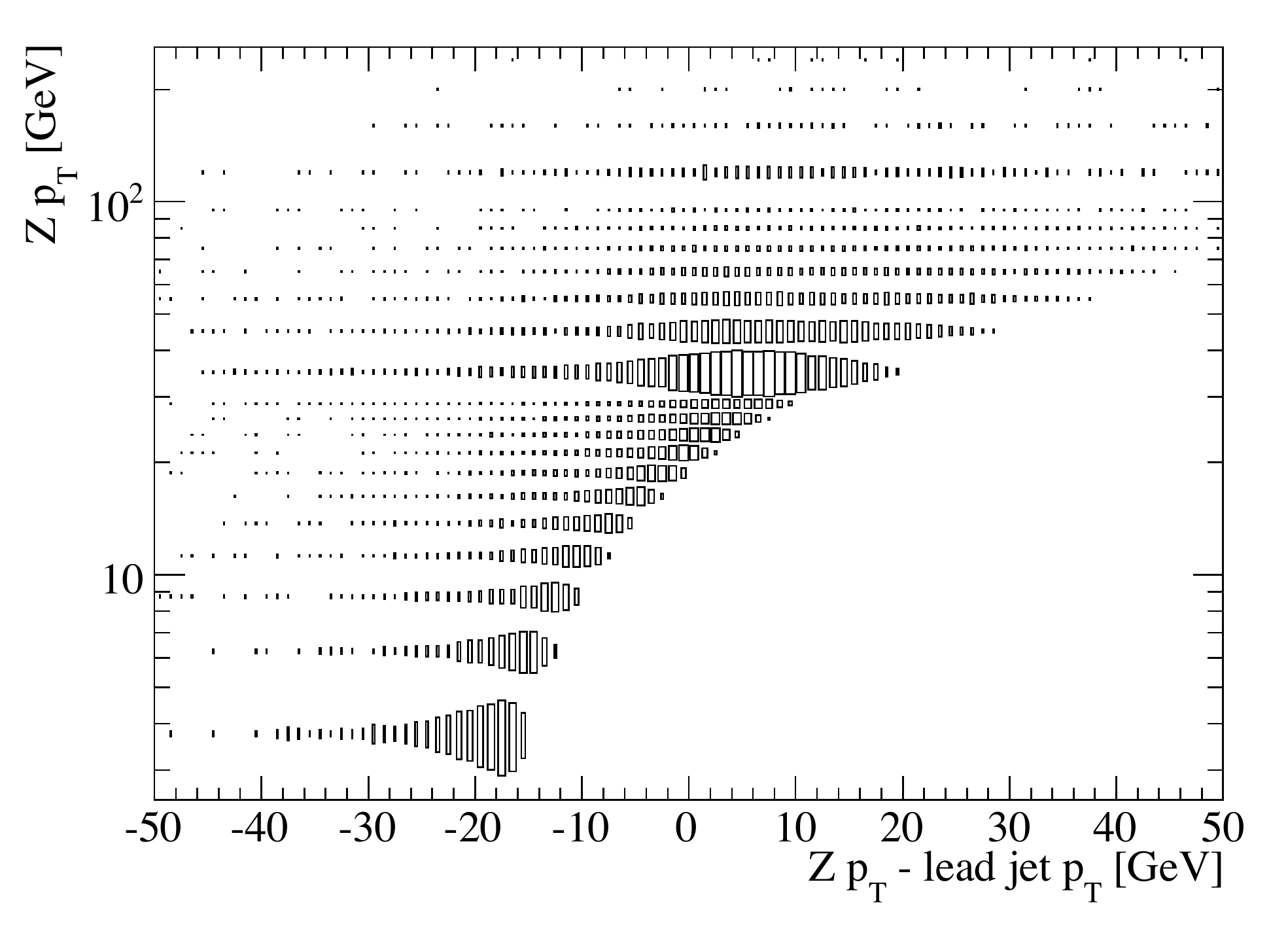}

\caption{\label{fig:3slicesMPIon}Difference between \emph{Z} $p_{\perp}$
and: sum of jets $p_{\perp}$ (left) and leading jet $p_{\perp}$
(right), as a function of bins of \emph{Z} $p_{\perp}$ for ${\tt HERWIG++}$ (upper) and ${\tt SHERPA}$ (lower), MPI
turned on in default for ${\tt HERWIG++}$ and with ${\tt CTEQ6L1}$
PDF, scale of 2.5 GeV and optmized ${\tt K\text{\_}PERP}$
in ${\tt SHERPA}$. The cuts on the analysis are, for the leading jet,  the same as Tevatron
analysis: $p_{\perp}(jet)>20\text{ GeV}$ and $|\eta|<2.1$, and for the sum of jets, $p_{\perp}(jet)>5\text{ GeV}$ and $|\eta|<10$. At least one jet is required.}

\end{figure}

\begin{figure}[!h]
\includegraphics[width=0.5\columnwidth]{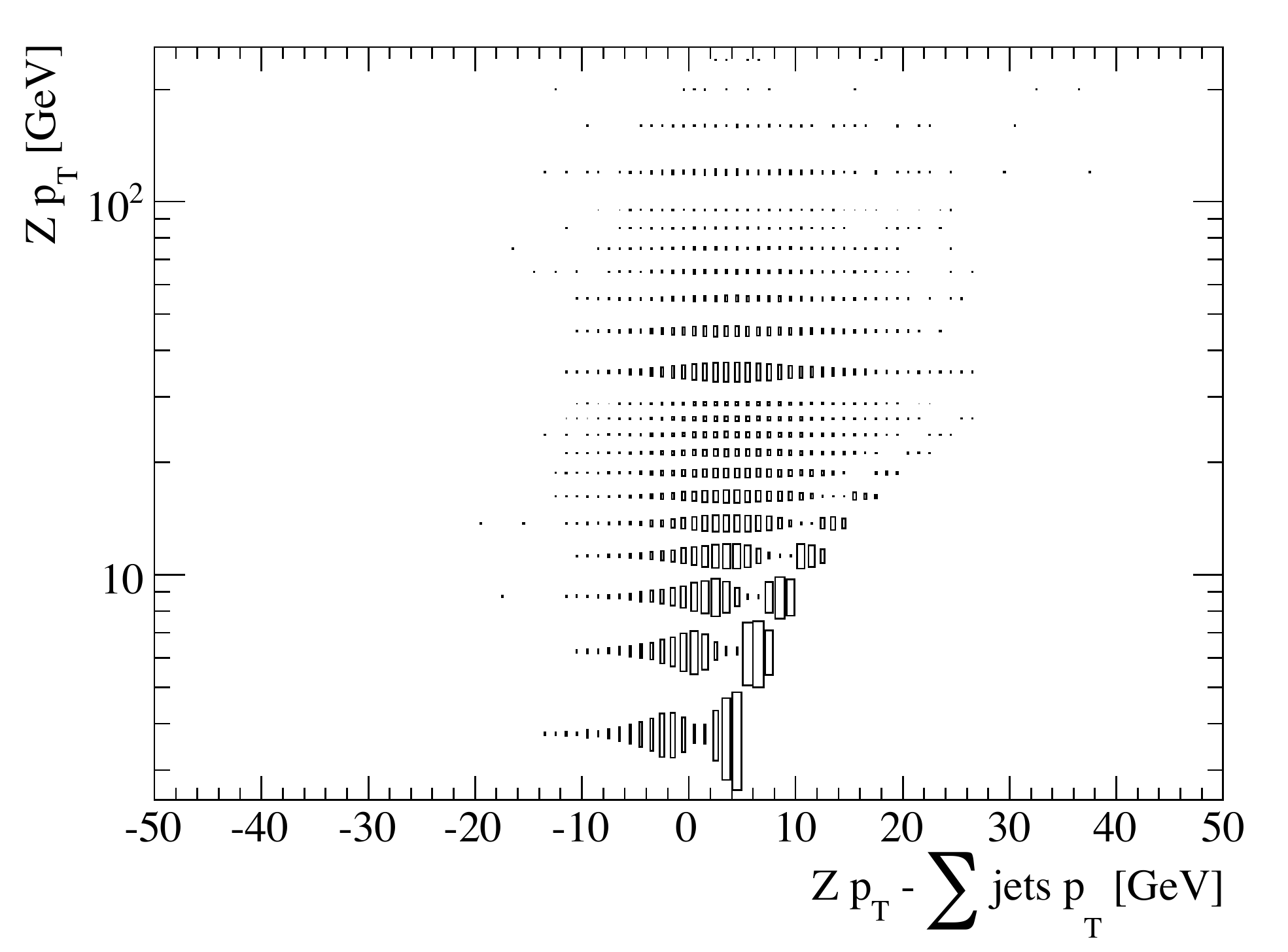}
\includegraphics[width=0.5\columnwidth]{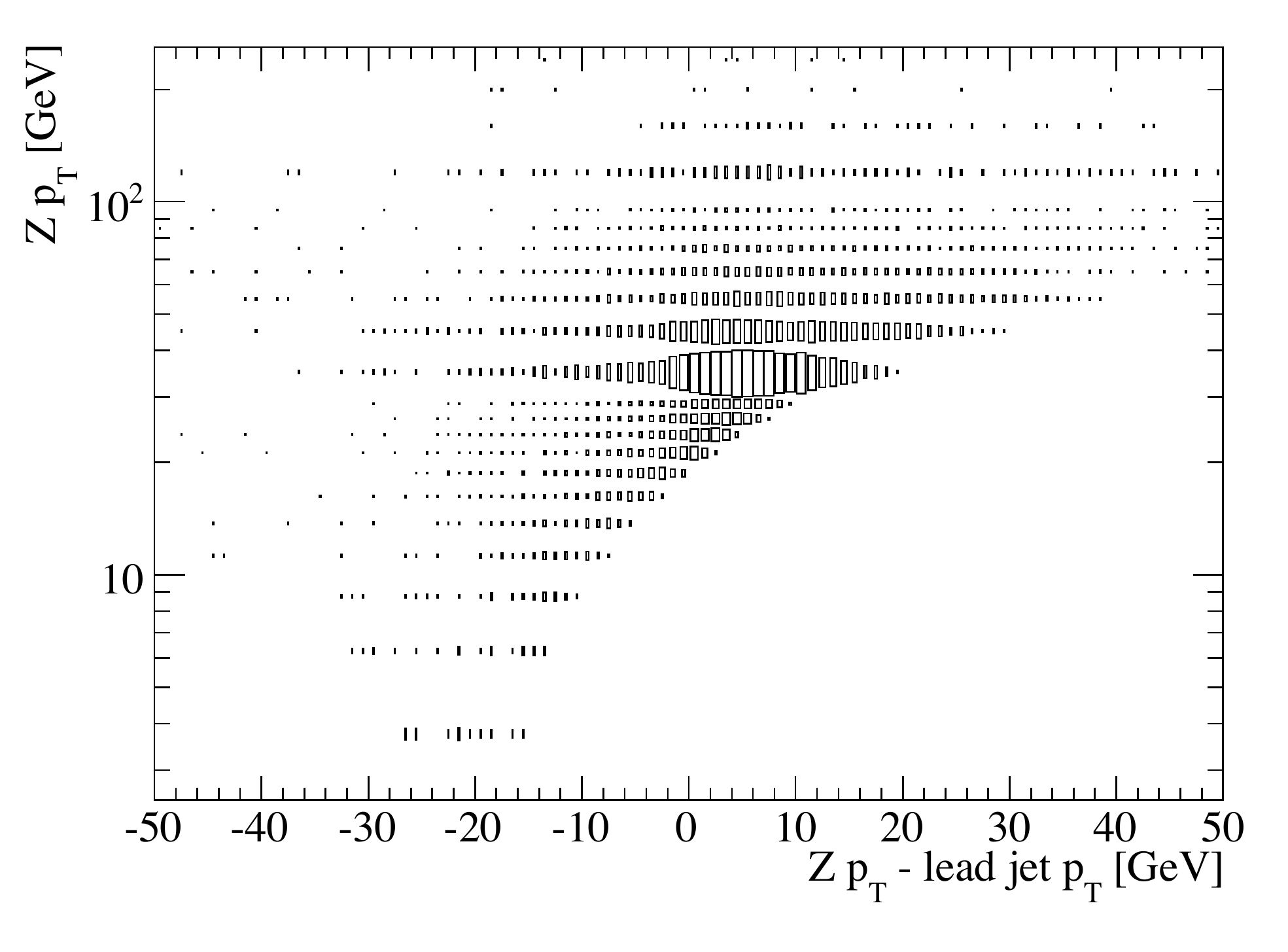}

\includegraphics[width=0.5\columnwidth]{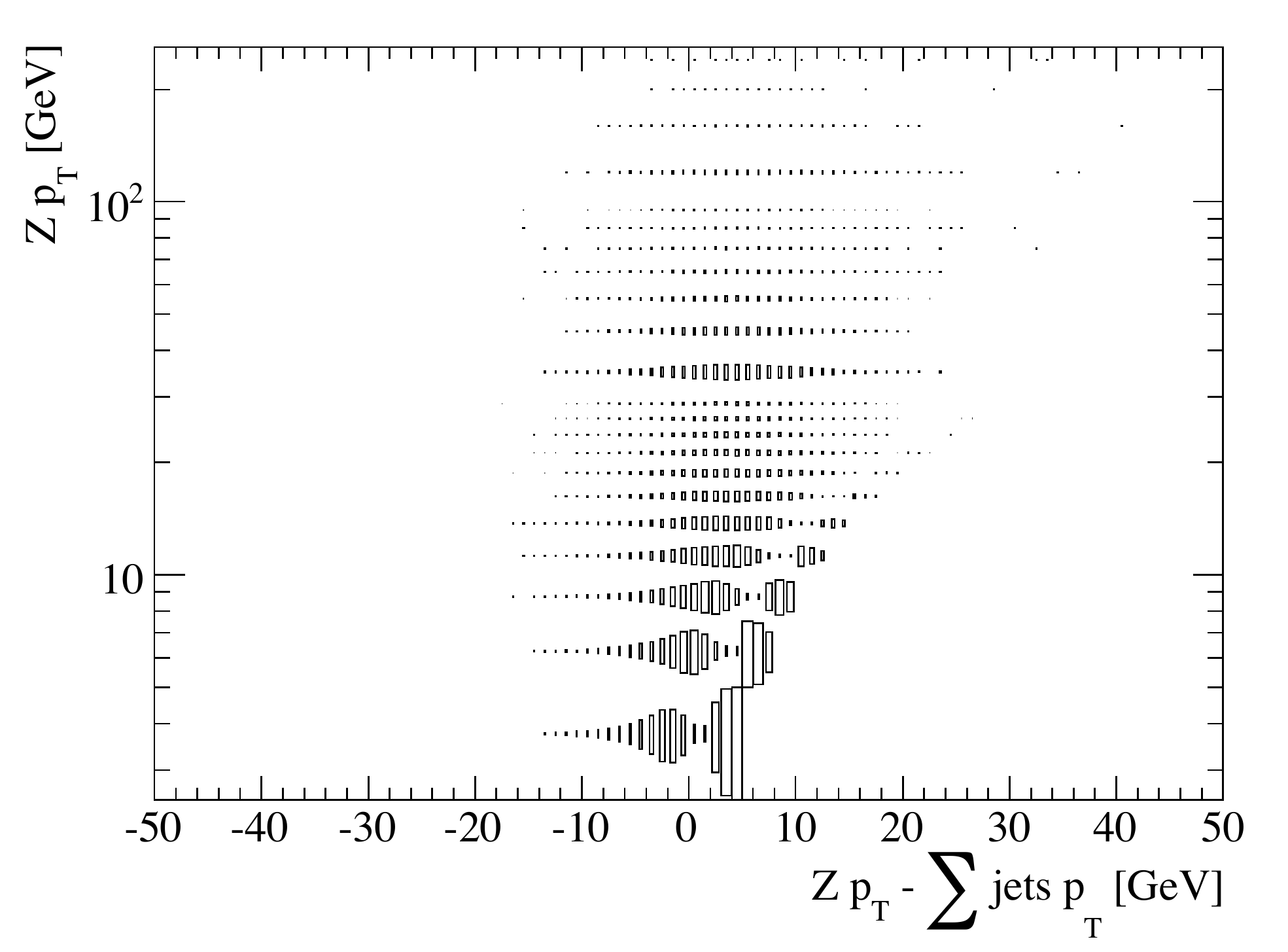}
\includegraphics[width=0.5\columnwidth]{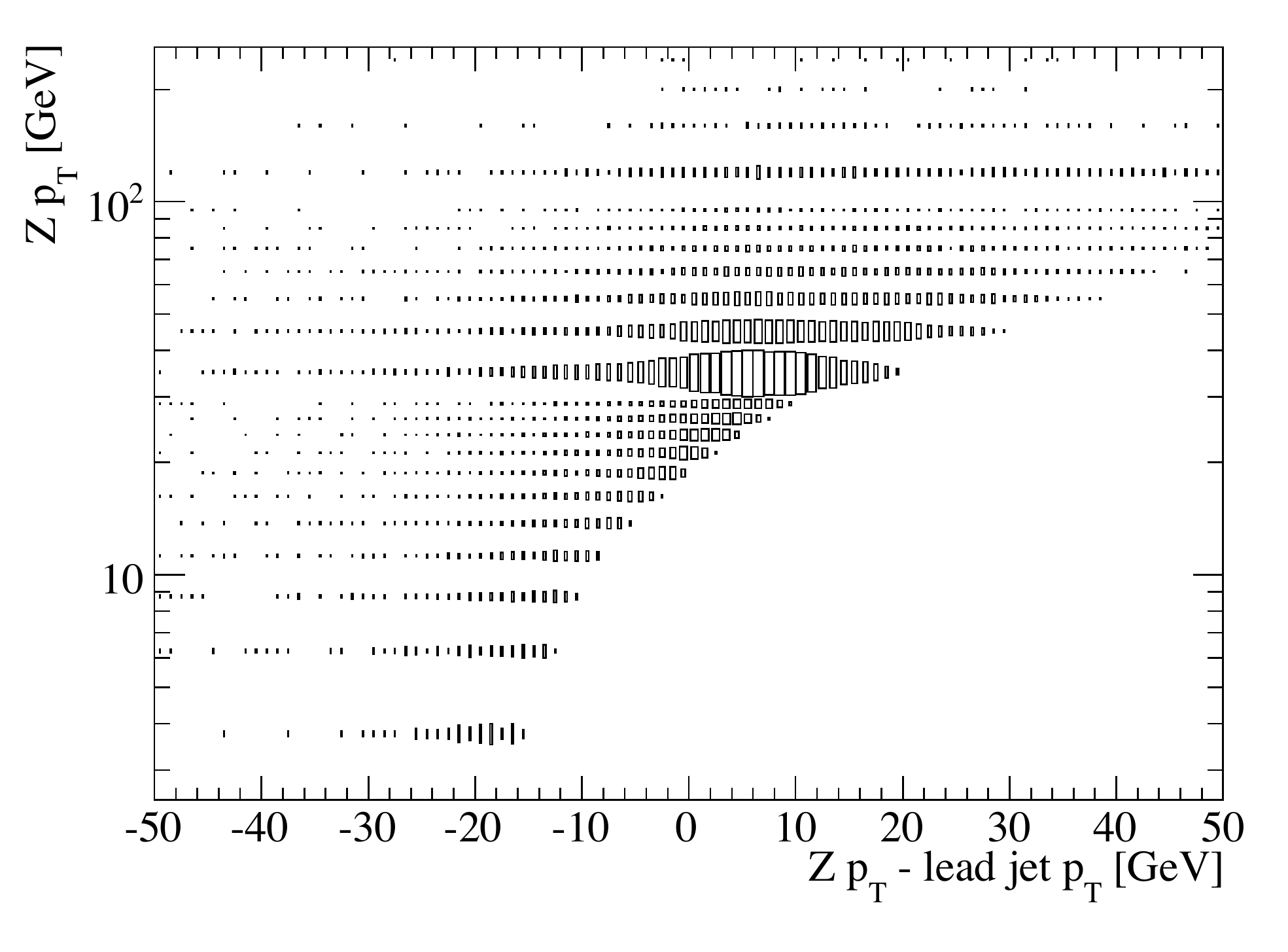}

\caption{\label{fig:3slicesMPIoff}Difference between \emph{Z} $p_{\perp}$
and: sum of jets $p_{\perp}$ (left) and leading jet $p_{\perp}$
(right), as a function of bins of \emph{Z} $p_{\perp}$ for ${\tt HERWIG++}$ (upper) and ${\tt SHERPA}$ (lower), MPI
turned off. ${\tt HERWIG++}$ has default PDF adjustment,
and ${\tt SHERPA}$ has ${\tt CTEQ6L1}$ PDF and optimized
${\tt K\text{\_}PERP}$. The cuts on the analysis are, for the leading jet,  the same as Tevatron
analysis: $p_{\perp}(jet)>20\text{ GeV}$ and $|\eta|<2.1$, and for the sum of jets, $p_{\perp}(jet)>5\text{ GeV}$ and $|\eta|<10$. At least one jet is required.}

\end{figure}

Fig. \ref{fig:3slicesMPIon} shows that with the MPI model
turned on, the ${\tt SHERPA}$ and ${\tt HERWIG++}$ generators have
different behaviour for low \emph{Z} $p_{\perp}$ region: ${\tt SHERPA}$
shows more events concentrated in the region of (\emph{Z} $p_{\perp}$ - jets $p_{\perp}$) = -17 for the leading jet and (\emph{Z} $p_{\perp}$ - jets $p_{\perp}$) in the range of [-5, 0] for the sum of jets,
while ${\tt HERWIG++}$ shows a behaviour that is very similar to that without the MPI model, 
in Fig. \ref{fig:3slicesMPIoff}, upper plots.

Fig. \ref{fig:3slicesMPIoff} shows that for the MPI model
off in the generators, the behaviour is similar. However in Fig. \ref{fig:3slicesMPIon} there is more $p_{\perp}$ in the jets, so this isn't balancing exactly all the transverse momentum 
that comes from the \emph{Z} boson. Specially in the low region of the boson $p_{\perp}$, we are probably seeing additional particles from MPI that don't contribute to the \emph{Z} $p_{\perp}$ but do contribute to the jet $p_{\perp}$.  

\begin{figure}[!h]
\includegraphics[width=0.32\columnwidth]{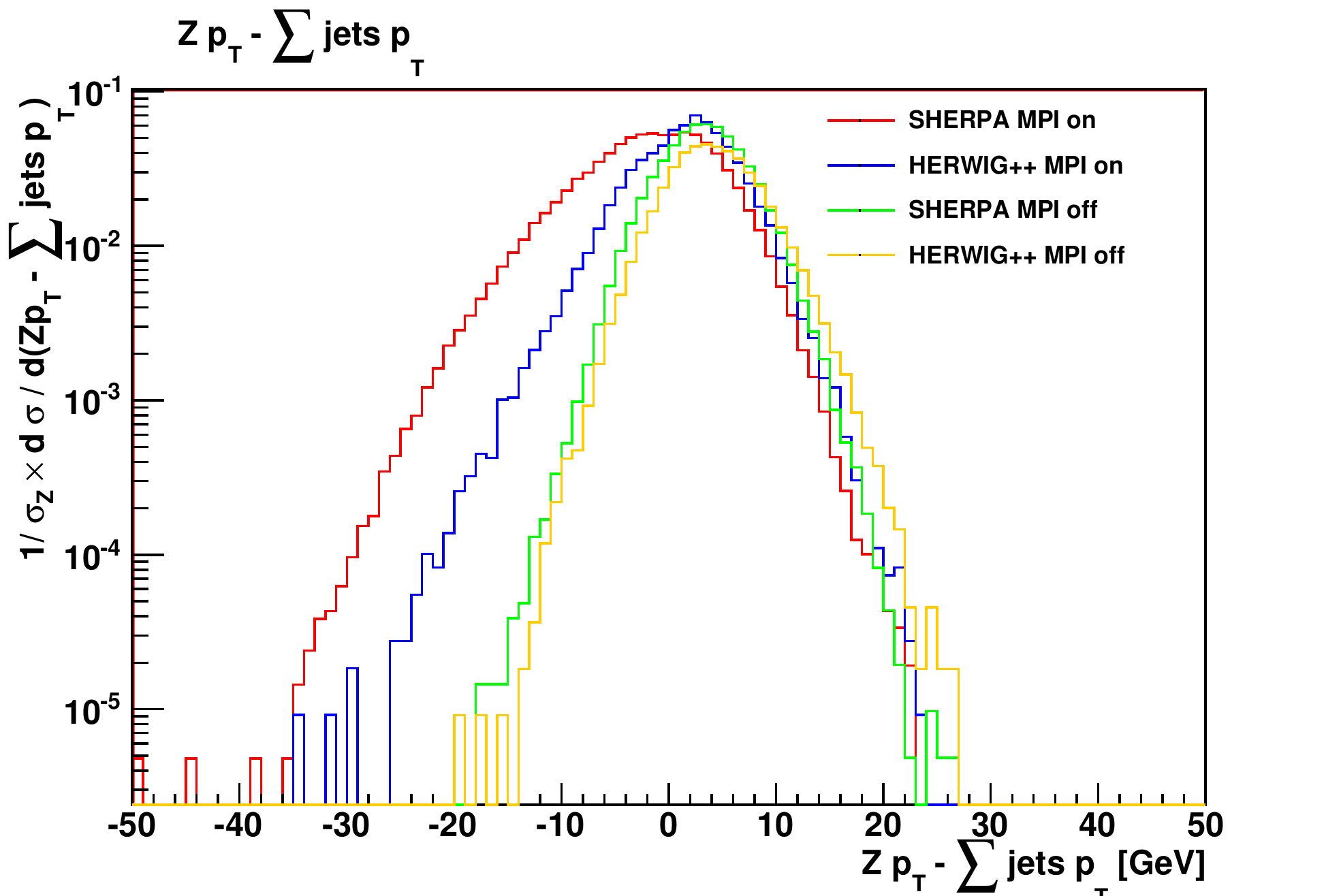}
\includegraphics[width=0.32\columnwidth]{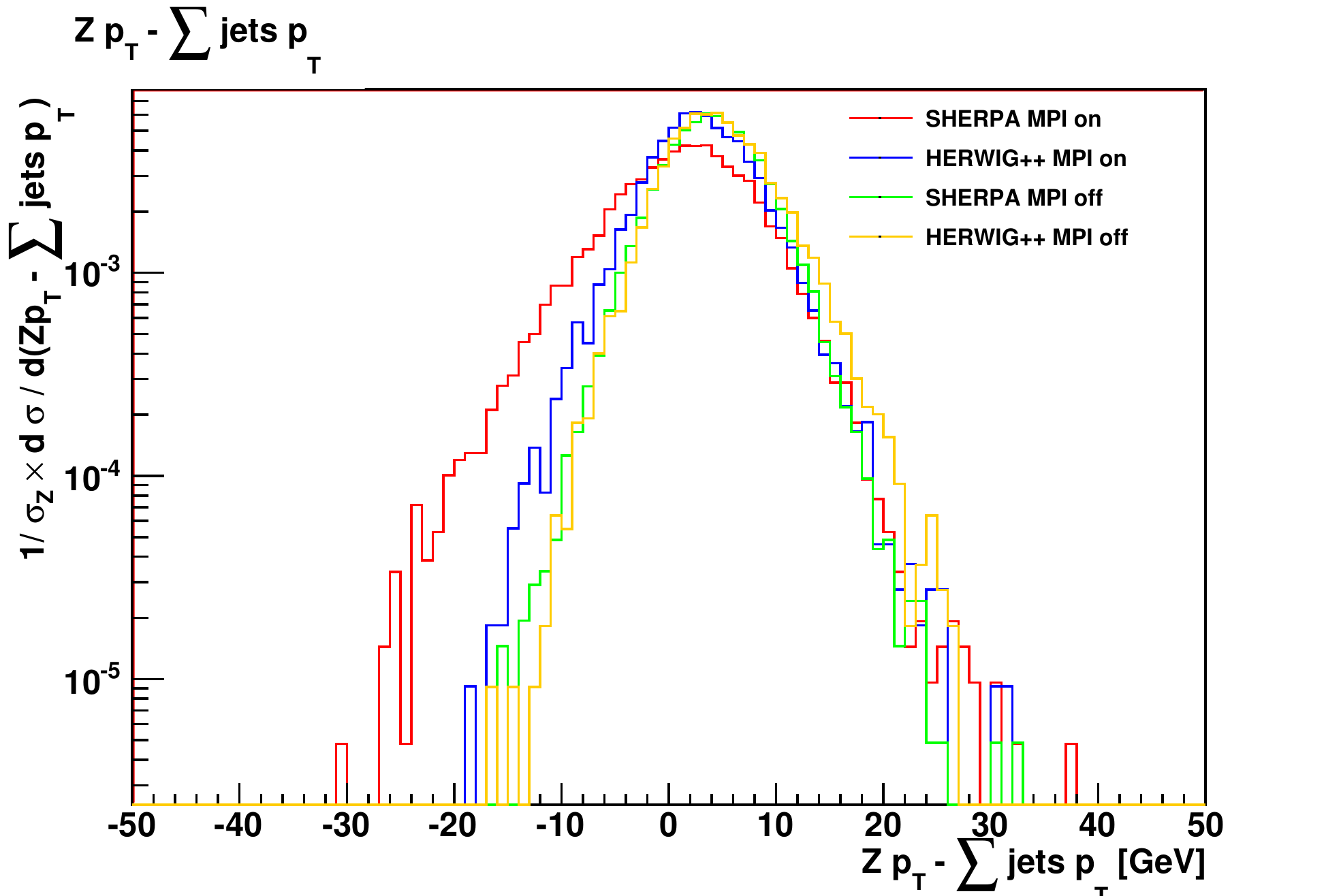}
\includegraphics[width=0.32\columnwidth]{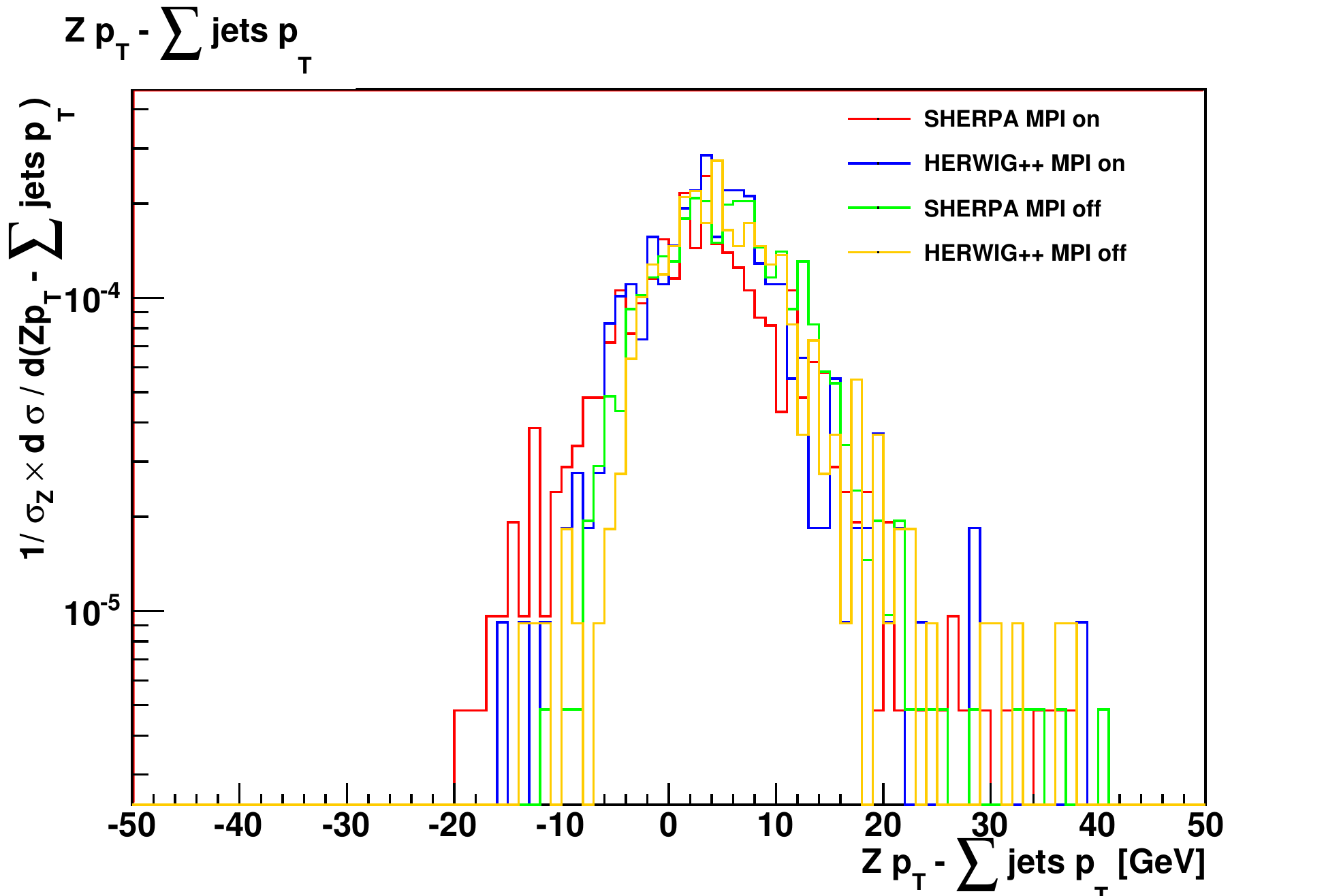}

\caption{\label{fig:slicesAllJets}The difference between \emph{Z} $p_{\perp}$
and the sum of all jets $p_{\perp}$ for three regions of boson $p_{\perp}$:
\emph{Z} $p_{\perp}$< 30 GeV, 30 GeV < \emph{Z} $p_{\perp}$ < 100
GeV, \emph{Z} $p_{\perp}$ > 100 GeV. The cuts on the analysis are $p_{\perp}(jet)>5\text{ GeV}$ and $|\eta|<10$. Generator
parameters are the same as in Figs. \ref{fig:3slicesMPIon} and \ref{fig:3slicesMPIoff}.}

\end{figure}

\begin{figure}[!h]
\includegraphics[width=0.32\columnwidth]{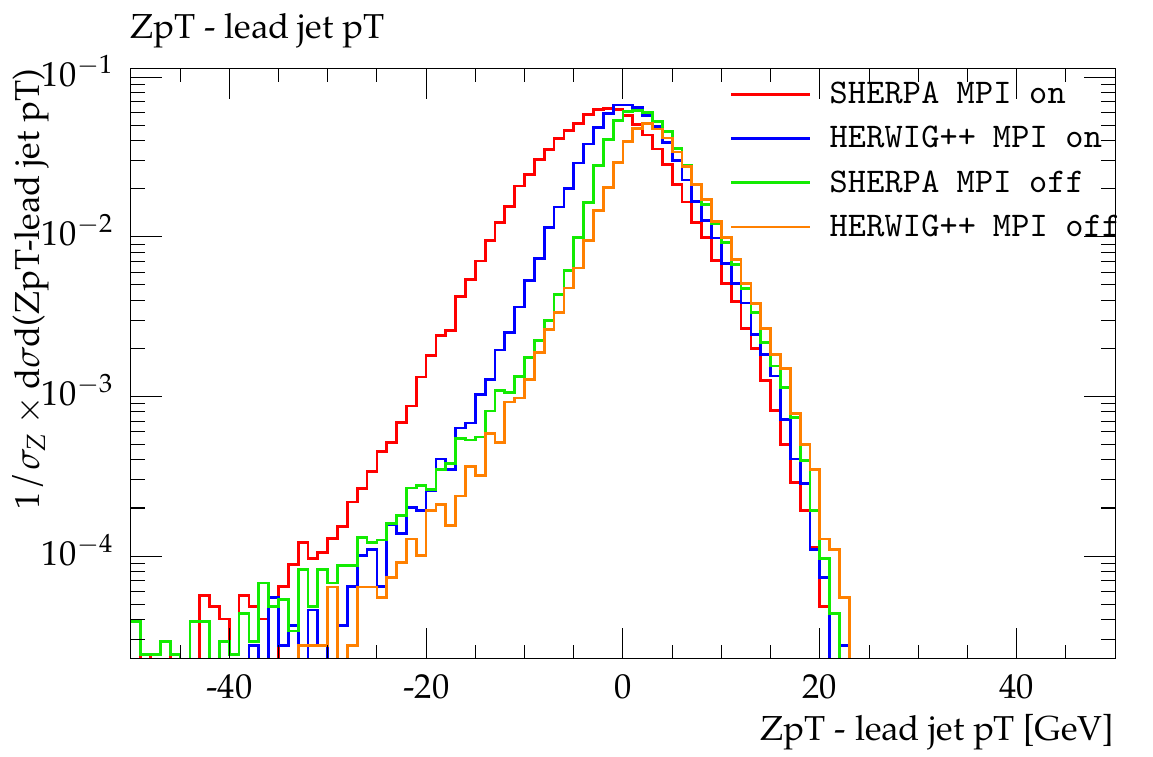}
\includegraphics[width=0.32\columnwidth]{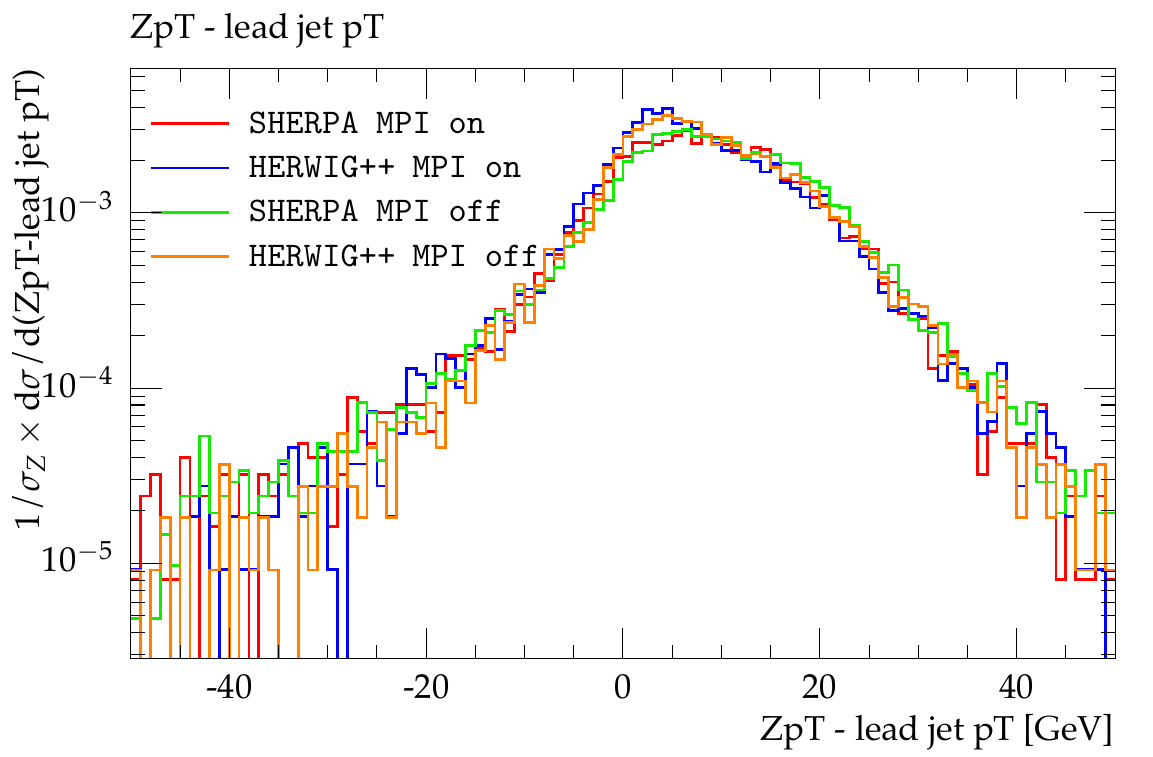}
\includegraphics[width=0.32\columnwidth]{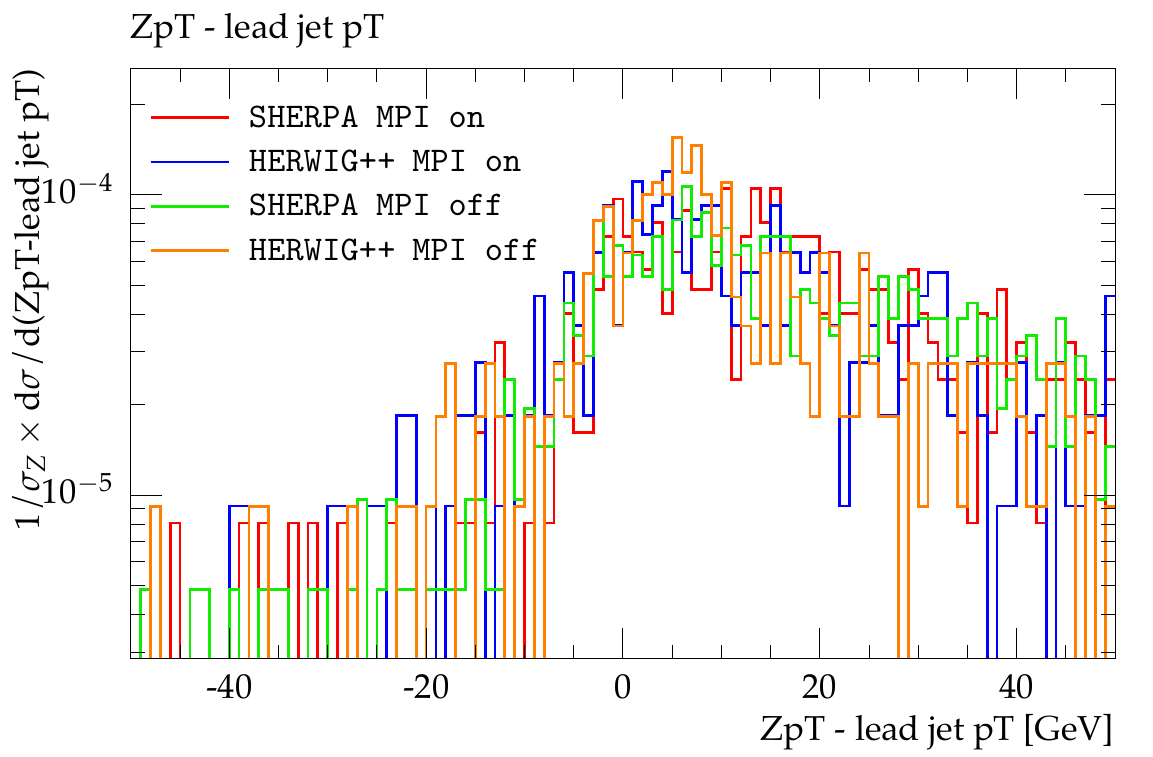}

\caption{\label{fig:slicesLeadJet}The difference between \emph{Z} $p_{\perp}$
and the leading jet $p_{\perp}$ for three regions of boson $p_{\perp}$:
\emph{Z} $p_{\perp}$< 30 GeV, 30 GeV < \emph{Z} $p_{\perp}$ < 100
GeV, \emph{Z} $p_{\perp}$ > 100 GeV. The cuts on the analysis are the same as Tevatron
analysis: $p_{\perp}(jet)>20\text{ GeV}$ and $|\eta|<2.1$. Generator
parameters are the same as in Figs. \ref{fig:3slicesMPIon} and \ref{fig:3slicesMPIoff}.}

\end{figure}

For a better comparison of the plots, we also make one-dimensional projections of the plots for
three different \emph{Z} $p_{\perp}$ regions (\emph{Z} $p_{\perp}$<
30 GeV, 30 GeV < \emph{Z} $p_{\perp}$ < 100 GeV, \emph{Z} $p_{\perp}$
> 100 GeV). These plots (Figs. \ref{fig:slicesAllJets} and \ref{fig:slicesLeadJet})
make more evident the difference that one can see, specially in the
lower $p_{\perp}$ region, between the generators with MPI
model turned on. ${\tt SHERPA}$ and ${\tt HERWIG++}$ disagree from
each other in the balance for both the sum of all jets $p_{\perp}$
and leading jet $p_{\perp}$. However, they agree reasonably when
there is no simulation of multiple parton interactions. This makes
more evident that these models should be better studied and implemented
in the context of the Drell Yan production, as have already been shown
concerning the ${\tt HERWIG++}$ generator in the UE analysis. 

\subparagraph{Leading Jet Rapidity for jet $p_{\perp}>\text{5}$ GeV }

\begin{figure}[!h]
\includegraphics[width=0.5\columnwidth]{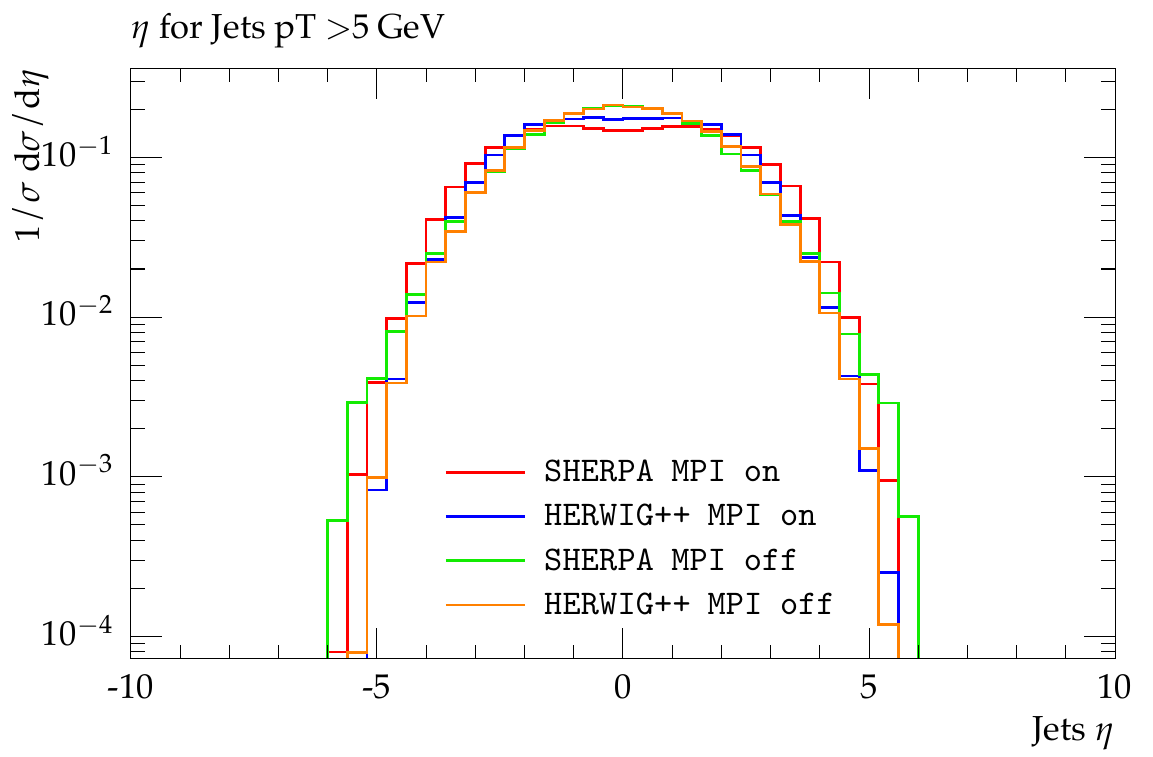}
\includegraphics[width=0.5\columnwidth]{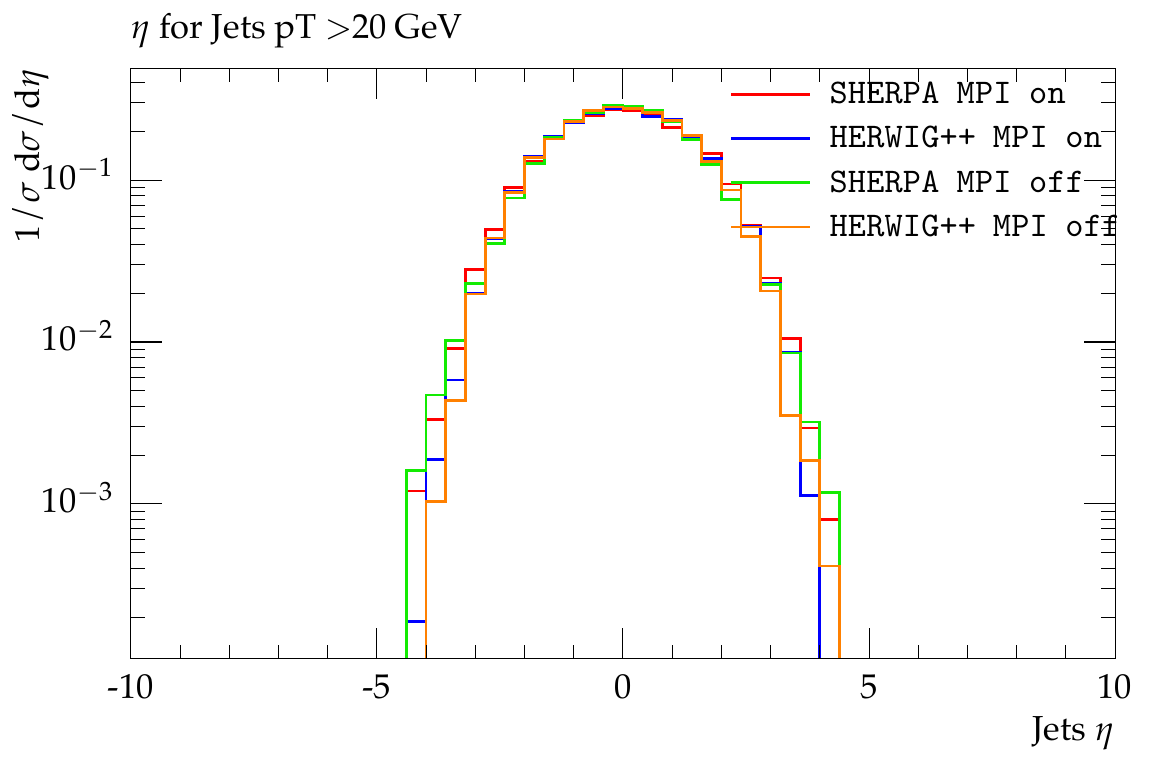}

\caption{\label{fig:Jet5GeVcut}Leading jet rapidity for jet cuts on tranverse
momentum of 5 GeV (left), and analysis cut of 20 GeV (right). ${\tt HERWIG++}$
has default parameters, ${\tt SHERPA}$ has ${\tt CTEQ6L1}$ PDF,
optimized ${\tt K\text{\_}PERP}$, and when MPI is on, scale
of 2.5 GeV.}

\end{figure}

The MPI model in each simulation is clearly producing additional
jets in \emph{Z} events, and we may expect these MPI jets
to have a different angular distribution to jets from the hard process.
So finally we check the rapidity distribution of all jets with $p_{\perp}$
> 5 GeV. It can be seen that, when the MPI
are turned on in the simulation, the shape is different in the central region:
around the pseudorapidity zero, we have a low bump, when compared to 
the plots without MPI simulation (Fig. \ref{fig:Jet5GeVcut}
left). 

The behaviour is almost gone when the cut on the jets is that
of the analysis (Fig. \ref{fig:Jet5GeVcut} right) - the MPI
model affects specially the production of particles in low transverse
momentum region. Once more, the MPI model in the generators
disagree between themselves. While such a low $p_{\perp}$ region
is experimentally difficult or impossible to access, it is clearly
a very sensitive region for MPI tuning and data with $p_{\perp}$
cuts as low as possible are very interesting.

\begin{thebibliography}{500}



\bibitem{Hoeche:2009rj}S.~Hoeche, F.~Krauss, S.~Schumann and F.~Siegert,
  ``QCD matrix elements and truncated showers,''
  JHEP {\bf 0905} (2009) 053
  [arXiv:0903.1219 [hep-ph]].
  
\bibitem{Gleisberg:2008ta}
  T.~Gleisberg, S.~Hoeche, F.~Krauss {\it et al.},
  ``Event generation with SHERPA 1.1,''
  JHEP {\bf 0902 } (2009)  007.
  [arXiv:0811.4622 [hep-ph]].
  
\bibitem{Bahr:2008pv}
  M.~Bahr, S.~Gieseke, M.~A.~Gigg {\it et al.},
  ``Herwig++ Physics and Manual,''
  Eur.\ Phys.\ J.\  {\bf C58 } (2008)  639-707.
  [arXiv:0803.0883 [hep-ph]].  
  
\bibitem{Seymour:1994df}
  M.~H.~Seymour,
  ``Matrix element corrections to parton shower algorithms,''
  Comput.\ Phys.\ Commun.\  {\bf 90 } (1995)  95-101.
  [hep-ph/9410414].

\bibitem{Frixione:2007vw}
  S.~Frixione, P.~Nason, C.~Oleari,
  ``Matching NLO QCD computations with Parton Shower simulations: the POWHEG method,''
  JHEP {\bf 0711 } (2007)  070.
  [arXiv:0709.2092 [hep-ph]].

\bibitem{Buckley:2010ar}
  A.~Buckley, J.~Butterworth, L.~Lonnblad {\it et al.},
  ``Rivet user manual,''

  [arXiv:1003.0694 [hep-ph]].


\bibitem{Affolder:1999jh}
  A.~A.~Affolder {\it et al.} [ CDF Collaboration ],
  ``The Transverse momentum and total cross-section of e+ e- pairs in the Z-boson region from p anti-p collisions at S**(1/2) = 1.8-TeV,''
  Phys.\ Rev.\ Lett.\  {\bf 84 } (2000)  845-850.
  [hep-ex/0001021].

\bibitem{:2007nt}
  V.~M.~Abazov {\it et al.} [ D0 Collaboration ],
  ``Measurement of the shape of the boson transverse momentum distribution in p anti-p ---> Z / gamma* ---> e+ e- + X events produced at s**(1/2) = 1.96-TeV,''
  Phys.\ Rev.\ Lett.\  {\bf 100 } (2008)  102002.
  [arXiv:0712.0803 [hep-ex]].


\bibitem{Abazov:2010kn}
  V.~M.~Abazov {\it et al.} [ D0 Collaboration ],
  ``Measurement of the normalized $Z/\gamma^* -> \mu^+\mu^-$ transverse momentum distribution in $p\bar{p}$ collisions at $\sqrt{s}=1.96$ TeV,''
  Phys.\ Lett.\  {\bf B693 } (2010)  522-530.
  [arXiv:1006.0618 [hep-ex]].


\bibitem{Martin:2009iq}
  A.~D.~Martin, W.~J.~Stirling, R.~S.~Thorne {\it et al.},
  ``Parton distributions for the LHC,''
  Eur.\ Phys.\ J.\  {\bf C63 } (2009)  189-285.
  [arXiv:0901.0002 [hep-ph]].


\bibitem{Martin:2004ir}
  A.~D.~Martin, R.~G.~Roberts, W.~J.~Stirling {\it et al.},
  ``Physical gluons and high E(T) jets,''
  Phys.\ Lett.\  {\bf B604 } (2004)  61-68.
  [hep-ph/0410230].


\bibitem{Affolder:2001xt}
  A.~A.~Affolder {\it et al.} [ CDF Collaboration ],
  ``Charged jet evolution and the underlying event in proton - anti-proton collisions at 1.8-TeV,''
  Phys.\ Rev.\  {\bf D65 } (2002)  092002.


\bibitem{Sjostrand:1987su}
  T.~Sjostrand, M.~van Zijl,
  ``A Multiple Interaction Model for the Event Structure in Hadron Collisions,''
  Phys.\ Rev.\  {\bf D36 } (1987)  2019.

\bibitem{Aaltonen:2010zza}
  T.~A.~Aaltonen {\it et al.} [ CDF Collaboration ],
  ``Measurement of $d\sigma/dy$ of Drell-Yan $e^+e^-$ pairs in the $Z$ Mass Region from $p\bar{p}$ Collisions at $\sqrt{s}=1.96$ TeV,''
  Phys.\ Lett.\  {\bf B692 } (2010)  232-239.
  [arXiv:0908.3914 [hep-ex], arXiv:0908.3914 [hep-ex]].

\bibitem{Lenzi:2009fi}
  P.~Lenzi, J.~M.~Butterworth,
  ``A Study on Matrix Element corrections in inclusive Z/ gamma* production at LHC as implemented in PYTHIA, HERWIG, ALPGEN and SHERPA,''
  
  [arXiv:0903.3918 [hep-ph]].


\bibitem{Abazov:2007jy}
  V.~M.~Abazov {\it et al.} [ D0 Collaboration ],
  ``Measurement of the shape of the boson rapidity distribution for p anti-p ---> Z/gamma* ---> e+ e- + X events produced at s**(1/2) of 1.96-TeV,''
  Phys.\ Rev.\  {\bf D76 } (2007)  012003.
  [hep-ex/0702025 [HEP-EX]].


\bibitem{Abazov:2006gs}
  V.~M.~Abazov {\it et al.} [ D0 Collaboration ],
  ``Measurement of the ratios of the Z/gamma* + >= n jet production cross sections to the total inclusive Z/gamma* cross section in p anti-p collisions at s**(1/2) = 1.96-TeV,''
  Phys.\ Lett.\  {\bf B658 } (2008)  112-119.
  [hep-ex/0608052].
  
\bibitem{Abazov:2009av}
  V.~M.~Abazov {\it et al.} [ D0 Collaboration ],
  ``Measurements of differential cross sections of Z/gamma*+jets+X events in proton anti-proton collisions at s**(1/2) = 1.96-TeV,''
  Phys.\ Lett.\  {\bf B678 } (2009)  45-54.
  [arXiv:0903.1748 [hep-ex]].
  
  
\bibitem{Abazov:2008ez}
  V.~M.~Abazov {\it et al.} [ D0 Collaboration ],
  ``Measurement of differential Z / gamma* + jet + X cross sections in p anti-p collisions at s**(1/2) = 1.96-TeV,''
  Phys.\ Lett.\  {\bf B669 } (2008)  278-286.
  [arXiv:0808.1296 [hep-ex]].

\bibitem{Campbell:2003hd}
  J.~M.~Campbell, R.~K.~Ellis, D.~L.~Rainwater,
  ``Next-to-leading order QCD predictions for W + 2 jet and Z + 2 jet production at the CERN LHC,''
  Phys.\ Rev.\  {\bf D68 } (2003)  094021.
  [hep-ph/0308195].


\end{thebibliography}
\end{document}